\UseRawInputEncoding
\documentclass[aps,pra,twocolumn,longbibliography]{revtex4-1}
\usepackage[colorlinks=true, citecolor=red, urlcolor=blue ]{hyperref}
 \usepackage[utf8]{inputenc}
\usepackage{graphicx}   
\usepackage{dcolumn}    
\usepackage{bm}         
\usepackage{hyperref}   
\usepackage{xcolor}     
\usepackage{amsmath}
\usepackage{relsize}
\usepackage{comment}
\usepackage[caption=false]{subfig} 
\usepackage{ragged2e}
\usepackage{booktabs}
\usepackage{indentfirst}
\usepackage{epsfig}
\usepackage[normalem]{ulem}
\usepackage{amssymb}
\usepackage{bm}

\begin{document}



\title{\texorpdfstring{Quantum refrigerator embedded in spin-star environments:\\Scalings of temperature and refrigeration time}{Quantum refrigerator embedded in spin-star environments: Scalings of temperature and refrigeration time}}

\author{Sukrut Mondkar}
\email{sukrutmondkar@hri.res.in}
\affiliation{Harish-Chandra Research Institute, A CI of Homi Bhabha National Institute, Chhatnag Road, Jhusi,
Prayagraj (Allahabad) 211019, India}

\author{Aparajita Bhattacharyya}
\email{aparajitabhattacharyya@hri.res.in}
\affiliation{Harish-Chandra Research Institute, A CI of Homi Bhabha National Institute, Chhatnag Road, Jhusi,
Prayagraj (Allahabad) 211019, India}

\author{Ujjwal Sen}
\email{ujjwal@hri.res.in}
\affiliation{Harish-Chandra Research Institute, A CI of Homi Bhabha National Institute, Chhatnag Road, Jhusi,
Prayagraj (Allahabad) 211019, India}


\begin{abstract}
We examine a quantum absorption refrigerator that comprises three qubits, each of which is connected with a separate spin-star environment{, with the three qubit-bath units coupled through an effective six-body interaction}. The refrigerator exhibits the feature of transient cooling, i.e., lowering of the temperature of the first qubit in sufficiently small timescales{, rather than steady-state refrigeration}. A key advantage of our model is that the symmetries of the Hamiltonian enable a {semi-analytic} solution of the reduced density matrices of the refrigerator qubits, even in the presence of a large number of environmental spins. We derive the condition for autonomous refrigeration and analyze how the optimal cold-qubit temperature scales with the number of bath {spins}. We find a power-law scaling towards a constant asymptotic value. We also find the scaling of the minimum time required for {optimal} cooling as a function of the number of bath spins. 
{Furthermore, we quantify the non-Markovianity of the cold-qubit dynamics using a restricted Breuer-Laine-Piilo information-backflow measure and observe that stronger backflow correlates with lower transient minimum temperatures across the sampled parameter regime. The transient-cooling performance is found to be robust under broad parameter variations. Compared to a conventional Markovian three-qubit refrigerator, the CSQAR achieves lower cold-qubit temperatures on shorter timescales. We further analyze the heat currents associated with the three qubits and their respective baths.}
\end{abstract}

\maketitle

\section{Introduction}

The commencement and progress of quantum thermodynamics~\cite{PhysRevLett.85.1799, Kosloff_2013, PhysRevE.92.042126, Gelbwaser_Klimovsky_2015, PhysRevE.92.042161, Millen_2016, Vinjanampathy_2016, Goold_2016, Benenti_2017, Binder:2018rix, deffner2019} has instigated the development of quantum thermal devices, which have become 
immensely significant in modern quantum technologies, especially in engineering small-scale devices and quantum circuits.
Extensive investigations in this arena have led not only to the extension of the fundamental principles of classical thermodynamics to the quantum domain, but also to insights into
the effects of quantum mechanical features such as entanglement and coherence in the performance of these miniaturized devices.
In recent years, the exploration and designing of quantum thermal devices  
such as quantum heat engines~\cite{PhysRevE.64.056130, PhysRevE.68.016101, PhysRevX.5.031044, Seah_2018, Mitchison_2019, three_stroke}, quantum batteries~\cite{PhysRevE.87.042123, campaioli2018, Bhattacharjee_2021}, quantum refrigerators~\cite{PhysRevLett.105.130401, Skrzypczyk_2011, PhysRevE.85.051117, PhysRevLett.108.070604, Kosloff_2014, PhysRevLett.123.170605}, quantum transistors~\cite{PhysRevLett.116.200601, Zhang_2018, su2018quantumcoherencethermaltransistors}, and quantum diodes~\cite{YUAN2021106086} have gained utmost importance, which focuses on providing advantages over their classical analogues. 

The introduction of \emph{quantum absorption refrigerators} (QAR) by Linden \emph{et al.}~\cite{PhysRevLett.105.130401} provided a significant impetus 
in the realm 
of 
miniaturization of quantum technologies. It was 
demonstrated in~\cite{PhysRevLett.105.130401, Skrzypczyk_2011} that quantum thermal devices of size as small as a mere three qubits 
can function as a cooling apparatus, with its mechanism being governed 
by the laws of quantum mechanics.
Such devices can 
operate not only without any external source of 
energy, but are also 
more efficient 
than their classical counterparts. Such a minimal model of 
QAR~\cite{PhysRevLett.105.130401, Skrzypczyk_2011, PhysRevE.85.051117, PhysRevE.87.042131, PhysRevE.89.032115, PhysRevE.92.062101} consists of three open qubits interacting with each other via energy-preserving transitions that facilitate the autonomous refrigeration process. It functions without any external 
energy input and relies solely on heat exchanges between different thermal reservoirs.

The performance of quantum thermal devices is inevitably influenced by the environments with which they are connected.
The environments can be broadly classified as Markovian or non-Markovian based on the validity of Born, Markov and secular approximations~\cite{BreuerOQS, AlickiOQS, RivasHuelgaOQS, lidar2020}. 
The works on QAR have hitherto been mostly focused on open qubits attached to Markovian environments~\cite{PhysRevE.90.052142, PhysRevE.92.012136, PhysRevE.96.052126, PhysRevE.96.012122, Du_2018, Das_2019, Naseem_2020, PhysRevE.101.012109,PhysRevA.104.042208, okane2022, PhysRevA.105.022214, ray2023kerr, PhysRevE.107.034128}. 
However, the assumptions of Markovianity impose restrictions on the system dynamics, which arise in specialized situations when the system does not possess any memory of its past events. In most of the 
realistic cases,
the most general quantum evolution of a system connected to an environment is non-Markovian in nature.
One of the prime examples of non-Markovian environments are spin environments, which act as a backbone of solid-state quantum technology platforms. Furthermore, not all non-Markovian environments admit a Markovian limit, spin-star environments being a notable example~\cite{PhysRevB.70.045323}. Since Markovian environments are 
prototypes of idealized situations and
are elusive in practical scenarios, it is important to investigate how non-Markovianity impacts the performance of quantum thermal devices such as quantum refrigerators. 
{However, non-Markovian generalization of QAR
is an arena that is less 
explored~\cite{Wiedmann2021,PhysRevA.110.022220,Bhattacharyya:2023pak}.}
Notably, non-Markovian environments have been shown to have an advantage over Markovian environments in various quantum information and quantum thermodynamics tasks~\cite{nm1,nm2,nm3,PhysRevE.106.014114, ELALLATI2024129316,nm4}, and in particular in quantum refrigeration~\cite{Bhattacharyya:2023pak}.
It is potentially possible to engineer specific non-Markovian environments which 
provide an advantage in certain quantum thermodynamic tasks. 

There exist scenarios where the attainment of equilibrium occurs at very large timescales, or sometimes, rapid cooling is necessary. In such situations,
transient cooling offers a more viable approach to refrigeration than steady-state cooling. Quantum systems often require rapid and localized cooling to suppress decoherence. Despite its practical relevance, short-time refrigeration in QAR remains largely unexplored. 
There are a few studies~\cite{PhysRevE.92.062101, Das_2019, Mitchison_2015, PhysRevE.87.042131, Bhattacharyya:2023pak} that have addressed this aspect, demonstrating that transient cooling can achieve lower temperatures than steady-state cooling. 

In this work, we introduce a QAR in a non-Markovian setting by incorporating a central-spin model~\cite{PhysRevB.70.045323} for environmental interactions. The central-spin model, also known as the spin-star model, describes a central quantum system (a qubit) coupled to a surrounding bath of spin degrees of freedom. We refer to this model of QAR as a central-spin quantum absorption refrigerator (CSQAR), where each of the three qubits of the quantum refrigerator is modeled as central spins of three different central-spin systems. The central-spin model is highly relevant to quantum dots~\cite{PhysRevLett.88.186802, PhysRevB.66.245303, PhysRevB.65.205309, PhysRevB.67.195329, Schliemann-JPhysCondMatt_2003, Schliemann_2003, PhysRevB.70.195340, PhysRevB.73.241303, RevModPhys.79.1217, PhysRevB.78.085315, PhysRevB.79.115320, PhysRevB.79.245314, PhysRevLett.102.057601, ActaPhysPolA119576, RevModPhys.85.79, PhysRevB.90.155117, Yang_2017}, nitrogen-vacancy (NV) centers in diamonds~\cite{doi:10.1126/science.1131871, doi:10.1126/science.1155400, PhysRevB.84.104301, PhysRevLett.106.217205, PhysRevB.85.115303, PhysRevLett.111.067601, PhysRevB.90.075201, doi:10.1126/sciadv.aat8978}, nuclear magnetic resonance (NMR)~\cite{PhysRevA.106.032435}, superconducting systems, and other solid-state quantum platforms that exhibit interactions with rotational symmetry about the central spin. 
Notably, central-spin models have been explored in various contexts, including quantum batteries~\cite{PhysRevB.104.245418}, quantum heat engines~\cite{T_rkpen_e_2017}, quantum channel capacities~\cite{PhysRevA.81.062353}, quantum state transfer~\cite{PhysRevA.73.062321, Deng-Hong-Liang_2008, Yung_2011}, quantum communication and cloning~\cite{Deng-Hong-Liang_2008, Deng_2008}, and quantum phase transitions~\cite{Deng_2008, Deng-Hong-Liang_2008-qpt}, but their role in quantum refrigeration remains 
{relatively less explored. 
Recent works have begun to address this direction. See, for example, Ref.~\cite{Spiecker:2022huj}, which provides an experimental demonstration of central-spin based thermalization and cooling effects, while Ref.~\cite{Janovitch:2025ncz} explores active reservoir engineering in central-spin models, including protocols that cool a spin environment by shifting its magnetization distribution toward lower-energy sectors, whereas the present work concerns autonomous refrigeration of a cold qubit in a CSQAR.}

One of the key advantages of our model is its ability to retain certain analytical tractability despite the presence of many environmental spins. The symmetries of the Hamiltonian allow us to derive a semi-analytical solution for the reduced density matrices of the system qubits, enabling us to study the cooling dynamics even in the presence of a large number of bath spins. 
{Spin-star environments can naturally give rise to memory effects because the central system interacts coherently with structured collective degrees of freedom of the bath. In the present work, we refer to non-Markovianity in the information-backflow sense. Importantly, this should not be understood as attributing the backflow merely to the  finite size of the environment. Indeed, the cooling behavior studied here persists for large bath sizes and admits a well-defined large-\(N\) scaling analysis. Our point is rather that the structured spin-star character of the environments supports non-Markovian memory effects relevant for transient cooling. The CSQAR considered here therefore exhibits transient refrigeration~\cite{Bhattacharyya:2023pak}, but it does not attain a steady-state refrigeration regime, potentially due to the absence of thermalization in central-spin models~\cite{Wang_2013}. This, in turn, provides a natural platform to investigate the relatively unexplored regime of transient cooling in QARs.}

We investigate the scaling of the cold-qubit temperature with the number of bath spins, utilizing the ability of our approach to handle large system sizes. By numerically optimizing the refrigerator Hamiltonian parameters, we study how the optimal cold-qubit temperature varies as a function of the bath size. Additionally, we find the scaling of the minimum time required to reach the minimum temperature as a function of the number of bath spins. 
{We further quantify the non-Markovianity of the cold-qubit dynamics using a restricted Breuer-Laine-Piilo (BLP)-type information-backflow measure and analyze its correlation with refrigeration performance. Within the sampled parameter regime, we find that stronger information backflow is correlated with lower transient minimum temperatures and, more weakly, with shorter cooling times. We also examine the robustness of transient refrigeration under broad variations of the model parameters. Furthermore, we compare the full CSQAR architecture with the conventional Markovian three-qubit QAR and show that it attains lower cold-qubit temperatures on shorter timescales.} 
{We also analyze the heat currents flowing through various components of the CSQAR and argue that the heat current through the cold-qubit environment can be regarded as an approximate indicator of cooling power of the CSQAR.}

The central-spin model is 
well investigated experimentally, even with a large number of 
environmental spins~\cite{SUTER201750, PhysRevB.95.195308, PhysRevLett.120.180602, PhysRevX.9.031045, Mahesh_2021}. Furthermore, it is demonstrated in~\cite{Arenz_2014} that in the central-spin model, the central spin is fully controllable independent of the number of bath spins. 
{Since the CSQAR is based on the central-spin model of open quantum systems, and since the effective six-body interaction is understood as an engineered collective interaction rather than a fundamental microscopic coupling, we regard the model as potentially experimentally realizable.}

{We remark that a related study of QAR embedded in spin-star environments was carried out in Ref.~\cite{Bhattacharyya:2023pak}. However, the focus there was different from that of the present work. In Ref.~\cite{Bhattacharyya:2023pak}, the emphasis was on how the performance of a three-qubit QAR changes when some of the local environments are non-Markovian and others are Markovian. In contrast, here we consider the regime in which all three local environments are non-Markovian spin-star baths, and focus on how the minimum cold-qubit temperature and the minimum time required to attain it scale with the bath size. In addition, the fully non-Markovian spin-star case considered in Ref.~\cite{Bhattacharyya:2023pak} retained the standard three-body interaction of the conventional QAR, whereas the CSQAR studied here involves the effective six-body interaction of Eq.~\eqref{eq:H_int-cs}, which couples the three qubit-bath units in the invariant collective-spin basis. Moreover, Ref.~\cite{Bhattacharyya:2023pak} was restricted to two spins in each local bath and relied on a fully numerical treatment of the dynamics. In the present work, by exploiting the symmetries of the CSQAR Hamiltonian, we develop a semi-analytical approach that allows us to access much larger bath sizes efficiently, and thereby to determine the scaling behaviour with $N$ as well as estimate the $N \rightarrow \infty$ limits of the minimum cold-qubit temperature and the corresponding minimum cooling time. }

The structure of the paper is as follows. In Section~\ref{sec:model}, we first review 
the conventional  QAR 
and then introduce CSQAR in the non-Markovian central-spin framework. In Section~\ref{sec:solution}, we first derive the reduced density matrix for a single central-spin system and then analyze the temperature dynamics of the cold qubit of the CSQAR. 
Section \ref{sec:heat-currents} presents the heat currents analysis. Section~\ref{sec:scaling} presents an analysis of the scaling behavior of the cold qubit temperature and refrigeration time with respect to the number of bath spins. {In Section~\ref{sec:non-Markov}, we quantify the non-Markovianity of the cold-qubit dynamics and analyze its correlation with refrigeration performance.} Section \ref{sec:Markov} is 
concerned with the comparison of {the full} CSQAR {architecture} with a QAR in the Markovian setting. Finally, we conclude with a discussion of our findings and their implications in Section~\ref{sec:conclusion}. 

\section{The Model}\label{sec:model}

\subsection{Quantum absorption refrigerator}
In this subsection, we introduce the concept of quantum absorption refrigerator (QAR) by reviewing the prototypical model introduced by Linden \emph{et} al. in the seminal paper~\cite{PhysRevLett.105.130401}.
A QAR~\cite{PhysRevLett.105.130401, Skrzypczyk_2011, PhysRevE.85.051117} comprises three qubits, each of which is in contact with a 
local environment. The first of the three qubits $(i=1)$, 
referred to as the \emph{cold qubit}, is the target 
of refrigeration.
The second qubit $(i=2)$, 
often specified as the \emph{room qubit}, takes heat from the cold qubit and dissipates it into its environment, much like a spiral heat exchanger in a conventional heat engine. The third qubit $(i=3)$, called the \emph{hot qubit}, plays the role of the \emph{engine}, enabling the quantum refrigerator to operate autonomously, i.e., without any external source of work. 

Each of the three qubits is in thermal equilibrium with its local thermal 
environment at the initial time. The three environments are maintained at temperatures $\tau_1, \; \tau_2, $ and $\tau_3$ respectively with $\tau_1 \leq \tau_2 \leq \tau_3$. The environment of the room qubit is maintained at room temperature $\tau_2=\tau_r$, whereas the environment of the hot qubit is maintained at a 
temperature, higher than the room temperature, i.e. $\tau_3=\tau_h > \tau_r$. 
The QAR operates without requiring external work because it harnesses free energy from two thermal baths maintained at different temperatures.
Note that $\tau_1, \; \tau_2, $ and $\tau_3$ are dimensionless temperatures which are related to the actual temperatures $\tilde{\tau}_1, \; \tilde{\tau}_2, $ and $\tilde{\tau}_3$ by $\tau_1 = \frac{ k_B \tilde{\tau}_1}{\hbar K}$, $\tau_2 = \frac{k_B \tilde{\tau}_2}{\hbar K}$, and $\tau_3 = \frac{k_B \tilde{\tau}_3}{\hbar K}$. $K$ is a constant of unit magnitude with dimensions of $time^{-1}$. Throughout the paper, various physical quantities are expressed in units of $K$.

\begin{figure*}[t]
    \centering
    \subfloat[Conventional three-qubit QAR with Markovian reservoirs.\label{fig:markovian_qar_schematic}]{
        \includegraphics[width=0.4\linewidth]{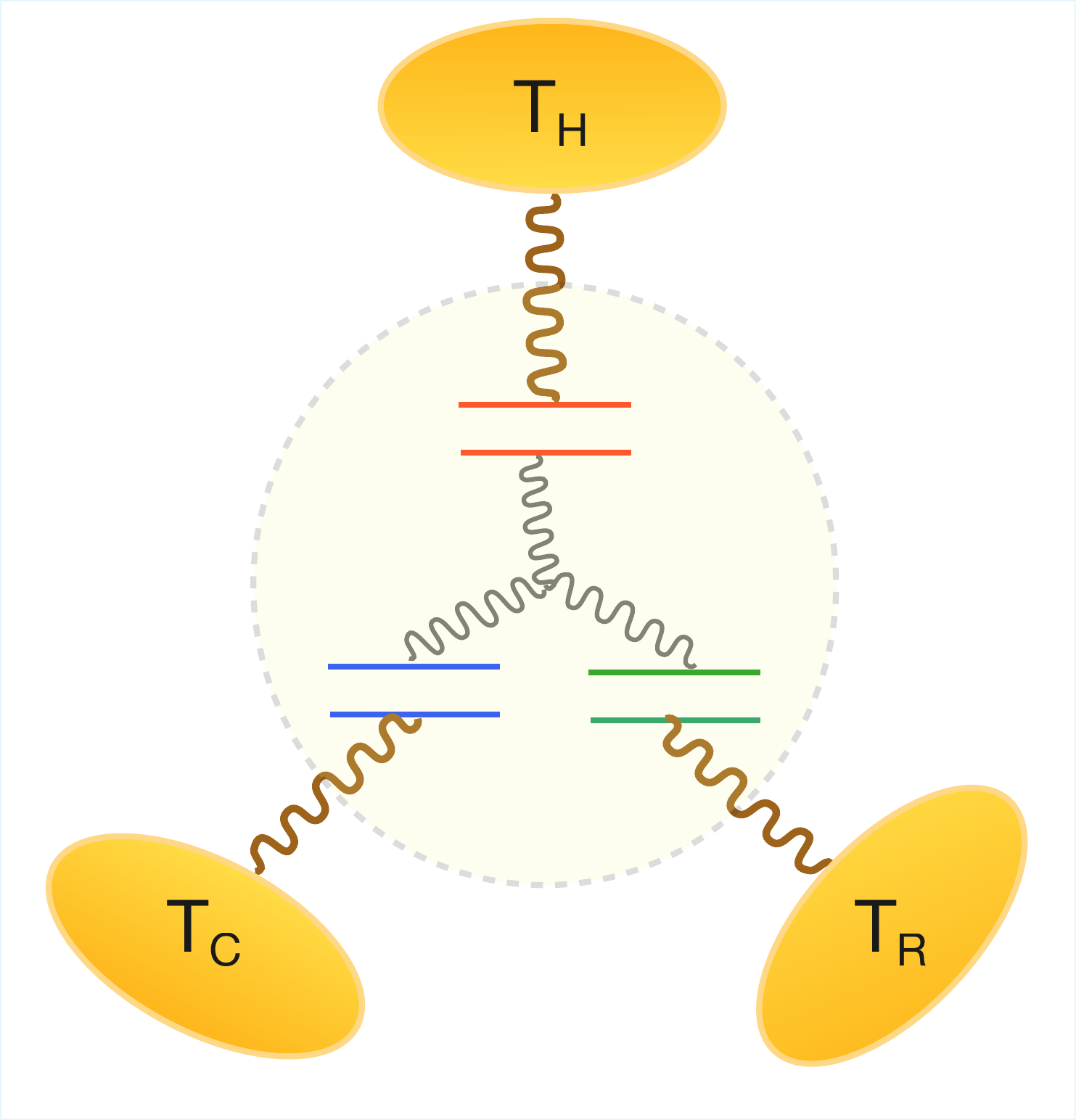}
    }
    \hfill
    \subfloat[Central-spin quantum absorption refrigerator with spin-star environments.\label{fig:csqar_schematic}]{
        \includegraphics[width=0.48\linewidth]{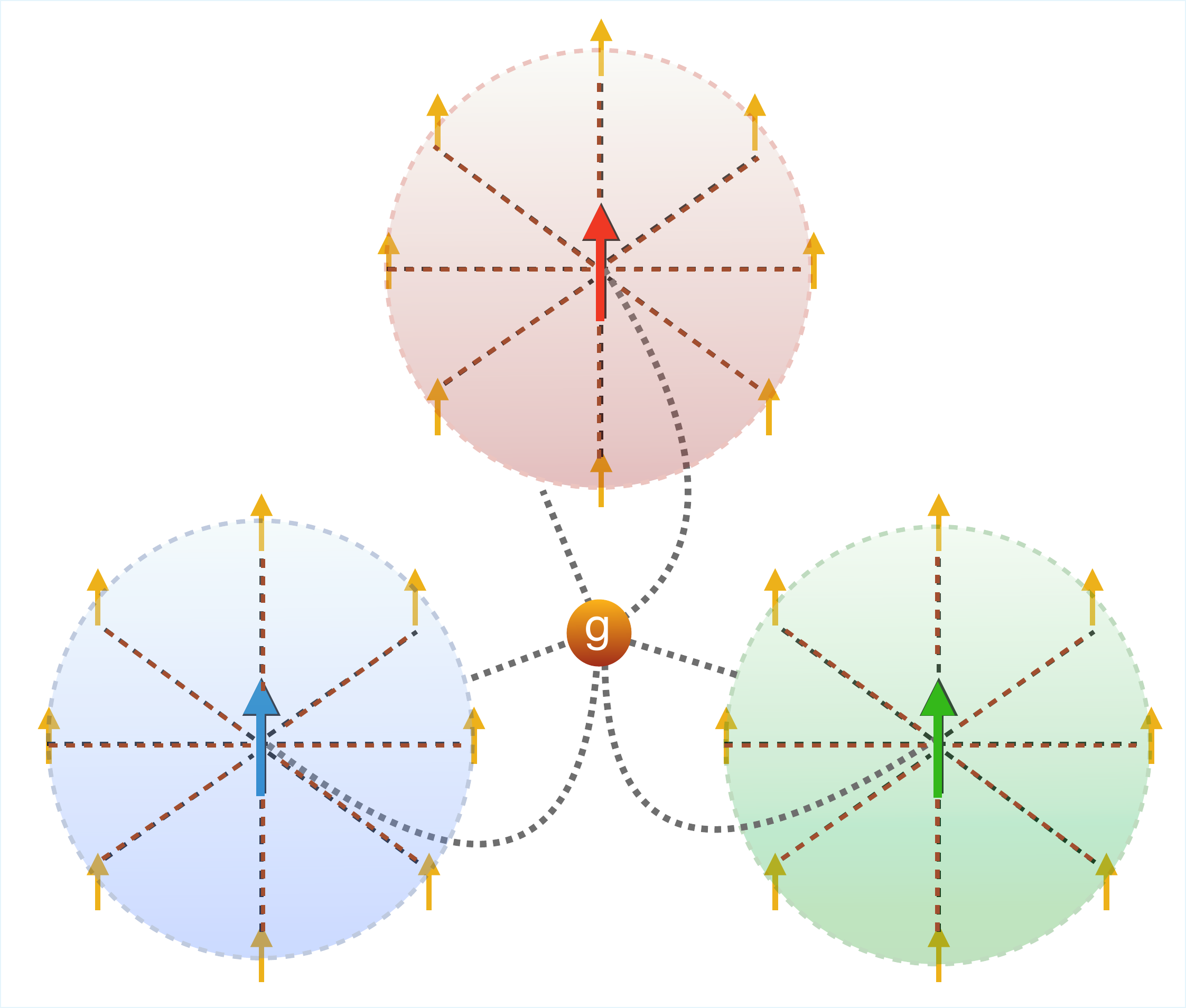}
    }
    \caption{{Schematic comparison of the two refrigerator models discussed in this work. 
    (a) Conventional three-qubit quantum absorption refrigerator (QAR), in which the cold (blue), room (green), and hot (red) qubits are coupled to local Markovian reservoirs at temperatures $T_C$, $T_R$, and $T_H$, respectively, and interact through the standard three-body transition, Eq.~\eqref{eq:HS-qref-M-third}, introduced in Ref.~\cite{PhysRevLett.105.130401}. 
    (b) Central-spin quantum absorption refrigerator (CSQAR), where each refrigerator qubit is embedded in an independent spin-star environment and the three qubit--bath units are coupled through an effective six-body interaction of Eq.~\eqref{eq:H_int-cs} of strength $g$. 
    The figure highlights the main structural difference between the two settings: in the conventional QAR the reservoirs are passive Markovian baths, whereas in the CSQAR the local environments are explicit dynamical spin systems that participate in the refrigerator dynamics.}}
    \label{fig:QAR_CSQAR_schematic}
\end{figure*}

The composite 
Hamiltonian comprising the system and environment of the three-qubit QAR is 
\begin{equation}\label{eq:H-qref-M}
    \widetilde{H} = \widetilde{H}_S + \widetilde{H}_B + \widetilde{H}_{SB} + \widetilde{H}_{\rm{int}},
\end{equation}
where $\widetilde{H}_S$ and $\widetilde{H}_B$ are the sum of the local Hamiltonians of the three system qubits and their local environments, respectively. 
$\widetilde{H}_{SB}$ includes interaction between the qubits of the system and their respective environments. 
The explicit form of  
$ \widetilde{H}_{\rm{int}}$, as 
introduced in~\cite{PhysRevLett.105.130401}, and 
$\widetilde{H}_S$ is given by 
{
\begin{subequations}\label{eq:HS-qref-M}
\begin{align}
    \widetilde{H}_S &= \sum_{i=1}^3 \widetilde{H}_S^{(i)}  \;\;\; \text{with} \;\;\; 
    \widetilde{H}_S^{(i)} = K  \varepsilon_i S^z_i.
    \;\;\; \text{and} \label{eq:HS-qref-M-second} \\
     \widetilde{H}_{\rm{int}} &= K  \hbar  \tilde{g}   \left( \left| -1/2; \, \, 1/2; \, \,  -1/2\right\rangle  \left\langle 1/2; \, \, -1/2; \, \, 1/2  \right| \right. \nonumber \\ 
     & + \left. \left|1/2; \, \, -1/2; \, \,  1/2 \right\rangle  \left\langle -1/2; \, \, 1/2; \, \, -1/2  \right| \right) \label{eq:HS-qref-M-third},
\end{align}
\end{subequations}
}
where
\noindent $\widetilde{H}^{(i)}_S$ is the
local Hamiltonian 
of each system qubit 
and $S^z_i= \hbar \sigma^z_i/2$, where $\sigma^z_i$ is the Pauli-$Z$ operator 
acting on the $i^{\rm{th}}$ qubit. The ground and excited states of each qubit are denoted by {$| -1/2 \rangle$} and {$| 1/2 \rangle$}, respectively. 
The coupling strength, ${\tilde{g}} $
of the three-body interaction, $\widetilde{H}_{\rm{int}}$,
is considered to be weak compared to the free Hamiltonian, i.e., $\varepsilon_i \gg {\tilde{g}} $~\cite{PhysRevLett.105.130401}. The parameters, $\varepsilon$ and ${\tilde{g}} $, are dimensionless quantities. 
The condition for autonomous refrigeration corresponding to the three-body interaction is $\varepsilon_2 = \varepsilon_1 + \varepsilon_3$. {This is because under this condition, the states {$| -1/2; \, \,  1/2; \, \,  -1/2 \rangle$ and $ | 1/2; \, \,  -1/2; \, \,  1/2 \rangle$} become degenerate. Therefore the transition {$| -1/2; \, \,  1/2; \, \,  -1/2 \rangle \leftrightarrow | 1/2; \, \,  -1/2; \, \,  1/2 \rangle$} costs no energy. {Furthermore, as $\varepsilon_3 > 0$, the autonomous refrigeration condition implies a biasing condition, $\varepsilon_2 > \varepsilon_1$, making the transition $ | 1/2; \, \,  -1/2; \, \,  1/2 \rangle \rightarrow | -1/2; \, \,  1/2; \, \,  -1/2 \rangle$ more favored than $ | -1/2; \, \,  1/2; \, \,  -1/2 \rangle \rightarrow | 1/2; \, \,  -1/2; \, \,  1/2 \rangle$.}} Together with the source of free energy provided by $\tau_h > \tau_r$, this ensures a self-sustained cooling process~\cite{PhysRevLett.105.130401, Skrzypczyk_2011}.

Initially, each of the three system qubits is in its respective thermal state, with the product initial state of the three-qubit refrigerator being given by
\begin{equation}\label{eq:qref-M-rhoi}
    \rho_{\text{ini}} = \rho_{1;\text{ini}} \otimes  \rho_{2;\text{ini}} \otimes \rho_{3;\text{ini}}
\end{equation}
\noindent where $\rho_{i;\text{ini}}$ is the thermal state of the $i^{th}$ qubit at temperature $T^{\text{ini}}_i =\tau_i$ given by $\rho_{i;\text{ini}} = Z_i^{-1} \text{exp} \left( - \beta_i \varepsilon_i \sigma_i^z / 2\right)$. The quantity, $Z_i = \text{Tr} \left[ \text{exp} \left( - \beta_i \varepsilon_i \sigma_i^z / 2\right)\right] $ is the 
partition function corresponding to the $i^{\text{th}}$ qubit, and $\beta_i = 1/\tau_i$ is the inverse temperature of the corresponding reservoir. We initially set the first qubit at room temperature, i.e., $\tau_1 = \tau_2 = \tau_r$.
The interaction between the qubits, $\widetilde{H}_{\rm{int}}$, is activated at time $t>0$. Note $t$ is dimensionless time which is related to the actual time $\tilde{t}$ by $t = K \tilde{t}$. {Once the interaction between the qubits is switched on for \(t>0\), the reduced dynamics of the three-qubit working medium is governed by the Gorini-Kossakolski-Sudarshan-Lindblad (GKSL) quantum master equation~\cite{Breuer2002,Alicki2007,Rivas2012,Lidar2019},}
\begin{align}\label{eq:QAR-Markov-Master-Eqn}
    \frac{\partial \rho_s(t)}{\partial t} = \mathcal{L}(\rho_s(t)) = - \frac{\mathrm{i}}{K} [\widetilde{H}_S(t), \rho_s(t)] + \sum_{i=1}^3 \frac{\hbar}{K} \mathcal{D}_i (\rho_s(t))
\end{align}
{Here, \(\rho_s(t)\) denotes the reduced density operator of the composite three-qubit system at the dimensionless time \(t\). The term \(\mathcal{D}_i(\rho_s(t))\) describes the dissipative contribution associated with the environment coupled to the \(i\)th qubit. Its explicit structure is determined by the physical nature of that environment, and therefore it can take different forms depending on whether the corresponding bath is Markovian or non-Markovian. In the standard formulation of a QAR, the three local reservoirs are most often modeled as Markovian environments~\cite{Das_2019}. We assume that, at the initial time, the reduced density matrices of the three individual qubits are diagonal in the eigenbases of their respective local Hamiltonians \(\widetilde{H}_{S_i}\). Since the Markovian reservoirs considered here do not generate coherences in these local energy bases, the single-qubit reduced states
\[
\rho_i(t)=\operatorname{Tr}_{j,k}[\rho_s(t)],\qquad j,k\neq i,\quad i,j,k\in\{1,2,3\},
\]
remain diagonal in the same bases throughout the evolution. This preservation of diagonality enables us to consistently assign effective local temperatures to the individual qubits, as described below.}

The time-evolved reduced density matrix of the cold qubit can be written as~\cite{PhysRevLett.105.130401}
{
\begin{equation}\label{eq:qref-M-rho-1}
    \rho_1(t) = r_1(t) | -1/2 \rangle \langle -1/2 | + \left( 1 - r_1(t) \right) | 1/2 \rangle \langle 1/2 |, 
\end{equation}
}
\noindent where $r_1(t)$ is the ground state population at time $t$. The local time-dependent temperature, $T_1(t)$ of the cold qubit is 
defined through the relation, $r_1(t) = Z_1(t)^{-1} \left( \text{exp} \left(  \varepsilon_1/ 2 T_1(t)\right) \right)$, with $Z_1(t) = \text{Tr} [  \text{exp} \left( - \varepsilon_1 \sigma_1^z/ 2 T_1(t)\right)  ] $. By inverting this 
equation, we obtain
\begin{equation}\label{eq:qref-M-T1}
    T_1(t) = \varepsilon_1 \left[ \text{ln} \left( \frac{r_1(t)}{1 - r_1(t)} \right)  \right]^{-1}
\end{equation}
$T_1(t)$ is dimensionless temperature and is related to the actual temperature $\tilde{T}_1(t)$  by $T_1(t) = \frac{k_B \tilde{T}_1(t)}{ \hbar K}$. The local temperatures of the other two qubits can be defined similarly. 
A lowering of the final temperature of the cold qubit at a later time $t$, compared to its initial temperature, i.e. $T_1(t) < \tau_1$, indicates refrigeration of the cold qubit.

\subsection{Central-spin quantum absorption refrigerator }\label{sec:model:subsec:csqar}

In this work, we analyze a three-qubit QAR where each qubit is connected to a spin-environment, specifically focusing on the scaling of the cold-qubit temperature as a function of the number of environment spins. Since there are no assumptions of Markovianity, the model depicts purely non-Markovian effects. We employ a semi-analytic approach to solve the dynamics of the three-qubit refrigerator to obtain the temperature of the cold qubit.

In the present work, we consider a specific model of a three-qubit QAR where 
each local environment
connected to every qubit is a
spin-star one.
Each qubit in the refrigerator acts as the central spin of an independent spin-star system, with its surrounding spins representing the corresponding local environment. The Hamiltonian of the central-spin quantum absorption refrigerator (CSQAR) is given by
{\small 
\begin{subequations}\label{eq:HS-qref}
\begin{align}
     H &= H_S + H_B + H_{SB} + H_{\rm{int}}, \; \text{with} \label{eq:H-qref-first} \\
     H_S &= \sum_{i=1}^3 H_S^{(i)}, \;  H_B = \sum_{i=1}^3 H_B^{(i)} \; \text{and} \;  H_{SB} = \sum_{i=1}^3 H_{SB}^{(i)}, \; \text{where}  \label{eq:HS-qref-second}\\
    H_S^{(i)} &= K \varepsilon_i S^z_i, \;    H_B^{(i)} = K E_i J^z_i, \;   H_{SB}^{(i)} = \frac{K}{\hbar} A_i \left( S^+_i J^-_i + S^-_i J^+_i \right). \label{eq:HS-qref-third} 
\end{align}
\end{subequations}
}
Here $S^\alpha_i = \hbar \sigma^\alpha_i /2$ are the spin operators of $i^{\rm{th}}$ qubit and $J^\alpha_i = \hbar \sum_{k=1}^{N_i\\} \sigma^\alpha_k/2$ are the collective spin operators of the local spin environment of the $i^{\rm{th}}$ qubit, consisting of $N_i$ spin-$\frac{1}{2}$s, with $\alpha = x,y,z$. 
The term, $H_{SB}^{(i)}$, 
is the Heisenberg XY coupling between the $i^{\rm{th}}$ qubit, and it's local bath, 
where $S^{\pm}_i = S^x_i \pm i S^y_i$, $J^{\pm}_i = J^x_i \pm i J^y_i$. Heisenberg XY coupling has been found to be an effective Hamiltonian for interactions of quantum dot systems~\cite{PhysRevLett.83.4204}. 
{$H_{\rm{int}}$ denotes an effective six-body interaction that plays, in the CSQAR, a role analogous to that of the three-body interaction in Eq.~\eqref{eq:HS-qref-M-third} for the conventional three-qubit QAR. Its explicit form is given later in Eq.~\eqref{eq:H_int-cs} of Sec.~\ref{sec:solution:subsec:dynamics-CSQAR}. We do not write this interaction explicitly at this stage because its structure is most naturally expressed in the invariant-subspace basis, which is introduced in Sec.~\ref{sec:solution:subsec:dynamics-CSQAR}. In particular, once this basis is introduced, it becomes transparent how the three-body transition in Eq.~\eqref{eq:HS-qref-M-third} is generalized to the effective six-body transition in Eq.~\eqref{eq:H_int-cs}.
The interaction is \textit{effectively} six-body because, the collective spin of each bath is treated as a single large spin. A key difference from Eq.~\eqref{eq:HS-qref-M-third} is that the effective interaction in the CSQAR couples not only the three refrigerator qubits, but also the corresponding collective states of their local spin environments.} 
{A schematic comparison between the 
CSQAR and a conventional 
QAR operating under a Markovian environment is presented in Fig.~\ref{fig:QAR_CSQAR_schematic}.}

The Hamiltonian in~\eqref{eq:HS-qref} has the symmetries 
\begin{subequations}\label{eq:symm}
\begin{align}
    [ H , S^z_i + J^z_i] &= 0,  \qquad  i=1,2,3 \label{eq:symm-first}\\
    [ H , J^2_i] &= 0,  \qquad  i=1,2,3 \label{eq:symm-second} 
\end{align}
\end{subequations}
\noindent where $J^2_i =  (J^x_i)^2 + (J^y_i)^2 + (J^z_i)^2$ is the collective total spin operator of $i^{\rm{th}}$ bath.
Therefore, the sum of eigenvalues of the $z$ component of the total spin operator ($ S^z_i + J^z_i$) of each qubit and that of its local bath, denoted by $\hbar m_i$ with $i=1,2,3$, is a conserved quantity. Additionally, the $J^2_i$ eigenvalue of the collective spin of each spin bath is also conserved.

{It follows from Eq.~\eqref{eq:HS-qref} that the baths enter the Hamiltonian only through
the collective spin operators \(J_i^\alpha\). For a bath consisting of
\(N_i\) spin-\(1/2\) particles, the bath Hilbert space decomposes into
sectors of definite total spin \(j_i\). A convenient basis is provided by the bath Dicke states which are simultaneous eigenstates \(|j_i,m_i\rangle_i\)
of \(J_i^2\) and \(J_i^z\), satisfying}
\begin{align}
J_i^2 |j_i,m_i\rangle_i
&= \hbar^2 j_i(j_i+1)|j_i,m_i\rangle_i, \nonumber \\
J_i^z |j_i,m_i\rangle_i
&= \hbar m_i |j_i,m_i\rangle_i . \nonumber
\end{align}
{The fully permutation-symmetric subspace of the \(N_i\) bath spins is the
maximal-total-spin sector \(j_i=N_i/2\). Equivalently, this subspace is
isomorphic to the Hilbert space of a single spin \(N_i/2\), and is spanned
by the symmetric Dicke states}
\[
|N_i/2,m_i\rangle_i,\qquad
m_i=-N_i/2,-N_i/2+1,\ldots,N_i/2 .
\]
{This sector has dimension \(N_i+1\), instead of the full bath dimension
\(2^{N_i}\).}

{The bath Hamiltonian \(H_B^{(i)}=K E_i J_i^z\) involves only collective
spin operators and is therefore symmetric under permutations of the bath
spins. Consequently, in the collective-spin description used here, the
thermal state of the \(i\)th bath is naturally written in the symmetric
Dicke sector \(j_i=N_i/2\), as}
\[
\rho_{B_i}^{\rm th}
=
\frac{1}{Z_{B_i}}
\sum_{m_i=-N_i/2}^{N_i/2}
e^{-\beta_i E_i m_i}
|N_i/2,m_i\rangle_i\langle N_i/2,m_i|,
\]
{where $Z_{B_i}$ is the partition function of \(i\)th bath. Furthermore, since the total Hamiltonian of the CSQAR involves only
collective spin operators of the baths, it is also symmetric under
permutations of the bath spins. Therefore, an initial bath state in the
fully symmetric Dicke sector remains in this sector throughout the
evolution. We refer to this \(j_i=N_i/2\) sector as the symmetric
subspace of bath Dicke states.}

{Initially, $(t = 0)$, the three refrigerator qubits are in thermal equilibrium with their respective environments. In the CSQAR model considered here, the spin-star environments are treated as explicit dynamical quantum subsystems rather than as stationary thermal reservoirs. Consequently, their reduced states evolve in time together with those of the refrigerator qubits.} As mentioned earlier, 
the six-body interaction also involves the states of the environments along with the system states necessitated by the symmetry~\eqref{eq:symm}. In this case, the condition for autonomous refrigeration is $E_2 - \varepsilon_2 = E_1 - \varepsilon_1 + E_3 - \varepsilon_3$ as will be shown in the 
succeeding section.

\section{Computing cold qubit temperature: a semi-analytic method}\label{sec:solution}

\subsection{
Dynamics of a single spin-star system}

We first illustrate the method to solve for the reduced density matrix of the central spin in a single central-spin system, which we will generalize to the CSQAR in the following subsection. The Hamiltonian of a central-spin system where the central spin is homogeneously coupled to a spin environment is given by 
\begin{subequations}\label{eq:cs-H}
\begin{align}
   H &= H_S + H_B + H_{SB} \label{eq:cs-H-first} \\
   &= K \varepsilon S^z + K E J^z + \frac{K}{\hbar} A \left( S^+ J^- + S^- J^+ \right) \label{eq:cs-H-second}
\end{align}
\end{subequations}
\noindent where $S^\alpha = \hbar \sigma^\alpha /2$ and $J^\alpha = \hbar \sum_{k=1}^{N} \sigma^\alpha_k/2$, $\alpha =x,y,z$. 
Each spin environment has $N$ spin-$\frac{1}{2}$s.
The total Hamiltonian commutes with the operators, $J^2$ and $S^z + J^z$, separately, i.e.
$[ H , S^z + J^z] = 0$ and 
   $ [ H , J^2] = 0$,
 which implies that the eigenvalues of operators $S^z + J^z$ and $J^2$, denoted by $ \hbar m$ and $ \hbar^2 j(j+1)$ respectively, 
remain invariant
under 
time evolution. 
These symmetries factorize the total Hilbert space of the central-spin system into superselection sectors labeled by a pair of quantum numbers $(j,m)$. Each $(j,m)$ sector forms an invariant subspace of the dynamics. The choice $j=N/2$ corresponds to the symmetric sector of the environment Dicke states, 
which are common eigenstates of operators $J^2$ and $J^z$.
The crux of our analyses lies in identifying the basis that harnesses the power of the invariant subspaces, enabling an exact analytic solution to the quantum dynamics.

We consider the joint initial state of the system and environment
to be a direct product of the thermal states of the central spin and the spins of the environment, i.e. $ \rho_{\text{ini}} = \rho_S^{\text{th}} \otimes \rho_B^{\text{th}}$.
Note that $\rho_B^{\text{th}}$ is the thermal state of the environment in the symmetric sector ($j=N/2$) of the environment Dicke states. Restricting to the symmetric subspace of bath Dicke states reduces the dimensionality of the bath Hilbert space from $2^N$ to $N+1$, which is an enormous simplification. This simplification is possible due to the symmetries $[ H, J^2 ]$ and $[H, S^z + J^z]$ of the central spin Hamiltonian.
The temperatures of the central spin and the environment are denoted by $T$ and $\tau$, respectively. At the initial time ($t=0$), the central spin is in thermal equilibrium with its environment, so $T(t=0) = \tau (t=0) = 1/\beta$, where $\beta$ is the inverse temperature of the initial thermal state of the environment.

We use the following notation throughout the rest of the paper.
The joint state of the system and environment is denoted by $| m_s \rangle_S| m_B \rangle_B $, where
the state of the system, given by $| m_s \rangle_S$, has $S^z$ eigenvalue $\hbar m_s$, and similarly, the 
state of the environment denoted by $| m_B \rangle_B $ has $J^z$ eigenvalue $\hbar m_B$. 
Henceforth, we will drop the $S$ and $B$ subscripts. The first ket or bra vector from the left will represent the system, and the second one will denote the bath.

The invariant basis that spans 
the invariant subspace, $(j=N/2,m)$, is given by
$\{|-1/2 \rangle | m  + 1/2 \rangle , |1/2 \rangle | m - 1/2 \rangle \}$ which is in general a two dimensional space. For $m = m_{max} = N/2 + 1/2$ and $m = m_{min} = -N/2 - 1/2$, the invariant basis is one dimensional and is $\{ |1/2 \rangle | m - 1/2 \rangle \}$ and $\{ |-1/2 \rangle | m + 1/2 \rangle \}$, respectively.

The thermal states of the system and environment are given by $\rho_S^{\text{th}} = Z_S^{-1}  e^{-\beta H_S / (\hbar K)} $ and  $\rho_B^{\text{th}} = Z_B^{-1}  e^{-\beta H_B  / (\hbar K)}$ respectively, where $Z_S = \sum_{m_s} e^{-\beta \varepsilon m_s}$ and $Z_B = \sum_{m_B} e^{-\beta E m_B}$ are the respective partition functions. The quantities, $\hbar m_s$ and  $\hbar m_B$, are the eigenvalues of $S^z$ and $J^z$ respectively. 
The efficacy of the 
invariant subspace structure can be exploited by recognizing that the joint initial state of the system and environment, given by $\rho_{\text{ini}} =   \rho_S^{\text{th}} \otimes \rho_B^{\text{th}} $, 
can be organized as a sum over $m = m_s + m_B$. 
%
The sum of the local Hamiltonians of the system and the environment is diagonal in the invariant basis, $\{|-1/2 \rangle | m  + 1/2 \rangle , |1/2 \rangle | m - 1/2 \rangle \}$. Using this, we calculate the exact expression of the joint initial state of the single qubit and its spin-star environment, which is provided below.
{\small
\begin{align}
    \rho_{\text{ini}} &=   \rho_S^{\text{th}} \otimes \rho_B^{\text{th}}  \nonumber \\
     &=  \frac{1}{\mathcal{N}}\sum_m \Big( e^{-\beta E m } \rho^m_{\text{ini}} \Big),
\end{align}
}
where

\begin{small}
\begin{align}
    & \rho^m_{\text{ini}} = \notag\\[-1ex]
  &\begin{cases}
    \begin{aligned}[t]
         e^{\frac{\beta \varepsilon}{2} } e^{- \frac{\beta E}{2} } | - 1/2 \rangle | m + 1/2 \rangle \langle - 1/2 | \langle m + 1/2 | 
    \end{aligned}
    , & m = m_{min},\\[1ex]
    \begin{aligned}[t]
        e^{\frac{-\beta \varepsilon}{2} } e^{\frac{\beta E}{2} } |  1/2 \rangle | m - 1/2 \rangle \langle  1/2 | \langle m - 1/2 |
    \end{aligned}
    , & m = m_{max}, \\[1ex]
    \begin{aligned}[t]
    & \Big( e^{\frac{\beta \varepsilon}{2} } e^{- \beta \frac{E}{2}} | - 1/2 \rangle | m + 1/2 \rangle \langle - 1/2 | \langle m + 1/2 |  \\
      &+ e^{\frac{-\beta \varepsilon}{2} } e^{ \beta \frac{E}{2}} |  1/2 \rangle | m - 1/2 \rangle \langle  1/2 | \langle m - 1/2 |  \Big),
    \end{aligned}
     & \text{other} \; m\\[1ex]
  \end{cases}
\end{align}
\end{small}
and the normalization factor is given by 
\begin{small}
\begin{align}
    & \mathcal{N} = \nonumber \\
    &\sum_{ \substack{m \\ m \neq m_{min},m_{max}}} e^{\frac{\beta \varepsilon}{2} } e^{- \beta E\left(m + \frac{1}{2}\right)} + e^{\frac{-\beta \varepsilon}{2} } e^{- \beta E\left(m - \frac{1}{2}\right)}   \nonumber \\
    &+  e^{\frac{\beta \varepsilon}{2} } e^{- \beta E\left(m_{min} + \frac{1}{2}\right)}  +   e^{\frac{-\beta \varepsilon}{2} } e^{- \beta E\left(m_{max} - \frac{1}{2}\right)}.
\end{align}
\end{small}
\noindent Sectors with different $m$ values do not mix under time evolution due to the symmetry, $[H, S^z + J^z]=0$. Consequently,
each $m$ sector can be evolved separately
by organizing the initial state in this manner.

The Hamiltonian of the central spin model~\eqref{eq:cs-H} takes the following form in the invariant basis, $\{|-1/2 \rangle | m  + 1/2 \rangle , |1/2 \rangle | m - 1/2 \rangle \}$, given by

\begin{subequations}\label{eq:Hm}
\begin{align}
     H &= \bigoplus_m H_m \label{eq:Hm:sub1}\\
     H_m  &= \hbar K \Big( b_{-1/2}  |-1/2 \rangle | m + 1/2 \rangle \langle -1/2| \langle  m + 1/2| \nonumber \\
     &  +  b_{1/2}  |1/2 \rangle | m - 1/2 \rangle \langle 1/2| \langle  m - 1/2| \nonumber \\
     &  + u
       |-1/2 \rangle | m + 1/2 \rangle \langle 1/2| \langle  m - 1/2| \nonumber \\
     &  +  u  |1/2 \rangle | m - 1/2 \rangle \langle -1/2| \langle  m + 1/2|  \Big) \label{eq:Hm:sub2},
\end{align}
\end{subequations}
where 
\begin{align}
    b_{-1/2}  &=  \left( - \frac{\varepsilon}{2} + E \left( m + \frac{1}{2} \right)  \right)   \nonumber \\
     b_{1/2}  &=  \left(  \frac{\varepsilon}{2} + E \left( m - \frac{1}{2} \right)  \right)  \nonumber \\
     u &=   \; A\sqrt{\left( \frac{N}{2} + m + \frac{1}{2} \right) \left( \frac{N}{2} - m + \frac{1}{2} \right)} 
\end{align}
%
%
The unitary operator for time evolution in the invariant  subspace labelled by $m$ is $U_m = e^{- \mathrm{i} \frac{H_m}{\hbar K} t}$,
with $H_m$ given in~\eqref{eq:Hm:sub2}.  The Hamiltonian, $H_m$, in the invariant subspace is a function of $m$, and so is the corresponding time-evolving unitary $U_m$. The total Hamiltonian, $H$~\eqref{eq:Hm:sub1}, is direct sum of the invariant subspace Hamiltonians, $H_m$~\eqref{eq:Hm:sub2}, and similarly the time evolution unitary $U(t)$ is also direct a sum of the invariant subspace unitaries, $U_m(t)$. Stated differently, both $H$ and $U(t)$ are block-diagonal with each block corresponding to a different value of $m$.
The 
state, $\rho(t)$, obtained after time evolution  is 
\begin{align}
    \rho(t) &= U \Big(  \rho_S^{th} \otimes \rho_B^{\text{th}} \Big) U^\dagger \nonumber \\
            &= \frac{1}{\mathcal{N}}\sum_m e^{- \beta E m} U_m  \rho^m_{\text{ini}} U_m^\dagger \nonumber \\
            &= \frac{1}{\mathcal{N}}\sum_m e^{- \beta E m} \rho_m(t).
\end{align}
Here, the composite unitary operator acting on the system and environment is a direct sum of $ U_m$'s belonging to each invariant subspace, i.e., $U=\bigoplus_m U_m$.
The matrix representation of $\rho_m(t)$ in the invariant basis, for each value of $m \; (\neq m_{min}, m_{max})$ is given by
\begin{align}
\label{rho}
    \rho_m(t) &= c(t)^m_{-1/2;-1/2} | -1/2 \rangle | m + 1/2 \rangle \langle - 1/2 | \langle m + 1/2 |  \nonumber \\
              & + c(t)^m_{1/2;-1/2} | -1/2 \rangle | m + 1/2 \rangle \langle  1/2 | \langle m - 1/2 |  \nonumber \\
              & + c(t)^m_{-1/2;1/2} | 1/2 \rangle | m - 1/2 \rangle \langle - 1/2 | \langle m + 1/2 |  \nonumber \\
              & + c(t)^m_{1/2;1/2} | 1/2 \rangle | m - 1/2 \rangle \langle  1/2 | \langle m - 1/2 |, 
\end{align}
where the analytic expressions for each matrix element of $\rho_m(t)$ 
is provided in the Appendix~\ref{matrix_ele}. For the edge values of $m$, namely $m_{min} = -N/2 - 1/2$ and $m_{max} = N/2 + 1/2$, $ \rho_m(t)$ is given by

\begin{align}
    \rho_{m_{min}}(t) &= | -1/2 \rangle | m_{min} + 1/2 \rangle \langle - 1/2 | \langle m_{min} + 1/2 | \nonumber \\
    \rho_{m_{max}}(t) &=  | 1/2 \rangle | m_{max} - 1/2 \rangle \langle  1/2 | \langle m_{max} - 1/2 |. 
\end{align}

The time-evolved reduced density matrix of the central spin $\rho^S(t)$ is obtained by tracing out the bath degrees of freedom. The orthonormality of the basis elements in the invariant subspace enables us to write down an exact analytic expression for $\rho^S(t)$ without the need to explicitly perform the partial trace.
\begin{small}
\begin{subequations}\label{eq:cs-rho-t}
    \begin{align}
    \rho^S(t) &= \text{Tr}_B [  \rho(t)], \label{eq:cs-rho-t-first}\\ 
     &=  \frac{1}{\mathcal{N}} \sum_m e^{-\beta E m} \rho_m^S (t), \; \; \text{where}  \label{eq:cs-rho-t-second} \\
        \rho_m^S (t) &=  \notag\\[-1ex]
         &\begin{cases}
    \begin{aligned}[t]
      {| - 1/2\rangle \langle - 1/2| }
    \end{aligned}
    , & m = m_{min},\\[1ex]
    \begin{aligned}[t]
        {| 1/2 \rangle \langle 1/2 |}
    \end{aligned}
    , & m = m_{max}, \\[1ex]
    \begin{aligned}[t]
    & c(t)^m_{-1/2;-1/2} | - 1/2\rangle \langle - 1/2|      \label{eq:cs-rho-t-third}  \\  
      &+ c(t)^m_{1/2;1/2} | 1/2 \rangle \langle 1/2 | 
    \end{aligned}
     & \text{other} \; m\\[1ex] 
  \end{cases}
    \end{align}
\end{subequations}
\end{small}
{For compactness in the expressions below, we extend the notation for the diagonal
coefficients to the one-dimensional edge sectors by the convention}
\[
c^{m_{\min}}_{-1/2;-1/2}(t)=1,\qquad
c^{m_{\min}}_{1/2;1/2}(t)=0,
\]
and
\[
c^{m_{\max}}_{-1/2;-1/2}(t)=0,\qquad
c^{m_{\max}}_{1/2;1/2}(t)=1.
\]
{With this convention, the subsequent sums over \(m\) can be written uniformly.}

Since the
reduced density matrix of the central spin after time evolution~\eqref{eq:cs-rho-t} is diagonal in the eigenbasis of the system Hamiltonian, $H_S = K \varepsilon S^z$, we can define the time-dependent temperature $T(t)$ of the central spin as
\begin{equation}\label{eq:cs-T-t}
     T(t) = \varepsilon \left[ \text{ln} \left( \frac{r(t)}{1 - r(t)} \right),  \right]^{-1}.
\end{equation}
where $r(t) = \left(\sum_m e^{-\beta E m} c(t)^m_{-1/2;-1/2}\right)/\mathcal{N}$ is the ground state population of $\rho^S(t)$.

The exact analytic expression for the reduced density matrix of the bath $\rho^B(t) $ can also be obtained similarly, given by
\begin{subequations}\label{eq:cs-rhoB-t}
    \begin{align}
    \rho^B(t) &= \text{Tr}_S [  \rho(t)], \label{eq:cs-rhoB-t-first} \\ 
     &= \frac{1}{\mathcal{N}} \Big(\sum^{N/2}_{m_B = -N/2} \Big( e^{- \beta (m_B - 1/2) E} c(t)^{m_B - 1/2}_{-1/2;-1/2}   \nonumber \\ & + e^{- \beta (m_B + 1/2) E} c(t)^{m_B + 1/2}_{1/2;1/2}  \Big) | m_B \rangle \langle m_B | \Big). \label{eq:cs-rhoB-t-second}
\end{align}
\end{subequations}

\subsection{Dynamics of the central-spin quantum absorption refrigerator}\label{sec:solution:subsec:dynamics-CSQAR}
\begin{figure*}
    \centering
    \subfloat[]{\includegraphics[width=0.32\textwidth]{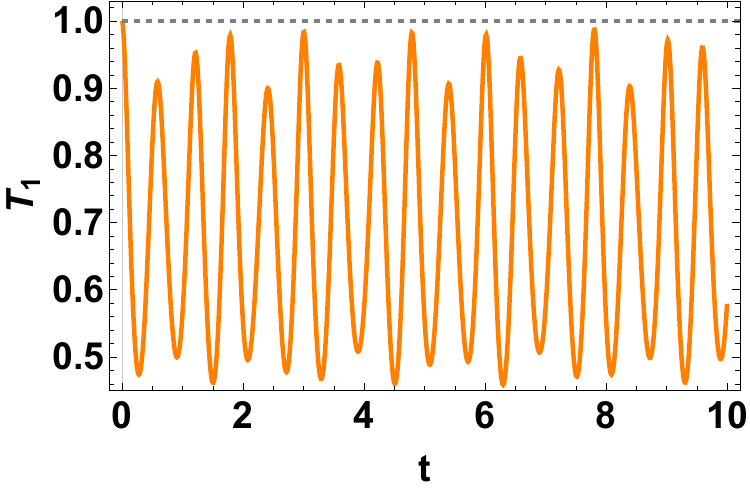} \label{fig:temp:sub1}}
    \hfill
    \subfloat[]{\includegraphics[width=0.32\textwidth]{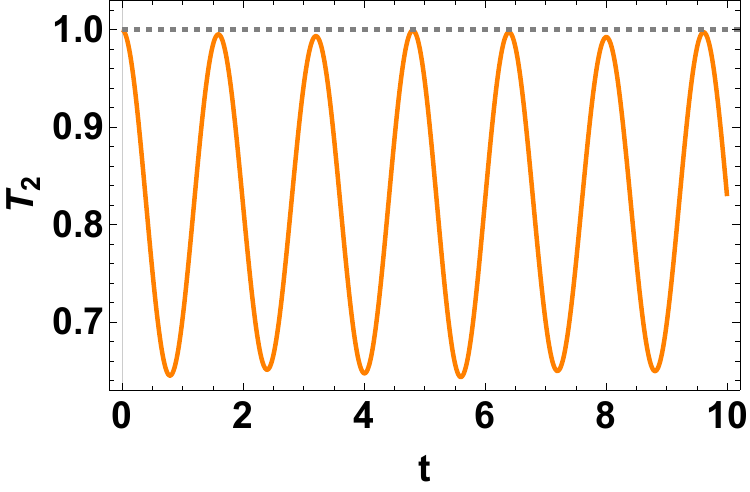} \label{fig:temp:sub2}}
    \hfill
    \subfloat[]{\includegraphics[width=0.32\textwidth]{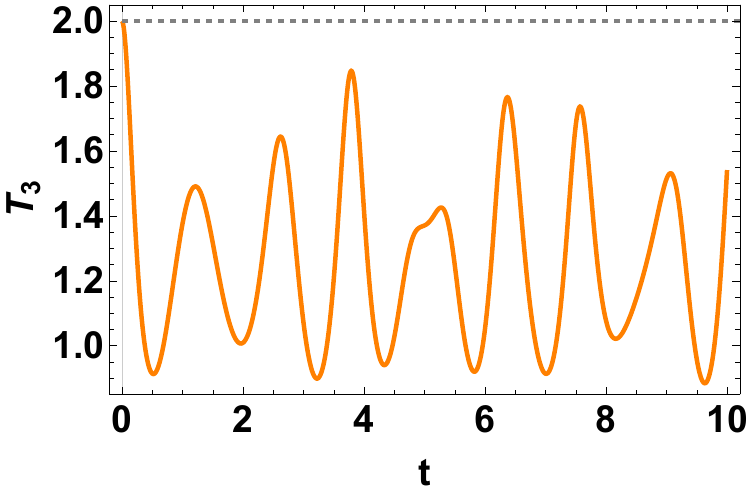} \label{fig:temp:sub3}}
    \caption{\justifying 
    Temperatures of the cold, room, and hot qubits, given by $T_1(t)$, $T_2(t)$, and $T_3(t)$, are plotted as functions of time in panels (a), (b), and (c), respectively. The energies of three qubits are $\varepsilon_1 = 1$, $\varepsilon_2 = 2$, and $\varepsilon_3 = 1$. The corresponding bath energies are $E_1 = 2$, $E_2 = 4$, and $E_3 = 2$. Note that they satisfy the autonomous refrigeration condition~\eqref{eq:autonomous-cs-refri}. The rest of the parameters in the CSQAR Hamiltonian~\eqref{eq:HS-qref} are optimized for the best possible refrigeration. {The values of the optimized parameters are $A_1= 0.950788$, $A_2= 0.308299$, $A_3= 0.440829$, $g= 0.0233622$.} The number of bath spins is chosen to be $N_1 = N_2 = N_3 =30$. The initial temperatures are {$T^{\text{ini}}_1 = 1$, $T^{\text{ini}}_2 =1$, and $T^{\text{ini}}_3 =2$}. The quantities plotted along the horizontal and vertical axes in each panel are dimensionless.
    }
    \label{fig:temp}
\end{figure*}

In this subsection, we solve the dynamics of the three-qubit CSQAR by extending the analysis in the preceding subsection to three qubits and their respective baths.
Let us initially digress a little to discuss the underlying symmetries in the system that will be useful for our analyses.
The symmetries~\eqref{eq:symm} of the total Hamiltonian~\eqref{eq:HS-qref}, ensure that the eigenvalues of operators $(S^z_i + J^z_i)$, $i=1,2,3$, denoted by
$\hbar m_1$, $\hbar m_2$, and $\hbar m_3$, respectively, are individually conserved. Thus, the Hilbert space factorizes into superselection sectors characterized by the {triple, $ \textbf{m}:=(m_1, m_2, m_3)$}. Different $\textbf{m}$ sectors do not mix with each other under time evolution due to the symmetry~\eqref{eq:symm}. The basis of such $\textbf{m}$ invariant subspaces is, in general, eight-dimensional spanned by the eight basis vectors, $\{|\psi_{ \pm 1/2; \pm 1/2;\pm 1/2} \rangle \}$, given in Eq.~\eqref{eq:qref-basis} of the Appendix~\ref{app:rho_m}. The eight-dimensional basis is a tensor product of three two-dimensional bases, $\{|-1/2\rangle |m_i + 1/2 \rangle, |1/2 \rangle | m_i - 1/2 \rangle \}$,
where each two-dimensional basis corresponds to the invariant subspace of the composite system of each
system qubit and its local bath. For the special cases when $m_i = m_{i;max} = N_i + 1/2$, the corresponding invariant basis for the $i^{\rm{th}}$ qubit-bath is one dimensional: $\{ |1/2 \rangle | m_{i;max} - 1/2 \rangle \}$. Similarly, when  $m_i = m_{i;min}= -N_i - 1/2$, the corresponding one dimensional basis is $\{ |-1/2 \rangle | m_{i;min} + 1/2 \rangle \}$. Therefore, in these spatial cases, the basis for the invariant subspace is less than eight-dimensional.

{Analogous to the three-body interaction in~\cite{PhysRevLett.105.130401}, the effective six-body interaction 
that we consider in the
Hamiltonian (refer to Eq.~\eqref{eq:H-qref-first}) of the
CSQAR is given by
\begin{subequations}\label{eq:H_int-cs}
\begin{align}
    H_{\rm{int}} &= \bigoplus_{\textbf{m}} H_{\rm{int}}^{\textbf{m}} \\
    H_{\rm{int}}^{\textbf{m}} &= \hbar K \; g \Big( | \psi_{-1/2;1/2;-1/2} \rangle  \langle \psi_{1/2;-1/2;1/2} | \nonumber \\
    & + |  \psi_{1/2;-1/2;1/2} \rangle \langle \psi_{-1/2;1/2;-1/2} |  \Big),
\end{align}
\end{subequations}
where recall the quantity, $\textbf{m}=(m_1,m_2,m_3)$, and the basis elements, as defined in~\eqref{eq:qref-basis}, are given by  
\begin{align}
   &  | \psi_{-1/2;1/2;-1/2} \rangle  \nonumber \\
   &= |-1/2 \rangle | m_1 + 1/2\rangle |1/2\rangle | m_2 - 1/2\rangle | - 1/2 \rangle | m_3 + 1/2\rangle  \nonumber \\
    & |  \psi_{1/2;-1/2;1/2} \rangle \nonumber \\
    &= | 1/2 \rangle | m_1 - 1/2 \rangle |- 1/2 \rangle | m_2 + 1/2\rangle |  1/2 \rangle | m_3 - 1/2\rangle. 
\end{align}
This is an effectively six-body interaction because each of the three baths can be treated as an effective single spin $N_i/2$ due to the homogeneous nature of the system-bath interactions. }
{In contrast to the Markovian scenario, the CSQAR treats the local spin environments as dynamical subsystems whose reduced states evolve with time.} The presence of an effective six-body interaction, $H_{\rm{int}}$, which couples not only the refrigerator qubits but also their corresponding local environments, is another distinct feature of the CSQAR compared to the original model given in~\cite{PhysRevLett.105.130401}. The requirement of such an effective six-body interaction arises from the choice of the invariant basis in which we perform our analyses.
For the CSQAR to function without any external source of work, the two states appearing in $H_{\rm{int}}$ of~\eqref{eq:H_int-cs} should be degenerate, i.e. 
\begin{align}\label{eq:cs-refri-degen}
    & \langle \psi_{-1/2;1/2;-1/2} | H_0 |  \psi_{-1/2;1/2;-1/2} \rangle \nonumber \\
    &=  \langle \psi_{1/2;-1/2;1/2} | H_0 |  \psi_{1/2;-1/2;1/2} \rangle,
\end{align}
Here the Hamiltonian $H_0$ is given by $H_0 = H_S + H_B + H_{SB}$.
The degeneracy constraint~\eqref{eq:cs-refri-degen} gives the following condition for the autonomous refrigeration
\begin{equation}\label{eq:autonomous-cs-refri}
    E_2 - \varepsilon_2 = E_1 - \varepsilon_1 + E_3 - \varepsilon_3.
\end{equation}
where $E_i$ and $\varepsilon_i$ are defined in~\eqref{eq:cs-H-second}. It is assumed that $g$, the strength of $H_{\rm{int}}$, is much smaller than all the parameters in $H_0$, i.e. $\varepsilon_i$, $E_i$ and $A_i$, $\forall i$.

Initially, the composite system is a tensor product of the thermal states of each of the three qubits and their respective baths. 
{As discussed in Sec.~\ref{sec:model:subsec:csqar}, for each bath we work in the fully permutation-symmetric Dicke sector \(j_i=N_i/2\), which is equivalent to treating the bath as a single collective spin \(N_i/2\). This is consistent with the Hamiltonian structure, since the bath Hamiltonians and the total CSQAR Hamiltonian involve only collective bath-spin operators and therefore preserve the \(j_i=N_i/2\) sectors under time evolution. The bath thermal states appearing below are therefore the Gibbs states written in these symmetric Dicke sectors. Such an initial state is given by}
\begin{align}\label{eq:csqar-ini-state}
   & \rho^{th}_{S_1} \otimes \rho^{th}_{B_1} \otimes \rho^{th}_{S_2} \otimes \rho^{th}_{B_2} \otimes \rho^{th}_{S_3} \otimes \rho^{th}_{B_3}  \nonumber \\  
   = & \frac{1}{\tilde{\mathcal{N}}}\sum_{\textbf{m} }   w_{\textbf{m}} \rho^{\textbf{m}}_{\text{ini}} ,
\end{align}
where $S_i$ denotes the $i^\text{th}$ qubit and $B_i$ denotes the corresponding bath, and {$w_{\textbf{m}}:=p_{m_1} p_{m_2} p_{m_2}$}, where $p_{m_i} = e^{-\beta_i E_i m_i }$, for  $i = 1,2$, $3$ are Boltzmann weight factors.
In the above equation, the operator, $\rho^{\textbf{m}}_{\text{ini}}$, is given by
\begin{equation}\label{eq:qref-rhoi-m123}
    \rho^{\textbf{m}}_{\text{ini}} = \rho^{m_1}_{1;\text{ini}}  \otimes \rho^{m_2}_{2;\text{ini}} \otimes \rho^{m_2}_{2;\text{ini}},
\end{equation}
with

\begin{small}
\begin{align}\label{eq:qref-rhoi-mi}
  &\rho^{m_i}_{i;\text{ini}} =   \notag\\[-1ex] 
  & \begin{cases}
    \begin{aligned}[t]
        e^{\frac{\beta_i \varepsilon_i}{2} } e^{- \frac{\beta_i E_i}{2} } | - 1/2 \rangle | m_i + 1/2 \rangle \langle - 1/2 | \langle m_i + 1/2 | 
    \end{aligned}
    , \\&  \hspace{-1cm} m_i = m_{i;min},\\[1ex]
    \begin{aligned}[t]
        e^{\frac{-\beta_i \varepsilon_i}{2} } e^{\frac{\beta_i E_i}{2} } |  1/2 \rangle | m_i - 1/2 \rangle \langle  1/2 | \langle m_i - 1/2 |
    \end{aligned}
    , \\& \hspace{-1cm}  m_i = m_{i;max}, \\[1ex]
    \begin{aligned}[t]
    & \Big( e^{\frac{\beta_i \varepsilon_i}{2} } e^{- \beta_i \frac{E_i}{2}} | - 1/2 \rangle | m_i + 1/2 \rangle \langle - 1/2 | \langle m_i + 1/2 |   \\
      &+ e^{\frac{-\beta_i \varepsilon_i}{2} } e^{ \beta_i \frac{E_i}{2}} |  1/2 \rangle | m_i - 1/2 \rangle \langle  1/2 | \langle m_i - 1/2 |  \Big)
    \end{aligned}
     \\& \hspace{-1cm} \text{other} \; m_i\\[1ex]
  \end{cases}
\end{align}
\end{small}
where the $p_{m_i}$s are defined previously.
The values of $\hbar m_i$ denote the sum of the eigenvalues of $S^z_i$ and $J^z_i$ operators corresponding to the $i^{\text{th}}$ qubit, where $i = 1,2$ and $3$.
The quantity, {$\tilde{\mathcal{N}} =  \prod_{i=1}^3  \mathcal{N}_{i}$} is the normalization factor, where
\begin{small}
\begin{align}
    & \mathcal{N}_i = \nonumber \\
    &\sum_{ \substack{m_i \\ m \neq m_{i;min},m_{i;max}}} e^{\frac{\beta_i \varepsilon_i}{2} } e^{- \beta_i E_i \left(m_i + \frac{1}{2}\right)} + e^{\frac{-\beta_i \varepsilon_i}{2} } e^{- \beta_i E_i\left(m_i - \frac{1}{2}\right)}   \nonumber \\
    &+  e^{\frac{\beta_i \varepsilon_i}{2} } e^{- \beta_i E_i\left(m_{i;min} + \frac{1}{2}\right)}  +   e^{\frac{-\beta_i \varepsilon_i}{2} } e^{- \beta_i E_i\left(m_{i;max} - \frac{1}{2}\right)}
\end{align}
\end{small}
Initially, i.e., at $t=0$, each of the refrigerator qubits is in thermal equilibrium with its local bath. Thus, the initial temperatures of each qubit are equal to the initial temperature of the corresponding spin-environment, i.e. $T_i(t=0) = \tau_i(t=0) = 1/\beta_i$. The quantities, $T_i$ and $\tau_i$,
denote the temperatures of the $i^{\text{th}}$ qubit and bath respectively.
The first two qubits are initially at room temperature $\tau_r$, whereas the third qubit is at a 
higher temperature $\tau_h$.

The composite state of the system and environment at a later time, $t$, can be written as 
\begin{align}\label{eq:rho-cs-refri}
    & \rho(t) =  \nonumber \\
    &   \frac{1}{\tilde{\mathcal{N}}}\sum_{\textbf{m} }  w_{\textbf{m}}  U_{\textbf{m}} \rho^{\textbf{m}}_{\text{ini}}  U^{\dagger}_{\textbf{m}}  \nonumber \\
    &=    \frac{1}{\tilde{\mathcal{N}}} \sum_{\textbf{m} }  w_{\textbf{m}}  \rho_{\textbf{m}}(t),
\end{align}
where the operator $\rho_{\textbf{m}}(t)=U_{\textbf{m}} \rho^{\textbf{m}}_{\text{ini}}  U^{\dagger}_{\textbf{m}}$.

The explicit form of the
time evolved density matrix, $\rho_{\textbf{m}}(t)$ within each $\textbf{m}$ sector, as given in~\eqref{eq:rho-cs-refri}, is provided in Appendix~\ref{app:rho_m}.

The reduced density matrix of the cold qubit is obtained by tracing out the bath of the first qubit, and the remaining system qubits and their respective environments.
The orthonormality of the basis elements in the invariant-subspace basis implies that the reduced state of the cold qubit is given by
\begin{subequations}\label{eq:qref-rhoC}
    \begin{align}
        {\rho^1(t)} &= {\text{Tr}_{B_1 S_2 B_2 S_3 B_3} [ \rho(t)  ]}   \label{eq:qref-rhoC-first}\\
          &= {\frac{1}{\tilde{\mathcal{N}}} \sum_{\textbf{m}} w_{\textbf{m}} \rho^1_{\textbf{m}}(t),} \label{eq:qref-rhoC-second} \\
         {\rho^1_{\textbf{m}}(t)} &= {r_1^{\textbf{m}}(t) |- 1/2\rangle \langle -  1/2 |}   \nonumber \\
         &{+ \left( 1 - r_1^{\textbf{m}}(t) \right)  | 1/2 \rangle \langle 1/2 | \;\;\;  \text{with}}  \label{eq:qref-rhoC-third} \\
         {r_1^{\textbf{m}}(t)} &= {\Big( \tilde{c}(t)^{\textbf{m}}_{-1/2;-1/2;-1/2} + \tilde{c}(t)^{\textbf{m}}_{-1/2;-1/2;1/2} } \nonumber \\
          &{+ \tilde{c}(t)^{\textbf{m}}_{-1/2;1/2;-1/2} +  \tilde{c}(t)^{\textbf{m}}_{-1/2;1/2;1/2}  \Big).}\label{eq:qref-rhoC-fourth},
    \end{align}
\end{subequations}
where $\tilde{c}$s are defined in Eq.~\eqref{eq:rho-cs-ref}.  These are matrix elements of $\rho_{\textbf{m}}(t)$ in the invariant basis~\eqref{eq:qref-basis}.
The time-dependent ground state population of the cold qubit $r_1(t)$, as observed from~\eqref{eq:qref-rhoC}, is given by
\begin{align}\label{eq:qref-gs-pop-cold}
    {r_1(t) } &= {\frac{1}{\tilde{\mathcal{N}}}\sum_{\textbf{m}} w_{\textbf{m}}  r_1^{\textbf{m}}(t),}
\end{align}
{where the sum over \(\mathbf m\) runs over all allowed triples
\((m_1,m_2,m_3)\), and \(w_{\mathbf m}\) is defined below Eq.~\eqref{eq:csqar-ini-state}.}
We 
numerically solve for $\rho_{\textbf{m}}(t)$ and thus $\rho(t)$ given in
Eq.~\eqref{eq:rho-cs-refri}.
Evaluating the matrix elements of $\rho_{\textbf{m}}(t)$, and thereby finding $\tilde{c}(t)^{\textbf{m}}$, we obtain the ground state population, $r_1(t)$, in terms of which we find the temperature of the cold qubit. 
Since $\rho^1(t)$, given in~\eqref{eq:qref-rhoC}, is diagonal in the eigenbasis of the cold qubit Hamiltonian, $K \varepsilon_1 S^z_1$, one can define the time-dependent temperature of the cold qubit in terms of its ground state population $r_1(t)$, given by 
\begin{equation}\label{eq:qref-T1-t}
     T_1(t) = \varepsilon_1 \left[ \text{ln} \left( \frac{r_1(t)}{1 - r_1(t)} \right)  \right]^{-1}.
\end{equation}

We plot the temperature of the three qubits of the CSQAR as a function of time in Fig.~\ref{fig:temp}.
The cold qubit temperature $T_1(t)$ shows an oscillatory behavior as a function of time, with its magnitude decreasing up to $0.5$ times the initial temperature, $1$.  Refer to Fig.~\ref{fig:temp:sub1} for this. 
In addition to $T_1(t)$, 
the behaviors of both $T_2(t)$ and $T_3(t)$ are depicted as a function of time in panels~\ref{fig:temp:sub2} and~\ref{fig:temp:sub3}, respectively. 
During the time evolution, the temperatures of both the room and hot qubits are significantly lowered compared to their initial temperatures.
In the conventional three-qubit QAR with 
three-body interactions among the qubits, 
the cold qubit temperature is lowered, 
{whereas that of the room qubit increases.}

{In the conventional three-qubit Markovian QAR, the three qubits play distinct thermodynamic roles: the first qubit is the object to be cooled, while the remaining qubits, together with their passive thermal reservoirs, provide the auxiliary room and hot terminals required for autonomous operation. The behavior in Fig.~\ref{fig:temp} is qualitatively different. In addition to the cooling of the cold qubit, the temperatures of the room and hot qubits can also decrease during the transient dynamics. 
It indicates that the usual Markovian separation of thermodynamic roles among the three qubits is modified when the environments are finite, dynamical, and non-Markovian. In the CSQAR, the spin-star environments are not merely passive reservoirs; their collective degrees of freedom participate in the effective interaction and can absorb energy during the transient cooling window. Therefore, part of the thermodynamic role normally played by the auxiliary qubits in the conventional Markovian QAR can, in a finite-environment non-Markovian setting, be played by the dynamical baths themselves. This observation is consistent with the possibility of autonomous refrigerators with a working substance of fewer than three qubits, as explicitly demonstrated for one- and two-qubit finite-environment refrigerators in Ref.~\cite{Bhattacharyya:2023pak}.}

Here we remark that although the temperatures of the three refrigerator qubits decrease simultaneously for 
significantly large timescales, this does not indicate any violation of the second law of thermodynamics. The precise statement of the second law of thermodynamics for Markovian open quantum systems is given by the Spohn's theorem \cite{10.1063/1.523789, doi:https://doi.org/10.1002/9780470142578.ch2}, which states that the entropy production rate defined as the sum of the rate of change of von Neuman entropy and the heat current for a single quantum system immersed in a heat bath is always positive. However, it is known that in the presence of non-Markovian effects, the entropy production rate can take negative values, leading to modified versions of Spohn's theorem \cite{Bylicka16, Marcatoni17, PhysRevA.95.012122, PhysRevA.98.012130, PhysRevE.99.012120, Naze_2020, PhysRevE.103.012109, PhysRevA.105.022424, PhysRevA.110.012451}. This indicates that in the presence of fully non-Markovian environments, like spin-star, Spohn's theorem would get modified, allowing for negative entropy production rates without contradicting the second law of thermodynamics. 

{We would like to clarify that the above discussion of negative entropy production pertains only to later-time thermodynamic features of the reduced non-Markovian dynamics, and not to the mechanism responsible for the onset of cooling at $t \approx 0$. The initial simultaneous decrease of all three qubit temperatures at very short times in Fig.~\ref{fig:temp} should not be attributed to negative entropy production or to information backflow. In particular, at $t \approx 0$ and in the absence of initial system–environment correlations, negative entropy production cannot occur. The simultaneous cooling of the three qubits at short times is connected to the fact that the environments are active parts of the interaction. This is consistent with Figs.~\ref{fig:heat} and~\ref{fig:QB}, which show that during the cooling window the corresponding bath heat currents are positive, i.e. the baths heat up while the qubits cool.}

One key advantage of our 
approach using the invariant basis is that we can obtain the reduced density matrix of the cold qubit even for a very large number of bath qubits, which is 
not possible using the usual methods due to an exponential increase in the dimensionality of the Hilbert space as the number of bath spins is increased. The symmetries~\eqref{eq:symm} together with the orthonormality of the invariant subspace basis~\eqref{eq:qref-basis} allow us to obtain the expression for the reduced density matrix of the cold qubit~\eqref{eq:qref-rhoC} without having to explicitly perform computationally demanding partial traces. Furthermore, note that restricting the 
Hilbert space of the environment to the corresponding symmetric Dicke subspaces reduces the dimensionality of each of the 
environment Hilbert spaces from $2^{N_i}$ to $N_i + 1$.

For completeness, we also provide the reduced density matrix of the environment corresponding to the cold qubit as a function of time.
It is given by 
{\small
\begin{subequations}\label{eq:qref-rhoB}
    \begin{align}
        & \rho^{B_1}(t) = \text{Tr}_{S_1 S_2 B_2 S_3 B_3 } [ \rho(t)  ]    \label{eq:qref-rhoB-first}\\
        &  = \frac{1}{\tilde{\mathcal{N}}} \Big(\sum_{m_2, m_3}  \sum^{N_1/2}_{m_{B_1} = -N_1/2} \nonumber  \\
&  \Big( p_{(m_{B_1} - 1/2)} p_{m_2} p_{m_3}  \tilde{c}(t)^{(m_{B_1} - 1/2) m_2 m_3}_{-1/2;-1/2;-1/2}   \nonumber \\ & + p_{(m_{B_1} + 1/2)} p_{m_2} p_{m_3}  \tilde{c}(t)^{(m_{B_1} + 1/2)m_2 m_3}_{1/2;1/2;1/2}  \Big) | m_{B_1} \rangle \langle m_{B_1} | \Big), \label{eq:qref-rhoB-SECOND}      
    \end{align}
\end{subequations}
}
where $\hbar m_{B_i}$ is the eigenvalue of collective bath spin operator $J^z_i$ with corresponding eigenstate $|m_{B_i}\rangle$, $i=1,2,3$. 
Similarly, the expressions for the reduced density matrices of the other two baths, $\rho^{B_2}(t)$ and $\rho^{B_3}(t)$, can be obtained.

In the model that we consider,
the presence of non-Markovianity in the environment introduces significant information backflow~\cite{PhysRevA.107.022209} to the system. The backflow can potentially defer reaching the stationary state, and hence thermalization, to very large times or can even prohibit reaching the stationary state at all. In fact, a single central-spin system has been shown to evade thermalization~\cite{Wang_2013}. Therefore, we expect only transient refrigeration and not steady-state refrigeration in CSQAR. Moreover, such a behavior is also reflected in the temperatures of the three qubits, as a function of time, as shown in Fig.~\ref{fig:temp}.

{We would like to make a remark on experimental feasibility of realizing the effective six-body interaction of Eq.~\eqref{eq:H_int-cs}. We do not interpret this effective six-body interaction in the CSQAR as a fundamental microscopic coupling among six bodies. Instead, it is more appropriately understood as an emergent, engineered interaction between two resonant collective states of the combined qubit-bath subsystems. A possible route toward realizing such a term is Hamiltonian engineering~\cite{Georgescu2014, PhysRevLett.128.190502}. Importantly, the key ingredients required to realize such an interaction have already been demonstrated experimentally in various quantum platforms. First, central-spin architectures are naturally realized in solid-state systems such as nitrogen-vacancy (NV) centers in diamond, where a single electronic spin interacts with a surrounding bath of nuclear or electronic spins and provides a canonical realization of the central-spin model~\cite{PhysRevLett.105.140502}. In our model, each bath enters only through collective spin operators, so within the symmetric Dicke sector each bath behaves as a single large spin, which makes the relevant transition effectively six-body in the reduced collective description. Notably, hybrid quantum systems have been experimentally realized in which a superconducting qubit is coherently coupled to a large ensemble of NV spins via a microwave resonator, enabling coherent exchange of excitations between the qubit and the spin ensemble~\cite{PhysRevLett.107.220501, Zhu:2011fkb}. In such systems, the ensemble dynamics is naturally described in terms of collective bright and dark spin modes with a collectively enhanced coupling, allowing the many microscopic spins of the bath to be manipulated effectively as a single collective spin degree of freedom~\cite{Zhu:2011fkb, PhysRevLett.105.210501, PhysRevLett.107.060502}. Furthermore, multi-body interactions beyond two-body have been synthesized experimentally, including five-body spin-exchange interactions in superconducting circuits~\cite{PhysRevLett.108.023005, Georgescu2014, PhysRevLett.119.173402, PhysRevLett.128.190502}. Therefore, the principal ingredients required for realizing the CSQAR interaction structure, namely, a central-spin architecture, collective manipulation of large spin ensembles, and experimentally engineered multi-body interactions, have already been achieved individually in existing platforms. We therefore regard the present interaction as an effective target Hamiltonian that is challenging, but not implausible, experimentally.}

\section{Heat currents}\label{sec:heat-currents}

\begin{figure*}
    \centering
    \subfloat[]{\includegraphics[width=0.32\textwidth]{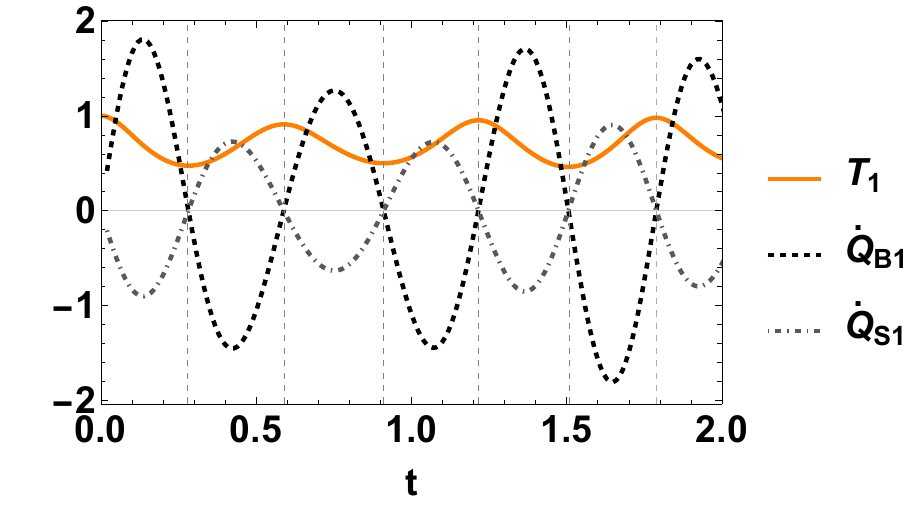} \label{fig:heat:sub1}}
    \hfill
    \subfloat[]{\includegraphics[width=0.32\textwidth]{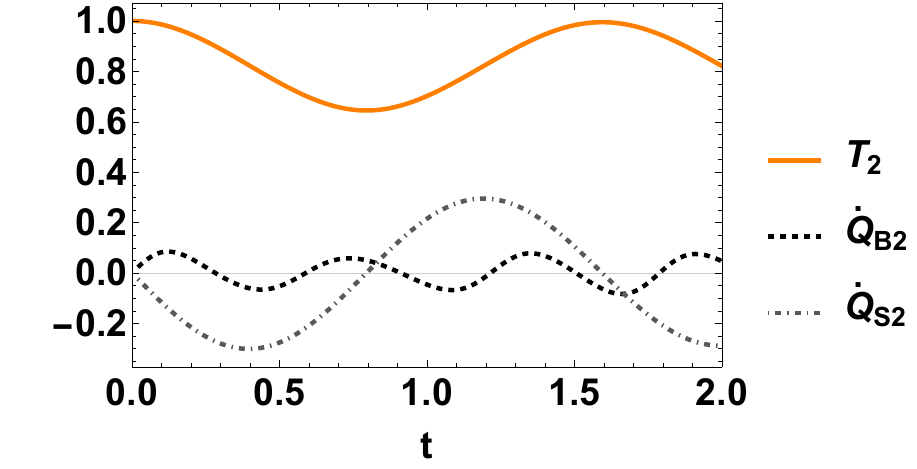} \label{fig:heat:sub2}}
    \hfill
    \subfloat[]{\includegraphics[width=0.32\textwidth]{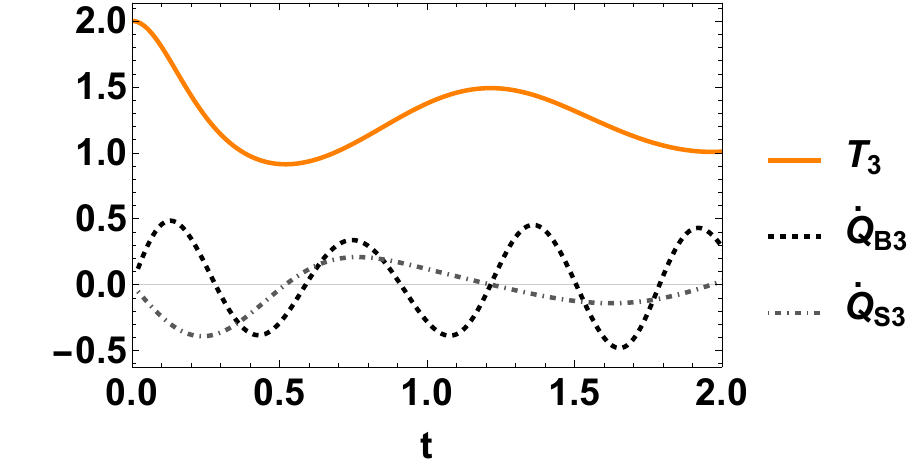} \label{fig:heat:sub3}}
    \caption{\justifying The heat current of the $i^{\text{th}}$ qubit and that of its bath, and the temperature of the qubit are plotted along the vertical axes as functions of time, for $i=1,2$ and $3$ in panels (a), (b) and (c) respectively.
    The system heat currents $\dot{Q}_{S_i}$ are {shown by dark-gray dot-dashed curves}, the bath heat currents $\dot{Q}_{B_i}$ by {black dashed curves}, while the temperature is plotted by orange solid curves. The Hamiltonian parameters are the same as in Fig.~\ref{fig:temp}. The number of bath spins is $N_1 = N_2 = N_3 =30$. The initial temperatures are {$T^{\text{ini}}_1 = 1$, $T^{\text{ini}}_2 =1$, and $T^{\text{ini}}_3 =2$}. The quantities plotted along the horizontal and vertical axes in each of the panels are dimensionless.}
    \label{fig:heat}
\end{figure*}

\begin{figure}
    \centering
    \includegraphics[width=0.8\linewidth]{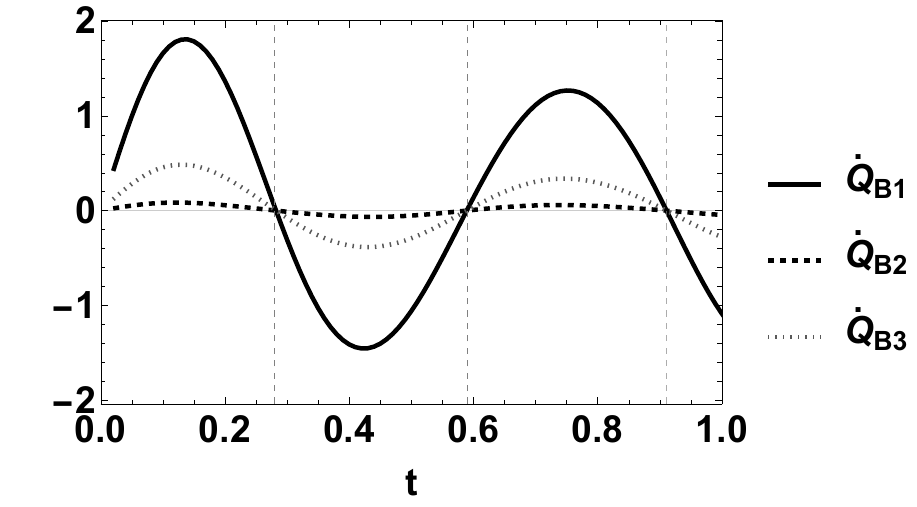}
    \caption{Time evolution of the bath heat currents corresponding to the cold, room, and hot qubits. {The curves for $\dot{Q}_{B_1}$, $\dot{Q}_{B_2}$, and $\dot{Q}_{B_3}$ are shown as black solid, black dashed, and black dotted lines, respectively.} The Hamiltonian parameters are the same as in Fig.~\ref{fig:temp}. The number of bath spins is $N_1 = N_2 = N_3 =30$. The initial temperatures are {$T^{\text{ini}}_1 = 1$, $T^{\text{ini}}_2 =1$, and $T^{\text{ini}}_3 =2$}. All the quantities plotted along both axes are dimensionless.}
    \label{fig:QB}
\end{figure}

The notion of heat current in the context of refrigerators with Markovian environments focuses on the amount of heat flowing through
the local environments connected to each qubit and is defined in terms of the Lindblad operators of the respective environments.
However, the heat currents that are relevant here are concerned with the amount of heat flowing through 
each individual system qubits of the CSQAR as well as through each of the local environments, and can be defined via the rate of change of populations of the respective reduced density matrices. 
Such types of heat currents have been previously explored in literature~\cite{bhargavatransistor}.
In the Markovian environment case, the heat current flowing into
the local environment of the cold qubit has an unambiguous interpretation as the \emph{cooling power} because the local environments are solely connected to their respective qubits. However, in CSQAR, 
due to the effective six-body interaction~\eqref{eq:H_int-cs}, the heat flow in and out of the local environments is not solely directed towards or away from the respective system qubit, 
but it has contributions from heat flowing throughout the entire refrigerator as well as through all the other local environments. 
The heat current through the $i^{\text{th}}$ qubit is given by 
\begin{equation}\label{eq:heat-S}
    \dot{Q}_{S_i} = \frac{1}{\hbar K} \text{Tr}[ \frac{\partial \rho^{S_i}(t)}{\partial t}  H^{(i)}_S  ],
\end{equation}
\noindent where $\rho^{S_i}(t)$ and $H^{(i)}_S$ are the reduced density matrix and Hamiltonian of the $i^{\text{th}}$ qubit respectively, and $t$ is the dimensionless time.
In a similar manner, the heat currents through the environments can also be defined.
The heat current through the environment connected to the $i^{\text{th}}$ qubit is given by 
\begin{equation}\label{eq:heat-B}
    \dot{Q}_{B_i} =   \frac{1}{\hbar K} \text{Tr}[ \frac{\partial \rho^{B_i}(t) }{\partial t} H^{(i)}_B  ],
\end{equation}
where $\rho^{B_i}(t)$ and $H^{(i)}_B$ refer to the reduced density matrix and the Hamiltonian, respectively, of the bath connected to the $i^{\text{th}}$ qubit.

The heat current of the cold qubit, given by $\dot{Q}_{S_1}$, heat current of its respective environment, $\dot{Q}_{B_1}$, and the local temperature of the cold qubit, $T_1$, are plotted as functions of time in  Fig.~\ref{fig:heat:sub1} in blue, green and orange curves respectively. 
We observe from Fig.~\ref{fig:heat:sub1} that the temperature of the cold qubit oscillates nearly periodically as a function of time. Let us consider one such period of oscillation of the cold qubit temperature. 
We find that when the temperature of the cold qubit decreases, the heat current of the cold qubit is negative, while that of its respective bath is positive, which suggests that heat is being deposited from the cold qubit into its bath. Meanwhile, when the temperature of the cold qubit increases, the heat current of the qubit is positive, whereas that of its respective bath is negative, which implies that the heat flows out of the bath and towards the qubit. The behavior of heat currents flowing through the other two qubits and their environments is shown in Fig.~\ref{fig:heat:sub2} and~\ref{fig:heat:sub3}, respectively. 
Intriguingly, we find that all the environment heat currents $\dot{Q}_{B_i}$ are positive during the cooling period of the cold qubit and all of them are negative during its heating period as shown in Fig.~\ref{fig:QB} together with Fig.~\ref{fig:heat:sub1}.

In the absence of interactions among the environments, heat current $\dot{Q}_{B_i}$ moves in and out of the $i^{th}$ qubit. Therefore, in this case $\dot{Q}_{B_1}$ has a clear interpretation as the heat current moving from the cold qubit into its local environment or vice versa. Positive value of $\dot{Q}_{B_1}$ implies that heat moves from the cold qubit to its local environment. The quantity, $\dot{Q}_{B_1}$, in this case, is called \emph{cooling power} of the refrigerator. In our model, however, the local environments do interact with each other through $H_{\rm{int}}$ of~\eqref{eq:H_int-cs}, and therefore the heat current $\dot{Q}_{B_1}$ not only flows in and out of the cold qubit but also to other two qubits and their environments. However, since the strength of interaction $H_{\rm{int}}$, $g$, is assumed to be very small compared to all other Hamiltonian parameters, we expect these leakages in heat current to be 
{small}. {In particular, for the parameter regime used in Figs.~\ref{fig:heat} and~\ref{fig:QB}, with bath size $N=30$ and weak six-body coupling, the leakage is comparatively small for the cold qubit and its local bath during the cooling window of interest. This is why, in Fig.~\ref{fig:heat:sub1}, the behavior of $\dot{Q}_{B_1}$ remains well correlated with that of the $\dot{Q}_{S_1}$ and $T_1$  and can be used as an approximate proxy for the cooling power. In contrast, Figs.~\ref{fig:heat:sub2}-\ref{fig:heat:sub3} show that such an interpretation is not equally valid for the room and hot qubits, where the leakage effects are more visible. 
Therefore $\dot{Q}_{B_1}$ can be regarded only as an approximate indicator of cooling power in CSQAR, rather than as an exact local heat current.}

{Moreover, as shown in Fig.~\ref{fig:QB}, the heat absorption and emission cycles of all three bath heat currents are synchronized. In contrast, the system heat currents of the room and hot qubits (Fig.~\ref{fig:heat}) are neither synchronized with their respective bath heat currents nor with each other. This behavior can be understood from the difference in size between the baths and the system qubits. Since the baths have a relatively large size ($N=30$), their heat currents are only weakly affected by leakage arising from the effective six-body interaction. In contrast, the {systems} are much smaller, and their heat currents are therefore more susceptible to such leakage effects. This explains why in Fig.~\ref{fig:heat} the periodic changes in the heat flow of the room and hot qubits' environments do not match those of their corresponding systems. For the cold qubit, however, the system and bath heat currents, as well as the qubit temperature, remain largely synchronized (Fig.~\ref{fig:heat:sub1}). We potentially attribute this to the fact that the interaction parameters are optimized with respect to the cooling of the cold qubit. This optimization effectively enforces synchronization between its system and bath dynamics. No such optimization is imposed on the room and hot qubits (Figs.~\ref{fig:heat:sub2} and~\ref{fig:heat:sub3}), which leads to the more pronounced mismatch observed in their respective heat currents.}

{In conventional Markovian three-qubit QARs, the local reservoirs are modeled as passive, time-independent thermal environments, and the three qubits are assigned distinct thermodynamic roles. In this setting, the hot terminal supplies the free-energy bias required for autonomous operation, while the room terminal provides the channel through which heat is dumped. In the CSQAR, the situation is different because the finite spin-star environments are dynamical quantum subsystems. Moreover, the effective interaction in Eq.~\ref{eq:H_int-cs} involves not only the three refrigerator qubits but also the collective states of their local environments. Consequently, the energy released by the qubits during the transient cooling window need not be accounted for solely by heating one of the other refrigerator qubits. This is reflected in Fig.~\ref{fig:QB}, where all three bath heat currents can be simultaneously positive over a finite time interval. Thus, the finite environments can temporarily store energy and participate in the refrigeration mechanism. The simultaneous cooling of all three refrigerator qubits
should therefore be understood as evidence that the operating mechanism differs from that of the conventional Markovian three-qubit QAR. It suggests, without by itself proving a new minimality result, that fewer-than-three-qubit autonomous refrigerators can become possible when finite non-Markovian environments are included as active dynamical components. This interpretation is consistent with Ref.~\cite{Bhattacharyya:2023pak}, where one- and two-qubit self-sustained refrigerators coupled to finite spin environments were explicitly analyzed.}

\section{Scaling of cold qubit temperature with the number of bath spins}\label{sec:scaling}

\begin{figure}
    \centering
    \includegraphics[width=0.45\textwidth]{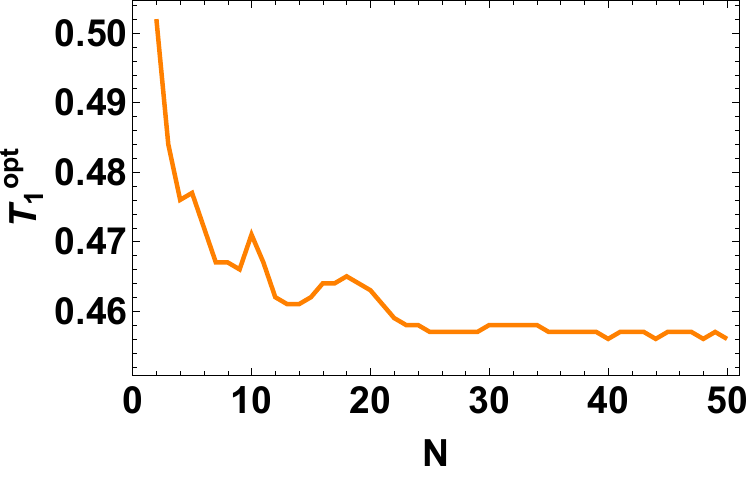} 
    \caption{Scaling of the optimal temperature of the cold qubit, $T_1^{\rm{opt}}$, with the number of environment spins $N =N_1 = N_2 = N_3$. The energies of three qubits are $\varepsilon_1 = 1$, $\varepsilon_2 = 2$, and $\varepsilon_3 = 1$. The corresponding bath energies are $E_1 = 2$, $E_2 = 4$, and $E_3 = 2$. The initial qubit temperatures are {$T^{\text{ini}}_1 = 1$, $T^{\text{ini}}_2 =1$, and $T^{\text{ini}}_3 =2$}. {The optimized values of the parameters  $A_1$, $A_2$, $A_3$, $g$, and $t$ together with the resulting optimal value $T_1^{\rm{opt}}$ for each value of $N$ plotted in this figure are provided in Table~\ref{tab:optimizer} of Appendix~\ref{app:otimizer}.} The quantities plotted along each of the horizontal and vertical axes are dimensionless.}
    \label{fig:T1opt-N-scaling}
\end{figure}

\subsection{Scaling of optimal cold qubit temperature}

Consideration of the invariant subspace approach in the CSQAR allows us  
to compute the temperature of the cold qubit in the presence of a large number of environment spins. We investigate how the optimal cold qubit temperature scales with the number of environment spins.  
Specifically, we optimize (minimize) the cold qubit temperature over the 
parameters of the Hamiltonian, $A_1$, $A_2$, $A_3$, $g$, and time $t$, for a fixed value of the number of environment spins for each of the three environments. Let the number of environment spins for each of the three environments be given by $N$.
The optimization ranges of the parameters are set to be $0 \leq A_i \leq 1$, $0 \leq g \leq 0.1$, and $0 \leq t \leq 10$. The energies of the three qubits and their environments are chosen such that they satisfy the autonomous refrigeration condition~\eqref{eq:autonomous-cs-refri}: $\varepsilon_1 = 1$, $\varepsilon_2 = 2$, $\varepsilon_3 = 1$, $E_1 = 2$, $E_2 = 4$, and $E_3 = 2$. The optimization range of $A_i$ is kept at the same order of magnitude as the energies of the three qubits. Since $g$ must be very small compared to any other parameter in the Hamiltonian, we keep the upper bound on the optimization range of $g$ to be within $10 \%$ of other Hamiltonian parameter values.

\begin{figure*}
    \centering
    \subfloat[]{\includegraphics[width=0.45\textwidth]{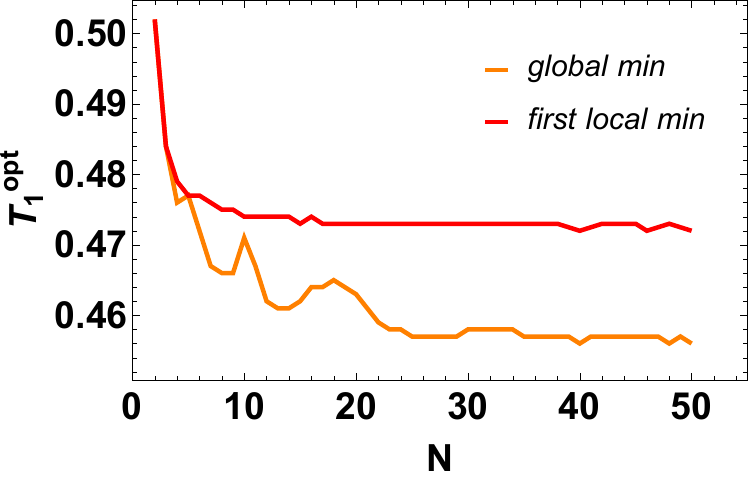} \label{fig:local-v-global:sub1}}
    \hfill
    \subfloat[]{\includegraphics[width=0.45\textwidth]{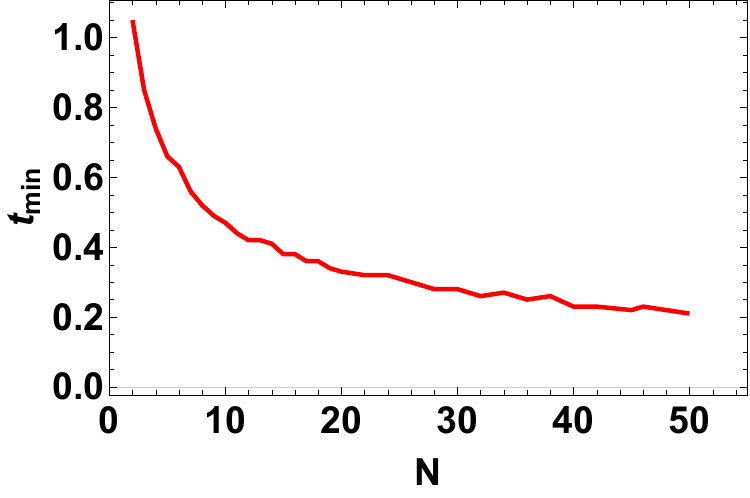} \label{fig:local-v-global:sub2}}
    \caption{Panel (a): Comparison of the global minimum of $T_1$ (given in orange) with the first local minimum in time (given in red) as functions of the number of environment spins $N =N_1 = N_2 = N_3$.
    Panel (b): Scaling of the minimum time, $t_{\text{min}}$, required to reach the local minimum of $T_1$ given as a function of $N$. The energies of three qubits are $\varepsilon_1 = 1$, $\varepsilon_2 = 2$, and $\varepsilon_3 = 1$. The corresponding bath energies are $E_1 = 2$, $E_2 = 4$, and $E_3 = 2$. The initial qubit temperatures are {$T^{\text{ini}}_1 = 1$, $T^{\text{ini}}_2 =1$, and $T^{\text{ini}}_3 =2$}. The quantities plotted along the horizontal and vertical axes in each of the panels are dimensionless.}
    \label{fig:local-v-global}
\end{figure*}

{Although the symmetric Dicke sector reduction changes the dimension of each local bath from $2^N$ to $N+1$, the computational cost of the optimized scaling analysis is not simply linear in $N$. For fixed Hamiltonian parameters, the dynamics can be evaluated for bath sizes much larger than those used in the representative plots. However, determining $T_1^{\rm{opt}}$ requires a five-parameter optimization over $A_1$, $A_2$, $A_3$, $g$, and $t$, for each value of $N$. In addition, each evaluation of $T_1(t)$ during the optimization run involves a sum over the invariant-sector triples $\textbf{m} = (m_1, m_2, m_3)$, whose number grows roughly as $(N+2)^3$ for $N_1 = N_2 = N_3 = N$. We therefore report the optimized scaling data up to $N=50$, which is sufficient for the present purpose because $T_1^{\rm{opt}}$ is already close to its large-$N$ limiting value by $N \simeq 30$ and becomes nearly flat for larger $N$.}

Figure~\ref{fig:T1opt-N-scaling} shows the behavior of optimal cold qubit temperature $T_1^{\rm{opt}}$ as a function of $N$.
Firstly, we observe that increasing the value of $N$ leads to 
an enhancement in cooling of the cold qubit. Furthermore,  $T_1^{\rm{opt}}$ reaches a limiting value of $T_1^{\infty}$ for $N \rightarrow \infty$. We estimated $T_1^{\infty}$ by finding the best non-linear curve fit to the data of Fig.~\ref{fig:T1opt-N-scaling} as well as by using Neville's polynomial extrapolation \cite{PressNum}. In the succeeding part, we describe these two methods to obtain the optimal temperature of the cold qubit in the limit of an infinite number of bath spins.

\emph{Power law fit}: We generated data points $\{ N, T_1^{\text{opt}} \}$ for several values of $N$ till $N=50$ which are plotted in Fig.~\ref{fig:T1opt-N-scaling} {and tabulated in Table~\ref{tab:optimizer} of Appendix~\ref{app:otimizer}}. We take an average of the last nine data points in the asymptotic flat region of Fig.~\ref{fig:T1opt-N-scaling} ($N \geq 41$) to get the limiting value $T_1^{\infty} = 0.457$. We then fit the $T_1^{\rm{opt}} (N)$ with the model $T_1^{\text{fit}}(N) := T_1^{\infty} + a N^{-b}$. The fit obtained has a standard deviation ($\sigma$) of $0.002$, with fitted parameters $a =0.097$ and $b = 1.096$. The standard deviation of the fit, $\sigma$, is defined as
\begin{equation}
  \sigma^2 = \frac{ 1}{d - p} \sum_{i =1 }^{d}  \left[ T_1^{\text{fit}} (N_i) - T_1(N_i) \right]^2
\end{equation}
\noindent where $d$ is the total number of data points and $p$ is the number of fitted parameters. In our case, 
{$d=49$}, and $p=2$.

\emph{Neville's extrapolation}: The Neville's polynomial extrapolation \cite{PressNum} is applied to estimate $T_1^{\infty}$. This is done by considering $T_1^{\rm{opt}}$ as a function of $(1/N)$ and extrapolating it numerically to $1/N \rightarrow 0$. This method gives an estimate of  $T_1^{\infty} = 0.454$. This is approximately the same as the independent estimate made by the power law fit. The details of the implementation of Neville's algorithm are provided in the Appendix~\ref{app:neville}.

\subsection{Scaling of minimum time required for refrigeration}

As demonstrated in Fig.~\ref{fig:temp:sub1}, the cold qubit temperature, $T_1$, varies periodically in time, and thus has many local minima with respect to time for fixed values of all other parameters. We observe that if we plot $T_1$ as a function of time, $t$, for the choice of optimal parameters $A_1$, $A_2$, $A_3$, and $g$ that minimize the cold qubit temperature, then the value of $T_1$ at the first local minimum with respect to time is approximately same as the global minimum value, $T_1^{\rm{opt}}$. This is demonstrated in Fig~\ref{fig:local-v-global:sub1}. 
We refer to this time of attaining the first local minimum as 
{$t_{\text{min}}$}.
We study the scaling of the time required, {$t_{\text{min}}$}, to reach the first local minimum of $T_1(N)$ 
with respect to $N$. The plot of {$t_{\text{min}}$} as a function of $N$ is shown in Fig.~\ref{fig:local-v-global:sub2}. We employ Neville's extrapolation algorithm to estimate {$t_{\text{min}}^{\infty}$}, the limiting value of {$t_{\text{min}}$} for large $N$. Furthermore, using this value of {$t_{\text{min}}^{\infty}$}, we fit a power law and extract the scaling exponent. We obtained {$t_{\text{min}}^{\infty} = 0.10$}. Using this value in the fitting model
{$t^{\rm{fit}}_{\text{min}}:=t_{\text{min}}^{\infty} + x N^{-y}$} fetches values of the fitting parameters $x=1.5$ and $y =0.62$ with a standard deviation of 0.01. The definition of the standard deviation is as given in the preceding subsection.

{Since both $T_1^{\rm{opt}}$ and {$t_{\text{min}}$} are obtained by optimizing over the interaction parameters for fixed system and bath energies, it is natural to ask how sensitive these quantities are to the choice of {model} parameters. We address this practical-feasibility question in Appendix~\ref{app:sys-params} by sampling the system and bath energies over broad parameter ranges and analyzing the resulting distributions of $T_1^{\rm{opt}}$ and {$t_{\text{min}}$}.}

{\section{Non-Markovianity of the 
cold-qubit dynamics in CSQAR
}\label{sec:non-Markov}}

{In Sec.~\ref{sec:scaling}, we analyzed how the cooling performance of the CSQAR scales with the bath size.
We now fix the bath size and sample the system and bath energies over a broad range to quantify the non-Markovianity of the cold-qubit dynamics and assess its correlation with the achievable transient cooling.}
{In this work, we assume that the initial state of the full system, comprising the three-qubit system together with the local environments, is a tensor product of the local thermal states. This choice is physically well motivated in our setting, as a local temperature can be consistently assigned to each refrigerator qubit only when its reduced state is diagonal in the corresponding local energy eigenbasis. Within this physically relevant class of initial conditions, we quantify non-Markovianity operationally using the Breuer-Laine-Piilo (BLP) information backflow measure~\cite{PhysRevLett.103.210401, PhysRevA.81.062115}. Specifically, we evaluate the BLP measure by restricting the optimization to pairs of thermal initial states of the cold qubit, and denote the resulting quantity by $\mathcal{N}_{\mathrm{BLP}}^{(\mathrm{th})}$.}

{\subsection{ Model parameter dependence of non-Markovianity}}

\begin{figure}[ht]
    \centering
    \includegraphics[width=0.45\textwidth]{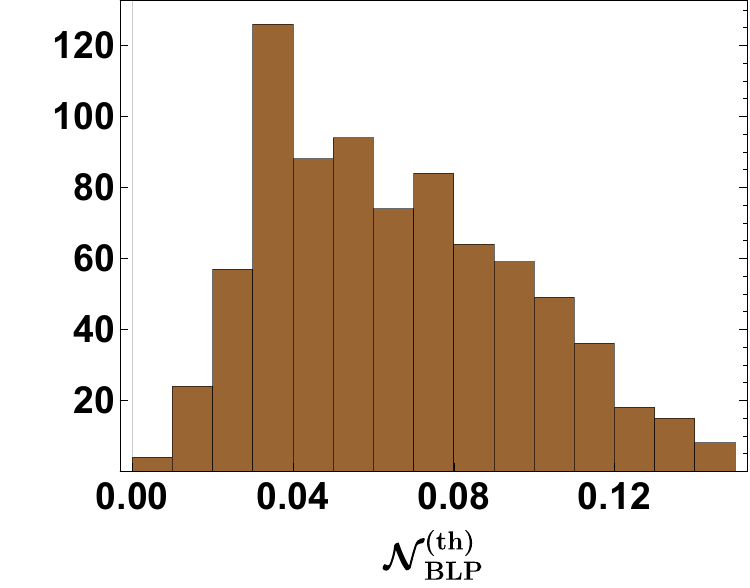}
    \caption{{Histogram of the restricted BLP quantifier $\mathcal{N}_{\mathrm{BLP}}^{(\mathrm{th})}$ for a sample of size $800$. The range of the data is $[0.005002, 0.1484]$ The quantity is computed from the reduced cold-qubit dynamics for $N_1=N_2=N_3=25$ and for randomly generated system and bath energies satisfying the autonomous refrigeration condition~\eqref{eq:autonomous-cs-refri}.}}
    \label{fig:BLP_histo}
\end{figure}

{For the dynamics of the cold qubit, the restricted BLP quantifier is defined as}
\begin{equation}
{\mathcal{N}_{\mathrm{BLP}}^{(\mathrm{th})}
=
\max_{\,T_{1,a}^{\mathrm{ini}},\,T_{1,b}^{\mathrm{ini}} \in [0,1]}
\int_{\sigma>0} dt \,
\sigma\!\left(t,\rho_{1,a}(0),\rho_{1,b}(0)\right),}
\label{eq:blp-restricted}
\end{equation}
{where}
\begin{equation}
{\sigma\!\left(t,\rho_{1,a}(0),\rho_{1,b}(0)\right)
=
\frac{d}{dt}
D\!\left(\rho_{1,a}(t),\rho_{1,b}(t)\right),}
\label{eq:sigma-restricted}
\end{equation}
{and $D(\rho_{1,a}(t),\rho_{1,b}(t))$ is the trace distance between two  cold-qubit states $\rho_{1,a}(t)$ and $\rho_{1,b}(t)$ evolved from two allowed thermal initial states $\rho_{1,a}(0)$ and $\rho_{1,b}(0)$ of the cold qubit. Here the optimization is over the initial cold-qubit temperatures $T_{1,a}^{\mathrm{ini}}$ and $T_{1,b}^{\mathrm{ini}}$ in the interval $[0,1]$, while the other two qubits' initial temperatures are kept fixed at}
\begin{equation}
{T_2^{\mathrm{ini}} = 1, \qquad T_3^{\mathrm{ini}} = 2.}
\end{equation}
{As in the rest of the paper, each qubit is initially in thermal equilibrium with its own local bath.}

{To investigate the dependence of $\mathcal{N}_{\mathrm{BLP}}^{(\mathrm{th})}$ on the model parameters, we generate an ensemble of $800$ parameter sets $(\varepsilon_1,\varepsilon_2,\varepsilon_3,E_1,E_2,E_3)$, sampled independently from uniform distributions subject to the autonomous refrigeration constraint~\eqref{eq:autonomous-cs-refri}. The sampling ranges coincide with those specified in Table~\ref{tab:ranges-random-energies} of Appendix~\ref{app:sys-params}. Throughout this section, we fix}
\begin{equation}
{N_1=N_2=N_3=25.}
\end{equation}
{For each sampled parameter set, $\mathcal{N}_{\mathrm{BLP}}^{(\mathrm{th})}$ is computed over the time interval $t\in[0,2]$. This time window is sufficient to capture the relevant transient information backflow in the considered parameter regime, as the characteristic oscillation period of the cooling dynamics (Figs.~\ref{fig:temp} and~\ref{fig:heat}) is smaller than $2$. For each parameter set, the interaction parameters $A_1$, $A_2$, $A_3$, and $g$ are fixed to the values obtained from the optimization of the minimum cold-qubit temperature.}

{The distribution of $\mathcal{N}_{\mathrm{BLP}}^{(\mathrm{th})}$ over the sampled parameter sets is shown in Fig.~\ref{fig:BLP_histo}. The values are confined to a bounded interval and are well captured by a Johnson $S_B$ distribution. As our primary objective here is to elucidate the relation between non-Markovianity and refrigeration performance, we defer the details of the fit, including the probability distribution function (PDF)/cumulative distribution function (CDF) overlays, quantile-quantile (Q-Q) plot, and goodness-of-fit (GOF) tests, to Appendix~\ref{app:blp-distribution}. We also note that the sample size employed in this analysis is smaller than that used in Appendix~\ref{app:sys-params}, as the evaluation of $\mathcal{N}_{\mathrm{BLP}}^{(\mathrm{th})}$ entails a significantly higher computational cost.}

\begin{figure*}
    \centering
    \subfloat[]{\includegraphics[width=0.45\textwidth]{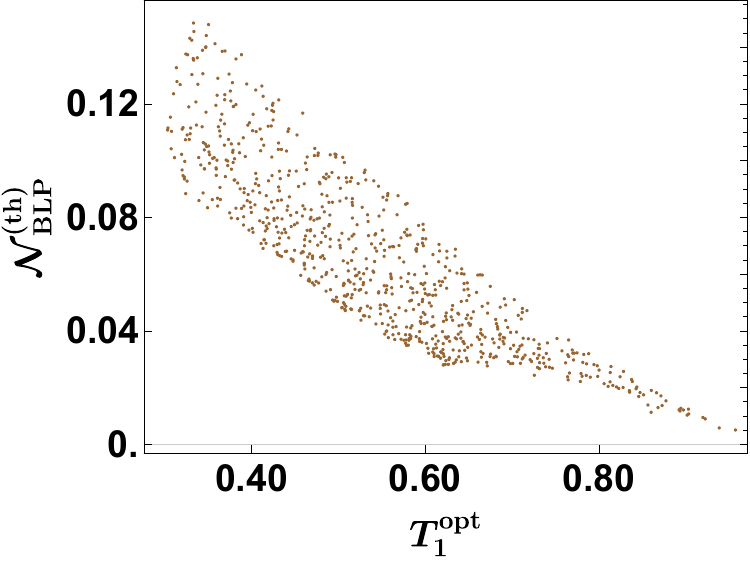} \label{fig:BLP-vs_T1opt}}
    \hfill
    \subfloat[]{\includegraphics[width=0.45\textwidth]{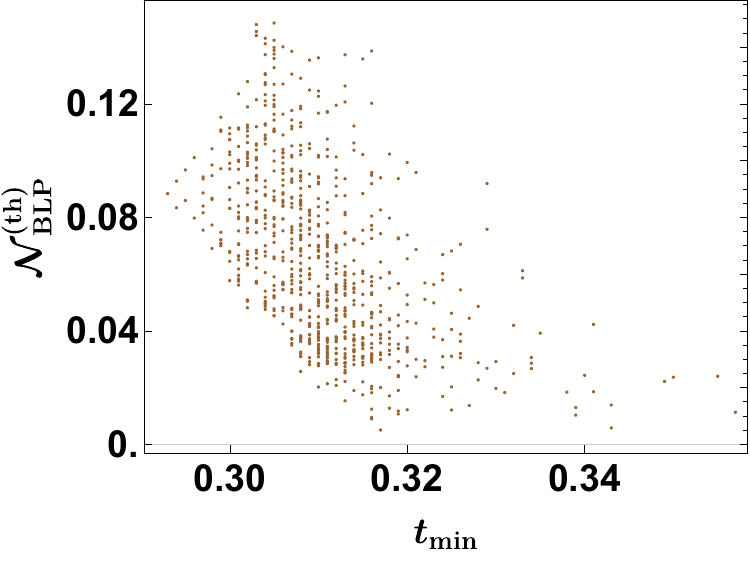} \label{fig:BLP-vs_tmin}}
    \caption{{Panel (a): The plot of $\mathcal{N}_{\mathrm{BLP}}^{(\mathrm{th})}$ vs $T_1^{\text{opt}}$ for a sample of 800 randomly generated sets of system and bath energies. Panel (b): The plot of $\mathcal{N}_{\mathrm{BLP}}^{(\mathrm{th})}$ vs $t_{\text{min}}$ for the same sample of 800 randomly generated sets of system and bath energies}}
    \label{fig:BLP-vs_T1opt-or_tmin}
\end{figure*}

{\subsection{Correlation of non-Markovianity with refrigerator performance}}

{We now investigate whether the non-Markovianity of the cold-qubit dynamics is correlated with the minimum cold-qubit temperature attained and with the minimum time required to attain it. To do so, for the same sample of $800$ randomly generated parameter sets used above, we plot  \\
(i) $\mathcal{N}_{\mathrm{BLP}}^{(\mathrm{th})}$ versus $T_1^{\mathrm{opt}}$ and \\
(ii) $\mathcal{N}_{\mathrm{BLP}}^{(\mathrm{th})}$ versus $t_{\min}$. \\
These plots are shown in Figs.~\ref{fig:BLP-vs_T1opt} and \ref{fig:BLP-vs_tmin}, respectively.}

{To quantify these correlations, we compute both the Pearson correlation coefficient (PCC)~\cite{Pearson1895} and the Spearman rank correlation coefficient (SCC)~\cite{Spearman1904}.  
The corresponding values are listed in Table~\ref{tab:PCC-and-SCC}. The PCC between any two lists $u$ and $v$ is defined as} 
\begin{align}
    {\mathcal{C}_{uv} := \frac{\text{Cov}(u,v)}{\sigma_u \sigma_v}}
\end{align}
{where}
\begin{align}
    {\text{Cov}(u,v) := \frac{1}{n - 1} \sum_{i=1}^n (u_i - \hat{\mu}_u)(v_i - \hat{\mu}_v)^*}
\end{align}
{is the covariance. Here $\hat{\mu}_u$ and $\hat{\mu}_v$ are means of lists $u$ and $v$, respectively, $n$ is the length of each list, and $\sigma_u$, $\sigma_v$ are respective standard deviations. The value of PCC lies between -1 and 1. The PCC value close to 1 indicates strong positive correlation between lists $u$ and $v$ whereas value close to -1 suggests strong negative correlation. SCC is defined as PCC between rank variables. It also lies between -1 and 1.}

{We find a strong negative correlation between $\mathcal{N}_{\mathrm{BLP}}^{(\mathrm{th})}$ and $T_1^{\mathrm{opt}}$, and a weaker yet still discernible negative correlation between $\mathcal{N}_{\mathrm{BLP}}^{(\mathrm{th})}$ and $t_{\min}$. These results indicate that, within the parameter regime explored here, stronger non-Markovianity in the reduced cold-qubit dynamics tends to be associated with lower transient minimum temperatures and, to a lesser extent, with faster cooling. We stress, however, that this should be understood as an empirical trend observed within the sampled ensemble employed in this work. Accordingly, the observed correlations should be interpreted as emergent features of the full CSQAR architecture, rather than as evidence of a direct causal role of non-Markovianity alone.}

{An analysis of the direct sensitivity of $T_1^{\mathrm{opt}}$ and $t_{\min}$ to the choice of  model parameters is presented in Appendix~\ref{app:sys-params}; this directly addresses the practical-feasibility question of whether the observed transient cooling is strongly fine tuned.}

\begin{table}[h!]
\centering
\begin{tabular}{|c|c|c|c|}
\hline
List 1 & List 2 & PCC & SCC \\ 
\hline
 $\mathcal{N}_{\text{BLP}}$ & $T_1^{\text{opt}}$ & -0.8684 &  -0.8919\\ 
\hline
 $\mathcal{N}_{\text{BLP}}$ & $t_{\text{min}}$ & -0.5227 & -0.5902 \\ 
\hline
\end{tabular}
\caption{{Pearson correlation coefficient (PCC) and Spearman rank correlation coefficient (SCC) between $\mathcal{N}_{\mathrm{BLP}}^{(\mathrm{th})}$ and the refrigeration performance indicators $T_1^{\mathrm{opt}}$ and $t_{\min}$ for a sample of $800$ randomly generated parameter sets.}}
\label{tab:PCC-and-SCC}
\end{table}

{\section{ Comparison between Markovian QAR and CSQAR
}\label{sec:Markov}}

\begin{figure*}
    \centering
    \subfloat[]{\includegraphics[width=0.45\textwidth]{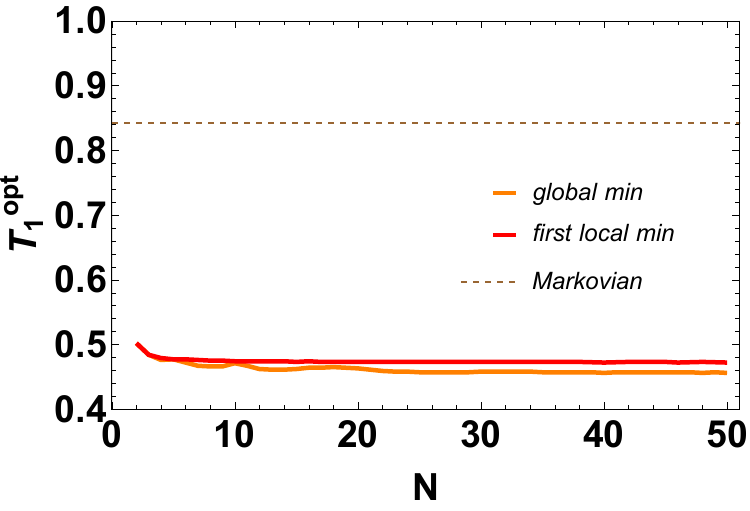} \label{fig:compare-markov:sub1}}
    \hfill
    \subfloat[]{\includegraphics[width=0.45\textwidth]{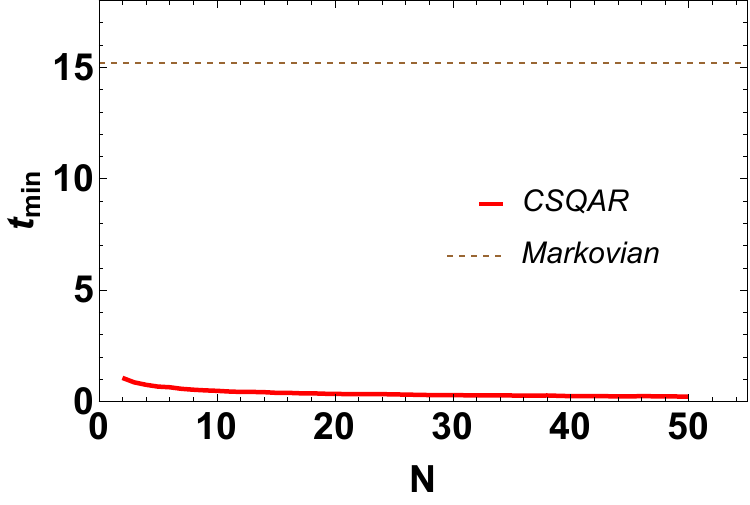} \label{fig:compare-markov:sub2}}
    \caption{Panel (a): Comparison of the global minimum of $T_1$ (given in orange) and the first local minimum in time (given in red) as functions 
    of the number of environment spins $N =N_1 = N_2 = N_3$, with the minimum $T_1$ obtained when all the environments are Markovian (given in brown dashed). Panel (b): Comparison of the minimum time, $t_{\text{min}}$,  required to reach the local minimum 
    of $T_1$ (given in red), with the corresponding minimum cooling time when all the environments are Markovian (given in brown dashed). The energies of three qubits are $\varepsilon_1 = 1$, $\varepsilon_2 = 2$, and $\varepsilon_3 = 1$. The corresponding bath energies are $E_1 = 2$, $E_2 = 4$, and $E_3 = 2$. The initial qubit temperatures are {$T^{\text{ini}}_1 = 1$, $T^{\text{ini}}_2 =1$, and $T^{\text{ini}}_3 =2$}. The quantities plotted along the horizontal and vertical axes in each panel are dimensionless.}
    \label{fig:compare-markov}
\end{figure*}

In this section, we compare the performance of 
the CSQAR
with that of a three-qubit QAR
where each of the three qubits is connected to a Markovian environment.
We demonstrate a significant advantage in cooling of the cold qubit of the CSQAR.
For the three-qubit QAR,
whose system Hamiltonian is given by Eq~\eqref{eq:HS-qref-M}, the Markovian baths corresponding to each qubit are composed of an infinite number of harmonic oscillators, with the $i^{\text{th}}$ bath Hamiltonian being given by
\begin{equation}
    \widetilde{H}_B^{(i)} =  \int_0^\Omega \hbar K  \eta_\omega^{i \dagger} \eta_\omega^i d \omega.
\end{equation}
\noindent Here 
$\Omega$ denotes the cut-off frequency, which is assumed to be identical for all three baths. It is chosen to be sufficiently high to ensure that the bath's memory time, $\sim \Omega^{-1}$, remains negligible. The operators $\eta_\omega^{i\ \dagger}$ ($\eta_\omega^{i}$), having the unit of $\frac{1}{\sqrt{\omega}}$, represent bosonic creation (annihilation) operators associated with the mode $\omega$ of the $i^{\text{th}}$ bath. The interaction between the $i^{\text{th}}$ system qubit and its Markovian environment is described by the following interaction Hamiltonian
\begin{equation}
    \widetilde{H}_{SB}^{(i)} =\int_0^\Omega  \frac{\hbar}{K}   d\omega a_i(\omega) \sigma^x_i (\eta_\omega^{i} +  \eta_\omega^{i\ \dagger})
\end{equation}
\noindent $a_i(\omega)$ is the coupling between $i^{\text{th}}$ system qubit and its environment. For harmonic oscillator environments $a_i^2(\omega) = J_i(\omega)$, where $J_i(\omega)$ is the spectral density function of $i^{\text{th}}$ bath. We take $J_i(\omega)$ to be Ohmic spectral density function of the form $J_i(\omega) = \alpha_i \omega \text{exp} (-\omega/\Omega)$. Here $\alpha_i$s stand for the dimensionless qubit-environment interaction strengths. The effect of Markovian baths on the reduced density matrix of the three-qubit quantum refrigerator can be described by the GKSL quantum master equation, given by 
\begin{equation}\label{eq:qref-M-qme}
     \frac{\partial \rho_s (t)}{\partial t} = \mathcal{L}\left( \rho_s(t) \right) = - \frac{ \mathrm{i} }{ \hbar K} [\widetilde{H}_S, \rho_s(t) ] + \frac{1}{K} \sum_{i=1}^3 \mathcal{D}_i \left( \rho_s(t) \right),
 \end{equation}
\noindent with the dissipative terms of the form

{ \small
\begin{equation}
\label{Lindblad}
    \mathcal{D}_i \left( \rho_s(t) \right) =  \sum_{\omega^\prime}  K \gamma_i (\omega^\prime) \left[  L_i^{\omega^\prime} \rho_s(t) L_i^{\omega^{\prime \dagger}}  - \frac{1}{2} \{ L_i^{\omega^{\prime \dagger}} L_i^{\omega^\prime} , \rho_s(t)  \}\right].
\end{equation}
}
\noindent Here $i=1,2,3$. $\tilde{\gamma}_i (\omega^\prime) =  K \gamma_i (\omega^\prime)$ are decay rates with units of inverse time, and $\gamma_i$ being their dimensionless versions. The operators $L_i^{\omega^{\prime}}$ are Lindblad or quantum jump operators linked to the transition frequency $\omega^{\prime}$ of the system. The expressions for $\gamma_i$ and $L_i^{\omega^{\prime}}$ are provided in the Appendix~\ref{app:linblad}.
To ensure the validity of the Born-Markov approximation, we work in the weak coupling regime, characterized by $ \text{max} \{\gamma_i (\omega^\prime) \} \ll \text{min} \{ \varepsilon_i, {\tilde{g}} \}$; where $\varepsilon_i$ are the system energies and ${\tilde{g}}$ is the strength of the three-body interaction~\cite{BreuerOQS, AlickiOQS, RivasHuelgaOQS, lidar2020}.
Note that the entire analysis in this work regarding Lindblad-type dynamics is based on a global Lindblad master equation. Specifically, the jump operators in the master equation are derived by decomposing the interaction Hamiltonian of the system and environment in the eigenbasis of the total three-qubit system Hamiltonian, along with the interactions between these qubits.

We optimize the cold qubit temperature $T_1$ over the parameters $\alpha_1$, $\alpha_2$, $\alpha_3$, ${\tilde{g}}$ and time $t$, and obtain a global minimum of $T_1^{\text{opt}} = 0.842$ at the minimum time $t_{min} = 15.2$, with initial cold qubit temperature of $T_1 = 1$. The corresponding values of parameters are $\alpha_1 = 7.98 \times 10^{-6}$, $\alpha_2 = 2.67 \times 10^{-5}$, $\alpha_3 = 3.13 \times 10^{-5}$, and ${\tilde{g}} = 0.0999197$. We chose the system energies to be $\varepsilon_1 = 1$, $\varepsilon_2=2$, and $\varepsilon_3=1$, which are the same as the values of corresponding system energies chosen for the CSQAR. The values of the initial temperatures for the other two qubits were also chosen to be the same as their chosen values in CSQAR, namely {$T^{\text{ini}}_2 =1$, and $T^{\text{ini}}_3 =2$}.

In Fig.~\ref{fig:compare-markov:sub1}, we compare the optimum cold qubit temperature obtained in the Markovian scenario {with three-body interaction} with that of the CSQAR {with effective six-body interaction}.
For the 
{CSQAR}, we plot two temperatures, the first of which is the global minimum plotted in orange, while the second one refers to the first local minimum, which is plotted in red. The method of calculating these two quantities is as mentioned in the preceding section. We find that $T_1^{\text{opt}}$ is significantly lower in the CSQAR
than in the QAR with Markovian environments {and three-body interaction}. This result is independent of the number of bath spins in the central-spin model.
We perform a similar comparison of the minimum time required to achieve the minimum cold qubit temperature, $t_{\text{min}}$, for the Markovian QAR and CSQAR in Fig.~\ref{fig:compare-markov:sub2}. We find that $t_{\text{min}}$ is an order of magnitude lower for the CSQAR than the Markovian QAR. So, the optimal temperature for the CSQAR is not only lower but also attained in significantly less time than the Markovian environment QAR.
These comparisons demonstrate the advantages of our quantum refrigerator model over the conventional model with Markovian environments {and three-body interaction between refrigerator qubits}. 

{In this section, the comparison is performed between two distinct models: the CSQAR with spin-star environments and an effective six-body interaction, and the conventional Markovian QAR with the standard three-body interaction introduced by Linden \textit{et al.}~\cite{PhysRevLett.105.130401}. The purpose of this comparison is to benchmark the overall cooling performance of the full CSQAR architecture against the standard Markovian QAR, rather than to attribute the observed advantage solely to non-Markovian effects. Accordingly, the lower minimum temperature and shorter cooling time should be interpreted as characteristics of the full CSQAR model, 
rather than a feature of either non-Markovianity or effective six-body interaction.}

{We further emphasize that the effective six-body interaction is not introduced ad hoc, but emerges naturally when the refrigerator qubits are embedded in spin-star environments and the dynamics is formulated in the invariant collective-spin basis. This structure also enables the exploitation of symmetries, allowing for a semi-analytic treatment even for large bath sizes.
} \\

\section{Conclusion}\label{sec:conclusion}

In this work, we scrutinized a quantum absorption refrigerator comprising three qubits, where each qubit is connected to a separate spin-star environment{, with the three qubit-bath units coupled through an effective six-body interaction}. Exploiting the symmetries of the {Hamiltonian}, we developed a semi-analytic method that enabled the study of transient cooling dynamics even in the presence of spin-star environments with a large number of spins. Since there are no assumptions of Markovianity, this model depicts non-Markovian behavior, and the attainment of a steady state is not guaranteed. Therefore, unlike conventional Markovian models that {often} focus on steady-state cooling, 
{our approach provides a platform to investigate transient refrigeration in the presence of structured non-Markovian spin-star environments}. We showed that the cold qubit's temperature varies periodically in time, characterized by alternate cooling and heating periods. During the cooling period, the heat moves from the cold qubit into its local environment, whereas, during the heating periods, the direction of this heat flow is opposite. 
{We argued that the heat current associated with the cold qubit's environment can be regarded as an approximate indicator of the cooling power of the CSQAR in the transient regime}.

We investigated the scaling behavior of the optimal temperature of the cold qubit with the number of environmental spins. Evaluation of the optimal temperature involves minimizing the temperature of the cold qubit over the system-bath coupling constants, the effective six-body interaction strength, and the time of evolution. Our results indicate that increasing the number of bath spins enhances cooling performance, with the cold-qubit temperature approaching a limiting value for a large number of bath spins following a power-law behavior with an exponent of 1.096. Our analysis demonstrated that the central-spin quantum refrigerator exhibits efficient cooling in the transient regime, which may or may not lead to steady-state cooling over a long period of time. Additionally, we obtained the scaling of the minimum time required to attain the optimal temperature of the cold qubit with the number of spins in the environment. {We also found that this transient-cooling performance is reasonably robust under broad variation of the system and bath energies, rather than being confined to a narrowly fine-tuned parameter set.
}

{We further quantified the non-Markovianity of the cold-qubit dynamics using a restricted Breuer-Laine-Piilo-type information-backflow measure. Within the sampled parameter regime, we observed a strong negative correlation between the restricted BLP quantifier and the optimal cold-qubit temperature, together with a weaker negative correlation with the minimum cooling time. These results indicate that, across the sampled ensemble, stronger information backflow is generally associated with lower transient minimum temperatures and, to a lesser extent, faster cooling. We emphasize, however, that this correlation should be interpreted as a characteristic feature of the full CSQAR architecture, rather than as evidence that non-Markovianity alone constitutes the unique causal mechanism underlying the enhanced cooling performance.}

{Finally, we compared the full CSQAR architecture with the conventional Markovian three-qubit quantum absorption refrigerator.} 
{We found that the CSQAR attains lower cold-qubit temperatures on shorter timescales than the conventional Markovian benchmark. This advantage should be understood as a property of the full CSQAR model, which combines non-Markovian spin-star environments with the effective six-body interaction, rather than as an effect of non-Markovianity alone.}

\acknowledgements
AB acknowledges support from ‘INFOSYS scholarship for senior students’ at Harish Chnadra Research Institute, India. US acknowledges financial support from the Anusandhan National Research Foundation (ANRF), Government of India, under the Grant No. ANRF/ARG/2025/004617/PS.

\appendix

\section{Matrix elements of the time-evolved density operator of a single central-spin system}
\label{matrix_ele}
Here we provide the exact analytic expressions for each matrix element of $\rho_m(t)$ given in Eq.~\eqref{rho}. The four coefficients, $c(t)^m_{-1/2;-1/2}, c(t)^m_{1/2;1/2}, c(t)^m_{-1/2;1/2}$ and $c(t)^m_{1/2;-1/2}$ , for a given value of $m$ are given by
\begin{align}\label{eq:cs-gs-t}
&  c(t)^m_{-1/2;-1/2} \nonumber \\
    & =\frac{1 }{4 \theta^2} \Bigg( 4 u^2 \Bigg(   \text{cos} \left( 2 \theta t \right) \text{sinh}\left( \frac{\beta}{2}  (\epsilon - E)  \right)  \nonumber  \\ 
    &+ \text{cosh}\left( \frac{\beta}{2}  (\epsilon - E)\right) \Bigg)  
    + e^{\frac{\beta}{2} (\epsilon - E)} \left( b_{-1/2} - b_{1/2} \right)^2 \Bigg)
\end{align}

\begin{align}\label{eq:cs-es-t}
 &  c(t)^m_{1/2;1/2} \nonumber \\
    & =\frac{1 }{4 \theta^2} \Bigg( 4 u^2 \Bigg(  - \text{cos} \left( 2 \theta t \right) \text{sinh}\left( \frac{\beta}{2}  (\epsilon - E)  \right)  \nonumber  \\ 
    &+ \text{cosh}\left( \frac{\beta}{2}  (\epsilon - E)\right) \Bigg)  
    + e^{\frac{\beta}{2} (-\epsilon + E)} \left( b_{-1/2} - b_{1/2} \right)^2 \Bigg)
\end{align}

\begin{align}\label{eq:cs-offdiag-t}
   & c(t)^m_{-1/2;1/2} =  \left(c(t)^m_{1/2;-1/2}\right)^* \nonumber \\
    &=  \frac{4 u \text{sin}\left(  \theta t\right) \text{sinh}\left( \frac{\beta}{2}  (\epsilon - E)  \right) }{4 \theta^2}   \times \nonumber \\
    & \Bigg( - 2 \mathrm{i} \theta  \text{cos} \left( \theta t \right) + \left( b_{-1/2} - b_{1/2} \right) \text{sin}\left(  \theta t\right)   \Bigg),
\end{align}

where 

\begin{equation}
    \theta = \sqrt{u^2  + \frac{1}{4} \left( b_{-1/2} - b_{1/2} \right)^2}.
\end{equation}

\begin{table}[htbp]
\centering
\caption{Optimized values of parameters corresponding to Fig.~\ref{fig:T1opt-N-scaling}}
\label{tab:optimizer}
\renewcommand{\arraystretch}{1.2}
\begin{tabular}{|c|c|c|c|c|c|c|}
\hline
$N$ & $A_1$ & $A_2$ & $A_3$ & $g$ & $t$ & $T_1^{\text{opt}}$ \\
\hline
2  & 0.999977 & 0.812687 & 0.838071 & 0.0974429 & 9.42432 &  0.502 \\
\hline
3  & 0.999974 & 0.994118 & 0.788688 & 0.094078 & 0.854837 & 0.484 \\
\hline
4  & 0.993917 & 0.54881 & 0.291878 & 0.00765429 & 6.88939 & 0.476 \\
\hline
5  & 1  & 0.0156779 & 0.428304 & 0.000611268 & 0.658423 &  0.477\\
\hline
6  & 0.953927 & 0.311895 & 0.270975 & 0.00228477 & 9.8334 &  0.472\\
\hline
7  & 0.983625 & 0.610295 & 0.655527 & 0.0154557 & 7.70913 &  0.467 \\
\hline
8  & 0.998244 & 0.30274 & 0.507758 & 0.0297916 & 7.11285 &  0.466 \\
\hline
9  & 0.993499 & 0.337532 & 0.364276 & 0.0423596 & 9.86206 &  0.466 \\
\hline
10 & 0.994992 & 0.404259 & 0.0593631 & 0.0723525 & 5.45032 &  0.471 \\
\hline
11 & 0.992858 & 0.490341 & 0.44579 & 0.0354257 & 5.20832 &  0.467 \\
\hline
12 & 0.998175 & 0.954794 & 0.630628 & 0.0495836 & 7.64982 &  0.462 \\
\hline
13 & 0.962915 & 0.646159 & 0.314688 & 0.0727223 & 7.62486 &  0.461 \\
\hline
14 & 0.950414 & 0.0693204 & 0.465566 & 0.0901092 & 7.43319 &  0.461 \\
\hline
15 & 0.998992 & 0.316213 & 0.526803 & 0.0595813 & 6.83722 &  0.462 \\
\hline
16 & 0.961793 & 0.235765 & 0.707898 & 0.066271 & 6.87238 &  0.464 \\
\hline
17 & 0.986025 & 0.175285 & 0.63437 & 0.0446051 & 4.21192 &  0.464 \\
\hline
18 & 0.965267 & 0.266784 & 0.80501 & 0.0294008 & 4.18774 &  0.465 \\
\hline
19 & 0.998464 & 0.524681 & 0.963726 & 0.024321 & 9.68238 &  0.464 \\
\hline
20 & 0.997259 & 0.348223 & 0.664864 & 0.0519767 & 7.36551 &  0.463 \\
\hline
21 & 0.9917 & 0.490369 & 0.0622195 & 0.0136257 & 7.22413 &  0.461 \\
\hline
22 & 0.983193 & 0.0713195 & 0.295551 & 0.0436792 & 7.11476 &  0.459 \\
\hline
23 & 0.982641 & 0.291765 & 0.781429 & 0.0414219 & 6.96474  &  0.458 \\
\hline
24 & 0.935233 & 0.303027 & 0.724068 & 0.0410827 & 7.16194 &  0.458 \\
\hline
25 & 0.981765 & 0.0887574 & 0.804871 & 0.0345816 & 6.68548 &  0.457 \\
\hline
26 & 0.952472 & 0.212375 & 0.331091 & 0.0341424 & 6.75026 &  0.457\\
\hline
27 & 0.98669 & 0.988809 & 0.410661 & 0.0618329 & 6.40057 &  0.457 \\
\hline
28 & 0.994281 & 0.259026 & 0.987737 & 0.036389 & 6.23297 &  0.457 \\
\hline
29 & 0.961515 & 0.800217 & 0.206622 & 0.000634969 & 6.33564 &  0.457 \\
\hline
30 & 0.950788 & 0.308299 & 0.440829 & 0.0233622 & 6.29494 &  0.458 \\
\hline
31 & 0.896336 & 0.228588 & 0.119953 & 0.0135877 & 6.57002 &  0.458 \\
\hline
32 & 0.998218 & 0.205587 & 0.246248 & 0.0456422 & 4.1634 &  0.458 \\
\hline
33 & 0.978284 & 0.401234 & 0.551214 & 0.0596998 & 4.18642 &  0.458 \\
\hline
34 & 0.926591 & 0.171625 & 0.299485 & 0.0654351 & 4.34676 & 0.458 \\
\hline
35 & 0.99874 & 0.495665 & 0.859154 & 0.0706887 & 3.9747 &  0.457 \\
\hline
36 & 0.997335 & 0.0841537 & 0.548582 & 0.0466855 & 3.9257 &  0.457 \\
\hline
37 & 0.958109 & 0.572043 & 0.344442 & 0.00954159 & 4.03376 &  0.457 \\
\hline
38 & 0.929821  & 0.323026 & 0.110755 & 0.0296691 & 4.09842 &  0.457 \\
\hline
39 & 0.970603 & 0.874637 & 0.451826 & 0.00439431 & 3.87506 &  0.457 \\
\hline
40 & 0.999859 & 0.408566 & 0.544846 & 0.0211024  & 3.71493 &  0.456 \\
\hline
41 & 0.935339 &  0.918743 & 0.958961 & 0.0904981 & 3.92168 &  0.457 \\
\hline
42 & 0.973247 & 0.66748 & 0.777264 & 0.0702673 & 3.7246 &  0.457 \\
\hline
43 & 0.897937 & 0.132162 & 0.595918 & 0.0411739 & 3.98592 & 0.457 \\
\hline
44 & 0.971576 & 0.855282 & 0.248196 & 0.031324 & 3.64451 & 0.456 \\
\hline
45 & 0.955114 & 0.793575 & 0.944425 & 0.02730 & 3.6642 & 0.457 \\
\hline
46 & 0.946234 & 0.324845 & 0.501082 & 0.00539495 & 3.66228 & 0.457 \\
\hline
47 & 0.968534 & 0.298291 & 0.614746 & 0.0110394 & 3.53671 & 0.457 \\
\hline
48 & 0.987951 & 0.664708 & 0.513699 & 0.0257276 & 3.43165 & 0.456 \\
\hline
49 & 0.985172  & 0.643441 & 0.761288 & 0.0910783 & 3.41036 & 0.457 \\
\hline
50 & 0.997862 & 0.91032 & 0.262157 & 0.0221283 & 3.33077 & 0.456 \\
\hline
\end{tabular}
\end{table}

\section{Form of the density operator within each \texorpdfstring{$(m_1, m_2, m_3)$}{(m1, m2, m3)} sector}
\label{app:rho_m}
The explicit form of the
time evolved density matrix, $\rho_{\textbf{m}}(t)$ within each $\textbf{m} :=(m_1, m_2, m_3 )$ sector, as given in~\eqref{eq:rho-cs-refri}, is provided below.

\begin{align}\label{eq:rho-cs-ref}
    &\rho_{\textbf{m}}(t) \nonumber \\
    &= \tilde{c}^{\textbf{m}}_{-1/2;-1/2;-1/2} |\psi_{-1/2;-1/2;-1/2} \rangle \langle \psi_{-1/2;-1/2;-1/2}| \nonumber \\
    &+ \tilde{c}^{\textbf{m}}_{-1/2;-1/2;1/2} |\psi_{-1/2;-1/2;1/2} \rangle \langle \psi_{-1/2;-1/2;1/2}| \nonumber \\
    &+ \tilde{c}^{\textbf{m}}_{-1/2;1/2;-1/2} |\psi_{-1/2;1/2;-1/2} \rangle \langle \psi_{-1/2;1/2;-1/2}| \nonumber \\
    &+ \tilde{c}^{\textbf{m}}_{-1/2;1/2;1/2} |\psi_{-1/2;1/2;1/2} \rangle \langle \psi_{-1/2;1/2;1/2}| \nonumber \\
    &+ \tilde{c}^{\textbf{m}}_{1/2;-1/2;-1/2} |\psi_{1/2;-1/2;-1/2} \rangle \langle \psi_{1/2;-1/2;-1/2}| \nonumber \\
     &+ \tilde{c}^{\textbf{m}}_{1/2;-1/2;1/2} |\psi_{1/2;-1/2;1/2} \rangle \langle \psi_{1/2;-1/2;1/2}| \nonumber \\
      &+ \tilde{c}^{\textbf{m}}_{1/2;1/2;-1/2} |\psi_{1/2;1/2;-1/2} \rangle \langle \psi_{1/2;1/2;-1/2}| \nonumber \\
       &+ \tilde{c}^{\textbf{m}}_{1/2;1/2;1/2} |\psi_{1/2;1/2;1/2} \rangle \langle \psi_{1/2;1/2;1/2}| \nonumber \\ 
       &+ \text{off-diagonal} \; \; \text{terms}
\end{align}
with
\begin{align}\label{eq:qref-basis}
     &|\psi_{ \pm 1/2; \pm 1/2;\pm 1/2} \rangle \nonumber \\
     &= |\pm 1/2 \rangle | m_1 \mp 1/2\rangle |\pm 1/2 \rangle | m_2 \mp 1/2\rangle | \pm 1/2 \rangle | m_3 \mp 1/2\rangle  \nonumber \\
\end{align}

\begin{figure*}
    \centering
    \subfloat[]{\includegraphics[width=0.45\textwidth]{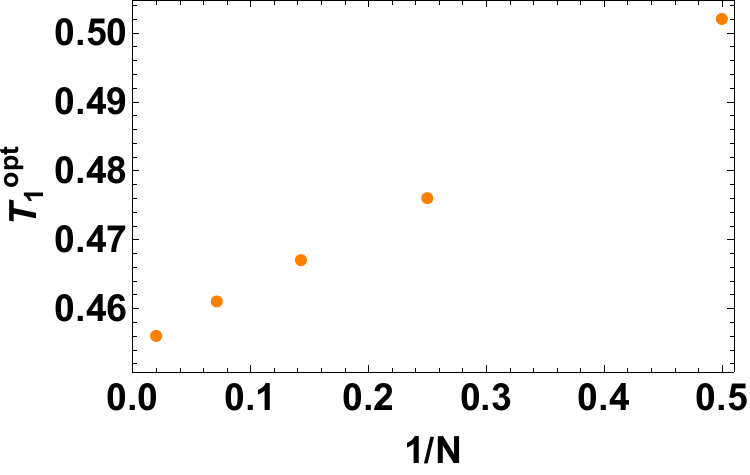} \label{fig:h-scaling-neville:sub1}}
    \hfill
    \subfloat[]{\includegraphics[width=0.45\textwidth]{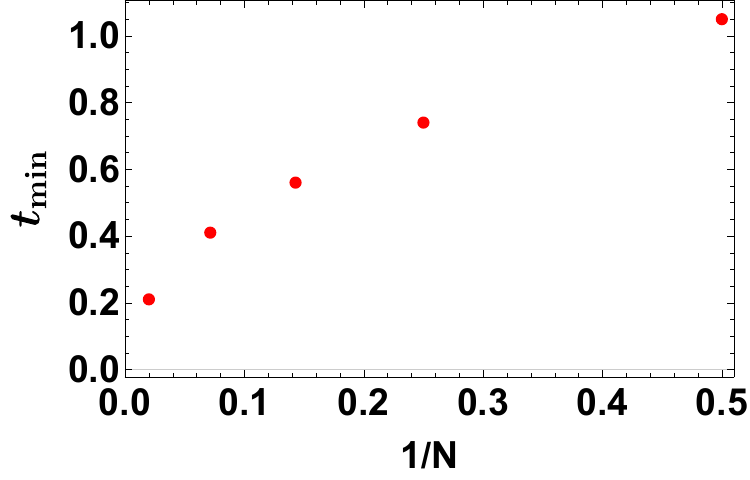} \label{fig:h-scaling-neville:sub2}}
    \caption{Scaling of $T_1^{\rm{opt}}$ and {$t_{\text{min}}$} with $h=1/N$; data points used for Neville's extrapolation to $N \rightarrow \infty$. The energies of three qubits are $\varepsilon_1 = 1$, $\varepsilon_2 = 2$, and $\varepsilon_3 = 1$. The corresponding bath energies are $E_1 = 2$, $E_2 = 4$, and $E_3 = 2$. The initial qubit temperatures are {$T^{\text{ini}}_1 = 1$, $T^{\text{ini}}_2 =1$, and $T^{\text{ini}}_3 =2$}. The quantities plotted along both the axes in each of the panels are dimensionless. }
    \label{fig:h-scaling-neville}
\end{figure*}

\section{Table of optimized parameter values}\label{app:otimizer}

{The optimized values of the parameters $A_1$, $A_2$, $A_3$, $g$, and $t$ for each value of $N$ plotted in Fig.~\ref{fig:T1opt-N-scaling} together with the resulting optimum value of cold-qubit temperature $T_1^{\rm{opt}}$ are provided in Table~\ref{tab:optimizer}.}

\section{Details of extrapolation}\label{app:neville}

\begin{table}
    \centering
    \begin{tabular}{c c c c c c }
       $x_1:$ & $y_1 = P_1$ & & & &  \\ 
           &   & $P_{12}$ & & &   \\ 
        $x_2:$   & $y_2 = P_2$   &   & $P_{123}$& &   \\ 
          &   & $P_{23}$ & &  $P_{1234}$ &   \\ 
           $x_3:$   & $y_3 = P_3$   &  & $P_{234}$ & & $P_{12345}$  \\ 
           &   & $P_{34}$ & &  $P_{2345}$ &   \\
     $x_4:$   & $y_4 = P_4$   & & $P_{345}$ & &   \\ 
      &   & $P_{45}$ & & &   \\
      $x_5:$   & $y_5 = P_5$   & & & &   \\ 
    \end{tabular}
    \caption{Tableu for Neville's algorithm with $n=5$}
    \label{tab:neville}
\end{table}

In this Appendix, we give details of Neville's extrapolation \cite{PressNum} performed to estimate $T_1^{\infty}$. Given a set of $n$ data points $(x_i,y_i)$, $i = 1,2, \cdots n$, one can extrapolate the value of $y = f(x)$ corresponding to $x$ which is outside of the range of data points $x_i$ using Neville's algorithm. There are a few caveats to be mentioned. Let the data points be arranged in an increasing order of $x$ so that $x_1 < x_2 < x_3< \cdots <x_n$ and say that the extrapolation point $x$ is greater than $x_n$. To ensure the stability of Neville's extrapolation, the difference $|x - x_n|$ should not be larger than the difference of any two consecutive data points, i.e. $|x - x_n| < |x_i - x_j| \; \; \forall \; \; 1 < i,j < n$, $i \neq j$. 

To perform Neville's extrapolation on our data, the appropriate variable to work with is $h = 1/N$. We use five data points $(h, T_1^{\rm{opt}})$ with $h = 1/2, 1/4, 1/7, 1/14, 1/50$ for the extrapolation. This choice of data points is made such that the difference between the consecutive $h$ values is larger than twice $1/50$, ensuring stability. Note that we extrapolate to $h = 0$.

The idea of Neville's algorithm can be illustrated through Table~\ref{tab:neville}. Each $P_i$ is a polynomial of degree zero passing through the point $(x_i,y_i)$. $P_{ij}$ is a polynomial of degree one passing through two points $(x_i, y_i)$ and $(x_j,y_j)$, and so on. The different $P$'s create a tableau, with ``ancestors'' on the left and converging toward a single ``descendant'' on the far right. Neville's algorithm recursively populates the tableau, filling in the numbers one column at a time from left to right. It is based on the following relationship between a ``daughter'' $P$, and its two ``parents''

\begin{align}\label{eq:neville}
    P_{i (i + 1) \cdots (i + m)} &= \frac{1}{x_i - x_{i + m}} \Big( (x - x_{i + m} P_{i (i + 1)\cdots(i+m-1)} \nonumber \\ 
    &+ (x_{i} - x)) P_{(i+1)(i+2)\cdots (i+m)} \Big)
\end{align}

The error is estimated by keeping track of the small differences between parents and daughters:

\begin{subequations}\label{eq:neville-error}
    \begin{align}
        C_{m,i} &\equiv P_{i \cdots (i + m)} - P_{i \cdots (i + m -1)} \label{eq:neville-error-first} \\
        D_{m,i} & \equiv P_{i \cdots (i+m)} - P_{(i+1)\cdots(i + m)} \\
        m = 1, 2, \cdots, n-1   \nonumber
    \end{align}
\end{subequations}

Since we are extrapolating to $x$ beyond $x_5 = 1/50$, only the $D_{m,i}$ are relevant for estimating extrapolation error in our case. We obtain $P_{12345} (h = 0) = 0.454$. The differences $D_{m,i}$ decrease as one approaches $P_{12345}$ along the lower diagonal in tableau of~\ref{tab:neville}: $P_5 \rightarrow P_{45} \rightarrow P_{345} \rightarrow P_{2345} \rightarrow P_{12345}$. The corresponding $D$ values are: $D_{1,4} = 0.002$, $D_{2,3} = 0.0002$, $D_{3,2} = 0.0001$, $D_{4,1}=0.00004$. The optimization data of Fig.~\ref{fig:T1opt-N-scaling} has been generated from NLopt to be accurate up to the third decimal place. Therefore, the errors $D_{2,3}$ onwards are essentially zero within the targeted accuracy. The data used for extrapolation is plotted in Fig.~\ref{fig:h-scaling-neville}.

{\section{Robustness of refrigeration performance under variation of {model} parameters
}\label{app:sys-params}}

{In this Appendix, we address the practical-feasibility question of whether the transient refrigeration predicted in the CSQAR requires strong fine tuning of the {model} parameters. To this end, we study how the minimum cold-qubit temperature and the minimum time required to attain it vary under broad changes of the system and bath energies. Throughout this Appendix, the number of spins in each bath is fixed to be $N=25$, and the initial temperatures of the three refrigerator qubits are chosen as}
\begin{equation}
{T^{\mathrm{ini}}_1=1,\qquad T^{\mathrm{ini}}_2=1,\qquad T^{\mathrm{ini}}_3=2.}
\end{equation}
{We have generated a sample of $12800$ sets of six {system and bath} parameters $(\varepsilon_1,\varepsilon_2,\varepsilon_3,E_1,E_2,E_3)$ drawn randomly from uniform distributions subject to the autonomous refrigeration condition~\eqref{eq:autonomous-cs-refri}. The ranges of the corresponding uniform distributions are specified in Table~\ref{tab:ranges-random-energies}.}

{For each sampled choice of $\{\varepsilon_i,E_i\}_{i=1,2,3}$, we compute the minimum cold-qubit temperature $T_1^{\mathrm{opt}}$ by optimizing over five parameters $A_1,A_2,A_3,g$, and $t$. Reacll that $A_i$, for $i=1,2,3$, denote the interaction strengths of the refrigerator qubits with their respective local spin baths, $g$ is the strength of the effective six-body interaction, and $t$ is the evolution time. The optimization ranges for these parameters are chosen to be}
\begin{equation}
{0\leq A_i\leq 1,\qquad 0\leq g\leq 0.1,\qquad 0\leq t\leq 2.}
\end{equation}
{Furthermore, for each sampled choice of $\{\varepsilon_i,E_i\}_{i=1,2,3}$, using the respective optimized parameter values $A_1,A_2,A_3$, and $g$ that minimize $T_1$, we determine the time of the first local minimum of $T_1$, which we denote by $t_{\min}$.}

{In what follows, we first discuss the distribution of $T_1^{\mathrm{opt}}$ and then that of $t_{\min}$. }

\begin{table}[h!]
\centering
\begin{tabular}{|c|c|}
\hline
parameter & range \\ 
\hline
$\varepsilon_1$ & $[1 , 3]$ \\ 
\hline
$\varepsilon_2$ & $[3, 5]$ \\ 
\hline
$\varepsilon_3$ & $[1, 3]$ \\ 
\hline
$E_1$ & $[3, 5]$ \\ 
\hline
$E_2$ & $[5, 10]$ \\ 
\hline
$E_3$ & $[3, 5]$ \\ 
\hline
\end{tabular}
\caption{{Ranges of parameters $\{ \varepsilon_i, E_i \}_{i=1,2,3}$ chosen randomly from uniform distributions over respective ranges}}
\label{tab:ranges-random-energies}
\end{table}

{\subsection{Distribution of \texorpdfstring{$\bm{T_1}^{\textbf{opt}}$}{T1opt}}}

\begin{figure}[ht]
    \centering
    \includegraphics[width=0.45\textwidth]{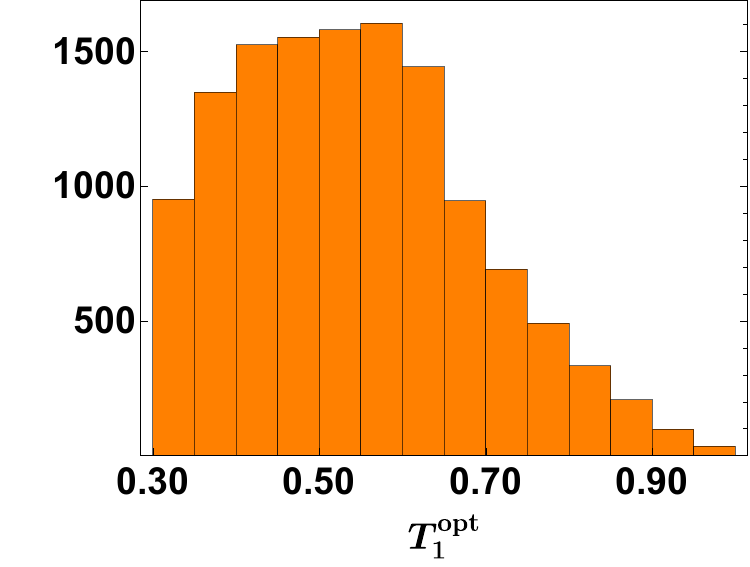}
    \caption{{Histogram of the optimized cold-qubit temperature \(T_1^{\rm opt}\) obtained from \(12800\) randomly sampled sets of system and bath energies satisfying the autonomous-refrigeration condition. The initial cold-qubit temperature is \(T_1^{\rm ini}=1\), and the sampled values lie in the interval \(T_1^{\rm opt}\in[0.3017,0.9940]\).}}
    \label{fig:T1opthisto}
\end{figure}

{We begin with the quantity of primary interest, namely the minimum temperature reached by the cold qubit. The histogram of $T_1^{\mathrm{opt}}$ for a sample of size $12800$ is shown in Fig.~\ref{fig:T1opthisto}. The range of the data is}
\begin{equation}
{T_1^{\mathrm{opt}}\in [0.3017,\,0.9940].}
\end{equation}
{Since the initial temperature of the cold qubit is $T_1^{\mathrm{ini}}=1$, the fact that the entire sampled distribution lies below $1$ shows that cooling of the cold qubit occurs for all randomly chosen {model}-parameter sets in the sample. Moreover, {the distribution is concentrated around $0.5$, with the peak occurring near $T_1^{\mathrm{opt}}\approx 0.5$.} This is close to the value $T_1^{\mathrm{opt}}\approx 0.457$ obtained in the main text for the representative choice $\varepsilon_1=1$, $\varepsilon_2=2$, $\varepsilon_3=1$, $E_1=2$, $E_2=4$, $E_3=2$, and $N=25$ (see Fig.~\ref{fig:T1opt-N-scaling} and Table~\ref{tab:optimizer}).  Therefore, the occurrence of substantial transient cooling is robust over a broad region of parameter space.} 

{In order to assess the dependence of $T_1^{\mathrm{opt}}$ on the model parameters, we analyze the distribution of $T_1^{\mathrm{opt}}$ obtained from ensembles of randomly generated parameter sets $\{\varepsilon_i, E_i\}_{i=1,2,3}$.
We first note that the distribution converges with increasing sample size. We quantify this convergence by means of the Kolmogorov--Smirnov (KS) test. The KS distance between two distributions $i$ and $j$ of a continuous random variable $x$ is defined as}
\begin{equation}
D_{ij}:=\max_x |F_i(x)-F_j(x)|,
\end{equation}
{where $F_{i(j)}$ is the cumulative distribution function (CDF) for distribution $i(j)$. We generated datasets with sample sizes $100,200,400,800,1600,3200,6400,12800$, and obtained the distribution of $T_1^{\mathrm{opt}}$ for each sample size. We then computed the KS distance between distributions for successive sample sizes and found that the KS distance progressively decreases. This is tabulated in Table~\ref{tab:KS-distance-convergence} below.}

\begin{table}[h!]
\centering
\begin{tabular}{|c|c|c|}
\hline
$i$ & $j$ & $D_{ij}$ \\ 
\hline
100 & 200 & 0.07 \\ 
\hline
200 & 400 &  0.0575\\ 
\hline
400 & 800 & 0.04 \\ 
\hline
800 & 1600 &  0.0375\\ 
\hline
1600 & 3200 &  0.01625\\ 
\hline
3200 & 6400 &  0.01610\\ 
\hline
6400 & 12800 &  0.007110\\ 
\hline
\end{tabular}
\caption{{KS distance between $T_1^{\text{opt}}$ distributions for various sample sizes. The first two columns denote sample sizes, whereas the third column denotes the KS distance $D_{ij}$ between the $T_1^{\text{opt}}$ distributions for sample sizes $i$ and $j$.}}
\label{tab:KS-distance-convergence}
\end{table}

{For completeness, we next characterize the shape of the  $T_1^{\mathrm{opt}}$ distribution. We find that both the beta distribution and the Kumaraswamy distribution describe the distribution of $T_1^{\mathrm{opt}}$ reasonably well, with the beta fit performing slightly better than the Kumaraswamy fit. We demonstrate this for the sample of size $12800$ shown in Fig.~\ref{fig:T1opthisto}.}

\begin{figure}[ht]
    \centering
    \includegraphics[width=0.45\textwidth]{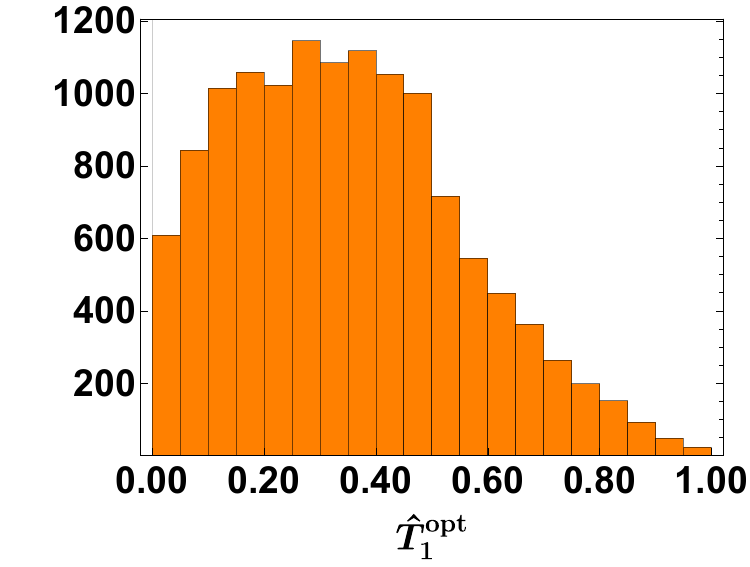}
    \caption{{Histogram of the rescaled optimized cold-qubit temperature \(\hat{T}_1^{\rm opt}\), defined in Eq.~\eqref{eq:T1opthat-defn}, for the same sample of \(12800\) parameter sets used in Fig.~\ref{fig:T1opthisto}.}}
    \label{fig:T1optHathisto}
\end{figure}

{We first scale the raw data of $T_1^{\mathrm{opt}}$, which lies in the interval $[0.3017,0.9940]$, as}
\begin{equation}
{\tilde{T}_1^{\mathrm{opt}}=\frac{T_1^{\mathrm{opt}}-0.3017}{0.9940-0.3017}.}
\end{equation}
{As a result, the scaled data $\tilde{T}_1^{\mathrm{opt}}$ lies in the interval $[0,1]$. Since the beta and Kumaraswamy distributions are defined on the open interval $(0,1)$, exact sample values at $0$ or $1$ can cause difficulties in likelihood-based fitting. To avoid this, we replace the scaled data $\tilde{T}_1^{\mathrm{opt}}\in[0,1]$ by~\cite{SV2006}}
\begin{equation}\label{eq:T1opthat-defn}
{\hat{T}_1^{\mathrm{opt}}=
\frac{\tilde{T}_1^{\mathrm{opt}}(n-1)+1/2}{n},}
\end{equation}
{where $n=12800$ is the sample size. This transformation maps the closed interval $[0,1]$ into the open interval $(0,1)$, sending $0$ to $1/(2n)$ and $1$ to $1-1/(2n)$, while perturbing all data points only by $\mathcal{O}(1/n)$. For our sample size, this correction is numerically negligible and does not materially affect the inferred distributional shape. The distribution of $\hat{T}_1^{\mathrm{opt}}$ is plotted in Fig.~\ref{fig:T1optHathisto}.}

\subsubsection{{Beta distribution fit}}

\begin{figure*}
    \centering
    \subfloat[]{\includegraphics[width=0.32\textwidth]{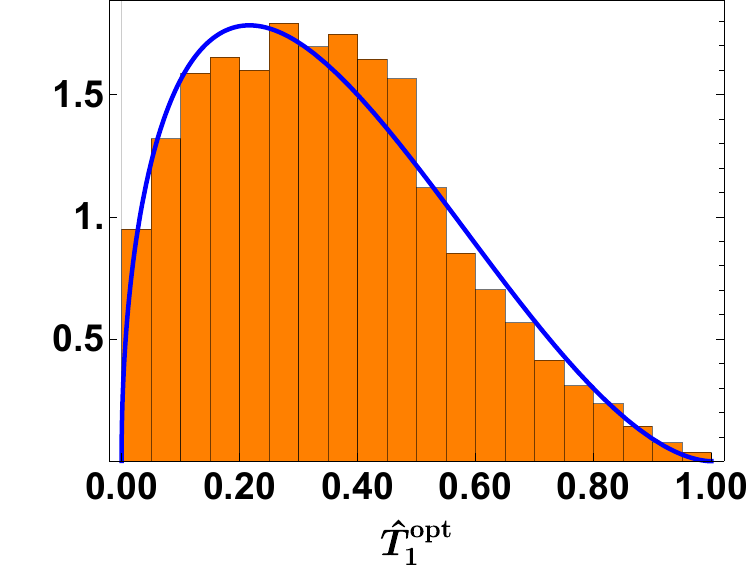} \label{fig:T1optHat-pdf-cdf-QQ-beta:pdf}}
    \hfill
    \subfloat[]{\includegraphics[width=0.32\textwidth]{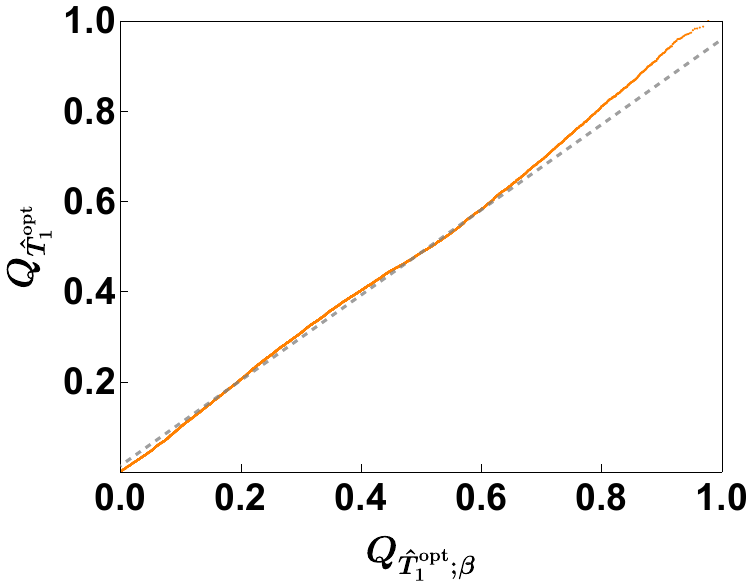} \label{fig:T1optHat-pdf-cdf-QQ-beta:QQ}}
    \hfill
    \subfloat[]{\includegraphics[width=0.32\textwidth]{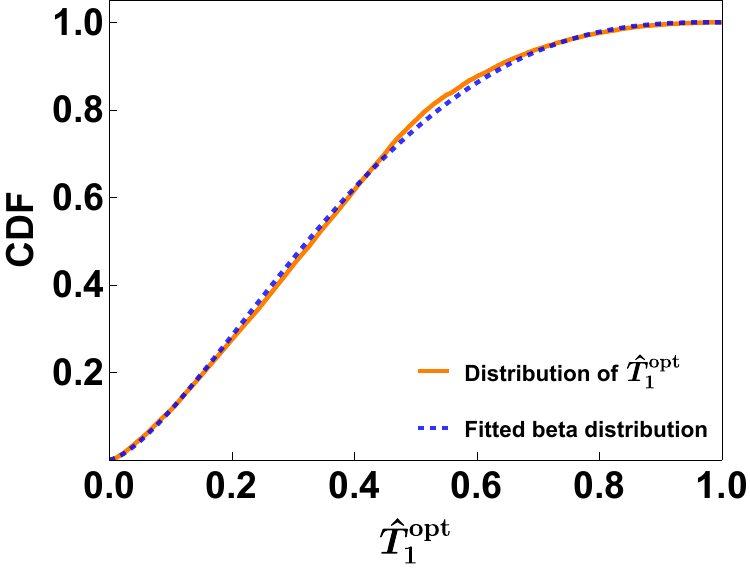} \label{fig:T1optHat-pdf-cdf-QQ-beta:cdf}}
    \caption{\justifying {Panel (a): Normalized histogram of \(\hat{T}_1^{\rm opt}\) together with the fitted beta distribution \(\mathcal{B}_{1.493,2.783}(\hat{T}_1^{\rm opt})\). Panel (b): Q-Q plot comparing the  quantiles of \(\hat{T}_1^{\rm opt}\) distribution denoted by $Q_{\hat{T}_1^{\rm opt}}$ with the quantiles of the fitted beta distribution \(\mathcal{B}_{1.493,2.783}(\hat{T}_1^{\rm opt})\) denoted by $Q_{\hat{T}_1^{\rm opt};\beta}$. The dashed diagonal is a reference line denoting perfect agreement between the actual and fitted quantiles. Panel (c): CDF of \(\hat{T}_1^{\rm opt}\) distribution together with the CDF of the fitted beta distribution \(\mathcal{B}_{1.493,2.783}(\hat{T}_1^{\rm opt})\).}
    }
    \label{fig:T1optHat-pdf-cdf-QQ-beta}
\end{figure*}

{The beta distribution for a continuous random variable $x$ in the range $0<x<1$ is given by}
\begin{equation}
{\mathcal{B}_{\alpha,\beta}(x)=\frac{x^{\alpha-1}(1-x)^{\beta-1}}{B(\alpha,\beta)},}
\end{equation}
{where $B(\alpha,\beta)$ is a normalization constant and $\alpha>0$ and $\beta>0$ are two real parameters of the distribution. Fitting the distribution of $\hat{T}_1^{\mathrm{opt}}$ defined in Eq.~\eqref{eq:T1opthat-defn} with a beta distribution gives the fit $\mathcal{B}_{1.493,2.783}(\hat{T}_1^{\mathrm{opt}})$, i.e., with best-fit values of parameters as $\alpha = 1.493$ and $\beta = 2.783$. The corresponding probability distribution function (PDF) overlay plot is shown in Fig.~\ref{fig:T1optHat-pdf-cdf-QQ-beta:pdf}, the quantile–quantile (Q-Q)~\cite{WilkGnanadesikan1968} plot in Fig.~\ref{fig:T1optHat-pdf-cdf-QQ-beta:QQ}, and the CDF overlay plot in Fig.~\ref{fig:T1optHat-pdf-cdf-QQ-beta:cdf}. The Q-Q plot is close to a straight line passing through the origin, which means that the beta distribution provides an overall good fit for the data. In particular, the fit is very good in the range $(0,0.65)$, with mild deviations appearing only near the tail for $\hat{T}_1^{\mathrm{opt}}>0.65$.}

\begin{figure}[ht]
    \centering
    \includegraphics[width=0.45\textwidth]{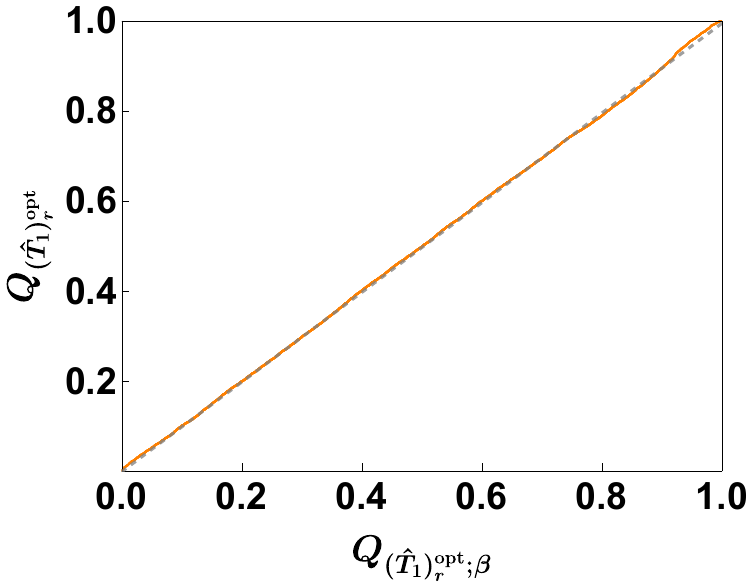}
    \caption{{Q-Q plot comparing the quantiles of the reduced bulk data \((\hat{T}_1)^{\rm opt}_{r}\) distribution denoted by $Q_{(\hat{T}_1)^{\rm opt}_{r}}$ with the quantiles of the fitted beta distribution \(\mathcal{B}_{1.313,1.334}((\hat{T}_1)^{\rm opt}_{r})\) denoted by $Q_{(\hat{T}_1)^{\rm opt}_{r};\beta}$. The dashed diagonal is a reference line denoting perfect agreement between the actual and fitted quantiles.}}
    \label{fig:QQplot-T1roptHat-Reduced}
\end{figure}

{Motivated by this observation, we also perform goodness-of-fit (GOF) tests for a reduced dataset describing the bulk of the distribution. To be conservative, we take the reduced data in the range $(0,0.6)$, which has sample size $11212$ and minimum and maximum values $0.00003906$ and $0.5999$, respectively. We rescale this reduced data from the interval $[0.00003906,0.5999]$ to $[0,1]$ as}
\begin{equation}
{(\tilde{T}_1)^{\mathrm{opt}}_r=
\frac{\hat{T}_1^{\mathrm{opt}}-0.00003906}{0.5999-0.00003906},}
\end{equation}
{and subsequently apply the transformation}
\begin{equation}\label{eq:THat1Reduced-defn}
{(\hat{T}_1)^{\mathrm{opt}}_r=
\frac{(\tilde{T}_1)^{\mathrm{opt}}_r(\tilde{n}-1)+1/2}{\tilde{n}},}
\end{equation}
{where $\tilde{n}=11212$ is the reduced sample size. This reduced data $(\hat{T}_1)^{\mathrm{opt}}_r$ is then fitted with the beta distribution $\mathcal{B}_{1.313,1.334}((\hat{T}_1)^{\mathrm{opt}}_r)$, i.e., by obtaining the best-fit parameter values as $\alpha = 1.313$ and $\beta = 1.334$. The Mathematica routine called DistributionFitTest[\textit{data},\textit{dist}] performs a GOF hypothesis test with null hypothesis $H_0$ that \textit{data} was drawn from a population with distribution \textit{dist} and alternative hypothesis $H_a$ that it was not; {see Refs.}~\cite{DAgostinoStephens1986,Stephens1974} {for standard discussions of GOF hypothesis tests.} A small probability value or p-value suggests that it is unlikely that the \textit{data} came from \textit{dist}. For a test for goodness-of-fit (GOF), a cutoff $k$ is chosen such that $H_0$ is rejected only if $p < k$. By default, $k$ is set to 0.05 in Mathematica. In the Table~\ref{tab:GOF-tests-epsScaled-tilde-beta} we report both the GOF test statistics and $p$-values for the Cram\'er--von Mises, Anderson--Darling, and Kolmogorov--Smirnov tests. We observe that the $p$-values are above $0.05$ for all three tests. {The corresponding Q-Q plot is displayed in Fig.~\ref{fig:QQplot-T1roptHat-Reduced}.}
}

\begin{table}[h!]
\centering
\begin{tabular}{|c|c|c|}
\hline
Test & Stastistic & p-value \\ 
\hline
Cram\'er-von Mises & 0.20391 & 0.260023 \\ 
\hline
Anderson-Darling & 1.77846 &  0.122006\\ 
\hline
Kolmogorov-Smirnov & 0.0143932 & 0.0646514 \\
\hline
\end{tabular}
\caption{{GOF test statistic and GOF p-value for various GOF hypothesis tests for fitting a beta distribution to the reduced data $(\hat{T}_1)_r^{\text{opt}}$ defined in Eq.~\eqref{eq:THat1Reduced-defn}.}}
\label{tab:GOF-tests-epsScaled-tilde-beta}
\end{table}

\subsubsection{{Kumaraswamy distribution fit}}

\begin{figure*}
    \centering
    \subfloat[]{\includegraphics[width=0.32\textwidth]{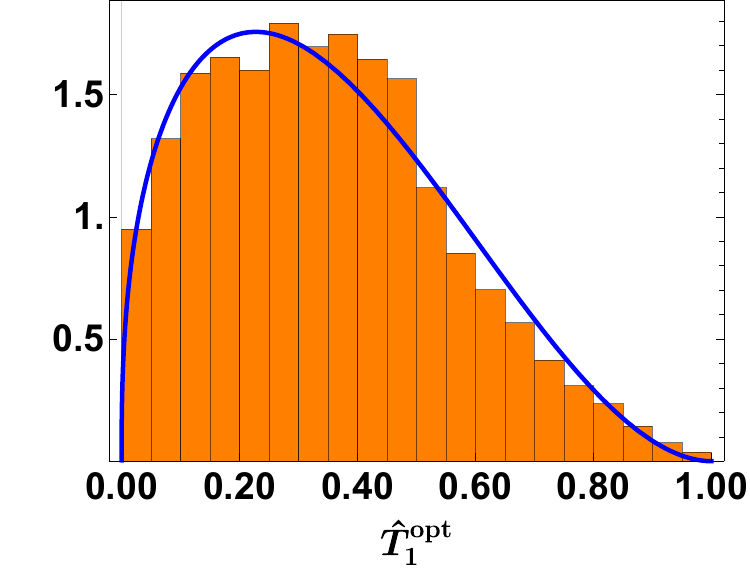} \label{fig:T1optHat-pdf-cdf-QQ-kumar:pdf}}
    \hfill
    \subfloat[]{\includegraphics[width=0.32\textwidth]{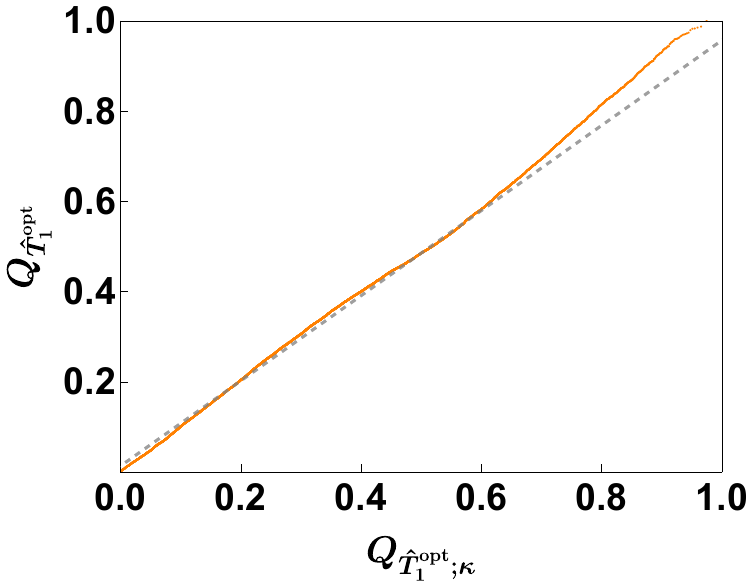} \label{fig:T1optHat-pdf-cdf-QQ-kumar:QQ}}
    \hfill
    \subfloat[]{\includegraphics[width=0.32\textwidth]{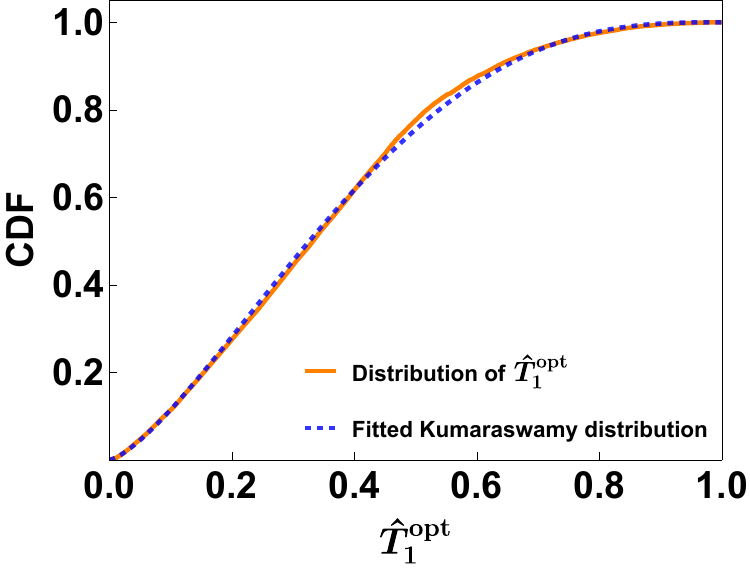} \label{fig:T1optHat-pdf-cdf-QQ-kumar:cdf}}
    \caption{\justifying {Panel (a): Normalized histogram of \(\hat{T}_1^{\rm opt}\) together with the fitted Kumaraswamy distribution \(\mathcal{K}_{1.393,2.934}(\hat{T}_1^{\rm opt})\). Panel (b): Q-Q plot comparing the quantiles of \(\hat{T}_1^{\rm opt}\) distribution denoted by $Q_{\hat{T}_1^{\rm opt}}$ with the quantiles of the fitted Kumaraswamy distribution \(\mathcal{K}_{1.393,2.934}(\hat{T}_1^{\rm opt})\) denoted by $Q_{\hat{T}_1^{\rm opt};\kappa}$. The dashed diagonal is a reference line denoting perfect agreement between the actual and fitted quantiles. Panel (c): CDF of \(\hat{T}_1^{\rm opt}\) distribution together with the CDF of the fitted Kumaraswamy distribution \(\mathcal{K}_{1.393,2.934}(\hat{T}_1^{\rm opt})\).}
    }
    \label{fig:T1optHat-pdf-cdf-QQ-kumar}
\end{figure*}

{The Kumaraswamy distribution for a continuous random variable $x$ in the range $0<x<1$ is given by}
\begin{equation}
{\mathcal{K}_{\alpha,\beta}(x)=\frac{x^{\alpha-1}(1-x^\alpha)^{\beta-1}}{\mathbb{K}(\alpha,\beta)},}
\end{equation}
{where $\mathbb{K}(\alpha,\beta)$ is a normalization constant and $\alpha>0$ and $\beta>0$ are two real parameters of the distribution. Fitting the distribution of $\hat{T}_1^{\mathrm{opt}}$ defined in Eq.~\eqref{eq:T1opthat-defn} with a Kumaraswamy distribution gives the fit $\mathcal{K}_{1.393,2.934}(\hat{T}_1^{\mathrm{opt}})$, i.e., with the best-fit values of parameters as $\alpha =1.393$ and $\beta=2.934$. The corresponding PDF overlay plot, Q-Q plot, and CDF overlay plot are shown in Figs.~\ref{fig:T1optHat-pdf-cdf-QQ-kumar:pdf}, \ref{fig:T1optHat-pdf-cdf-QQ-kumar:QQ}, and \ref{fig:T1optHat-pdf-cdf-QQ-kumar:cdf}, respectively. As in the beta distribution fit case, the Q-Q plot is close to a straight line through the origin for most of the data, with slight deviations appearing near the tail.}

\begin{figure}[ht]
    \centering
    \includegraphics[width=0.45\textwidth]{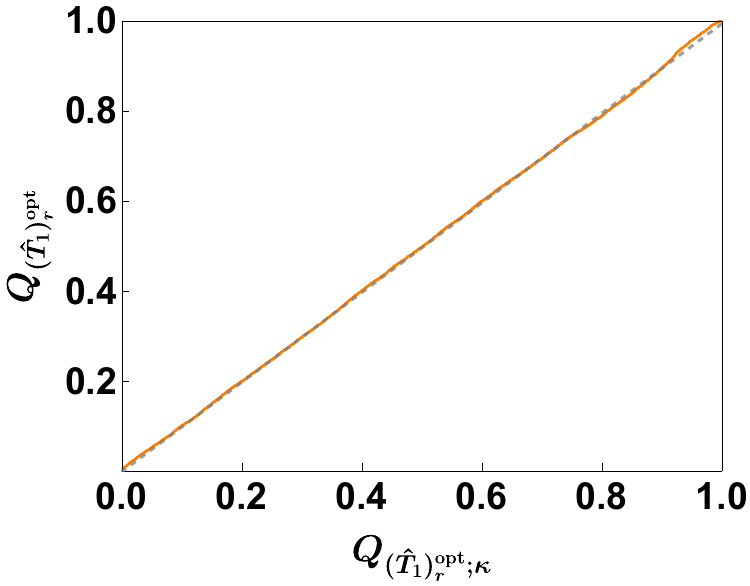}
    \caption{{Q-Q plot comparing the quantiles of the reduced bulk data \((\hat{T}_1)^{\rm opt}_{r}\) distribution denoted by $Q_{(\hat{T}_1)^{\rm opt}_{r}}$ with the quantiles of the fitted Kumaraswamy distribution \(\mathcal{K}_{1.292,1.340}((\hat{T}_1)^{\rm opt}_{r})\) denoted by $Q_{(\hat{T}_1)^{\rm opt}_{r};\kappa}$. The dashed diagonal is a reference line denoting perfect agreement between the actual and fitted quantiles.} }
    \label{fig:QQplot-T1roptHat-Reduced-Kumar}
\end{figure}

{We furthermore perform GOF tests for the Kumaraswamy distribution fitted to the reduced bulk data $(\hat{T}_1)^{\mathrm{opt}}_r$ defined in Eq.~\eqref{eq:THat1Reduced-defn}. Fitting this reduced data with a Kumaraswamy distribution gives $\mathcal{K}_{1.292,1.340}((\hat{T}_1)^{\mathrm{opt}}_r)$. The resulting GOF test statistics and $p$-values are listed in Table~\ref{tab:GOF-tests-epsScaled-tilde-kumar}. We observe that the $p$-values are above $0.05$ for the Cram\'er--von Mises, Anderson--Darling, and Kolmogorov--Smirnov tests. {The corresponding Q-Q plot is displayed in Fig.\ref{fig:QQplot-T1roptHat-Reduced-Kumar}.}}

\begin{table}[h!]
\centering
\begin{tabular}{|c|c|c|}
\hline
Test & Stastistic & p-value \\ 
\hline
Cram\'er-von Mises & 0.216145 & 0.238261 \\ 
\hline
Anderson-Darling & 2.16308 &  0.0748494\\ 
\hline
Kolmogorov-Smirnov & 0.0118321 & 0.0866184 \\
\hline
\end{tabular}
\caption{{GOF test statistic and GOF p-value for various GOF hypothesis tests for fitting a Kumaraswamy distribution to the reduced data $(\hat{T}_1)_r^{\text{opt}}$ defined in Eq.~\eqref{eq:THat1Reduced-defn}.}}
\label{tab:GOF-tests-epsScaled-tilde-kumar}
\end{table}

{Finally, we compare the beta and the Kumaraswamy fits for $\hat{T}_1^{\mathrm{opt}}$ using (i) the log-likelihood, (ii) the Akaike information criterion (AIC), and (iii) the Bayesian information criterion (BIC). Larger log-likelihood indicates a better fit, whereas smaller AIC/BIC indicates a better fit. The results are shown in Table~\ref{tab:beta-vs-kumar}. We conclude that both fits are comparable, with the beta distribution performing marginally better than the Kumaraswamy distribution.}

\begin{figure}[ht]
    \centering
    \includegraphics[width=0.45\textwidth]{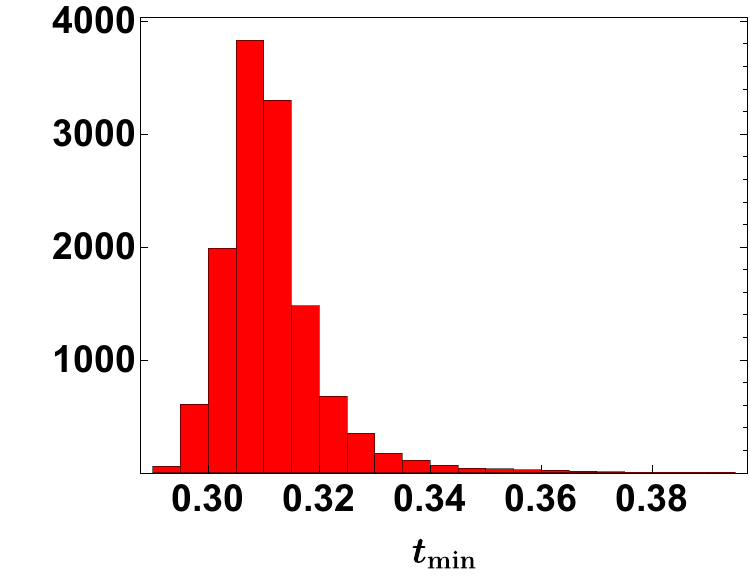}
    \caption{{Histogram of the first-local-minimum time \(t_{\min}\) obtained from the same \(12800\) randomly sampled parameter sets used in Fig.~\ref{fig:T1opthisto}. The sampled values lie in the interval \(t_{\min}\in[0.292,0.393]\).}}
    \label{fig:tminhisto}
\end{figure}

{The above fitting analysis is included to characterize the shape of the sampled distribution; the main physical conclusion is the robust occurrence of transient cooling across the sampled parameter region.}

\begin{table}[h!]
\centering
\begin{tabular}{|c|c|c|}
\hline
Test & beta  & Kumaraswamy \\ 
\hline
Log-likelihood & $3.038 \, \times \, 10^3$ & $3.036 \, \times \, 10^3$\\ 
\hline
AIC & $-6.072 \, \times \, 10^3$ &  $-6.068 \, \times \, 10^3$\\ 
\hline
BIC & $-6.057 \, \times \, 10^3$ & $-6.053 \, \times \, 10^3$ \\
\hline
\end{tabular}
\caption{{Comparsion of beta and Kumaraswamy distribution fits for data of $\hat{T}_1^{\text{opt}}$. Both the beta and Kumaraswamy fits are comprable with beta distribution fit performing marginally better than Kumaraswamy distribution fit.}}
\label{tab:beta-vs-kumar}
\end{table}

{\subsection{Distribution of \texorpdfstring{$\bm{t}_{\textbf{min}}$}{tmin}}}

{We now turn to the minimum time required to attain the first local minimum of the cold-qubit temperature for the optimized values of parameters $A_1$, $A_2$, $A_3$, and $g$. The histogram of $t_{\min}$ for a sample of size $12800$ is shown in Fig.~\ref{fig:tminhisto}. The range of the data is}
\begin{equation}
{t_{\min}\in [0.292,\,0.393].}
\end{equation}
{Thus, although the refrigeration minimum is attained in the transient regime and therefore depends on the evolution time, the corresponding time scale varies only within a comparatively narrow interval over the sampled parameter space. This indicates that, within the sampled ranges, the cooling time is only weakly sensitive to the system and bath energies.}

{As in the case of $T_1^{\mathrm{opt}}$, we first verify convergence of the sampled distribution with increasing sample size. We generated datasets with sample sizes $100, \, 200,  \, 400,  \, 800,  \, 1600,  \, 3200,  \, 6400,  \, 12800$, obtained the distribution of $t_{\min}$ for each sample size, and computed the KS distance between $t_{\min}$ distributions for successive sample sizes. The resulting distances are listed in Table~\ref{tab:KS-distance-convergence-tmin}, and show a progressive decrease with increasing sample size.}

\begin{table}[h!]
\centering
\begin{tabular}{|c|c|c|}
\hline
$i$ & $j$ & $D_{ij}$ \\ 
\hline
100 & 200 &  0.07\\ 
\hline
200 & 400 &  0.055\\ 
\hline
400 & 800 &  0.0375\\ 
\hline
800 & 1600 &  0.04268\\ 
\hline
1600 & 3200 &  0.02384\\ 
\hline
3200 & 6400 &  0.01891\\ 
\hline
6400 & 12800 &  0.008329\\ 
\hline
\end{tabular}
\caption{{KS distance between $t_{\text{min}}$ distributions for various sample sizes. The first two columns denote sample sizes, whereas the third column denotes the KS distance $D_{ij}$ between the $t_{\text{min}}$ distributions for sample sizes $i$ and $j$.}}
\label{tab:KS-distance-convergence-tmin}
\end{table}

\begin{figure*}
    \centering
    \subfloat[]{\includegraphics[width=0.32\textwidth]{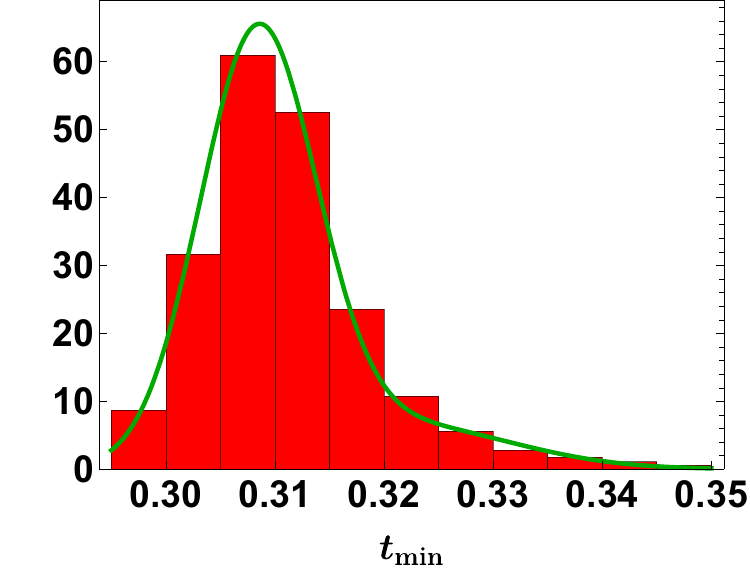} \label{fig:tmin-bulk-pdf-cdf-QQ-two-normal-mix:pdf}}
    \hfill
    \subfloat[]{\includegraphics[width=0.32\textwidth]{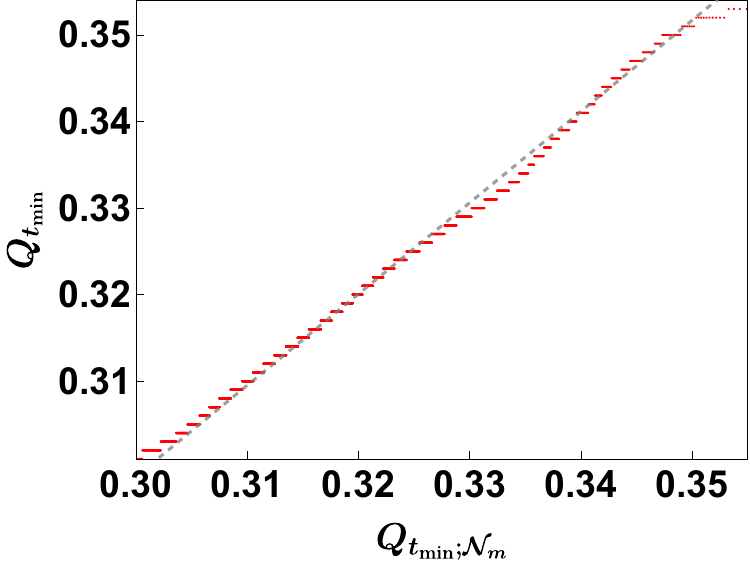} \label{fig:tmin-bulk-pdf-cdf-QQ-two-normal-mix:QQ}}
    \hfill
    \subfloat[]{\includegraphics[width=0.32\textwidth]{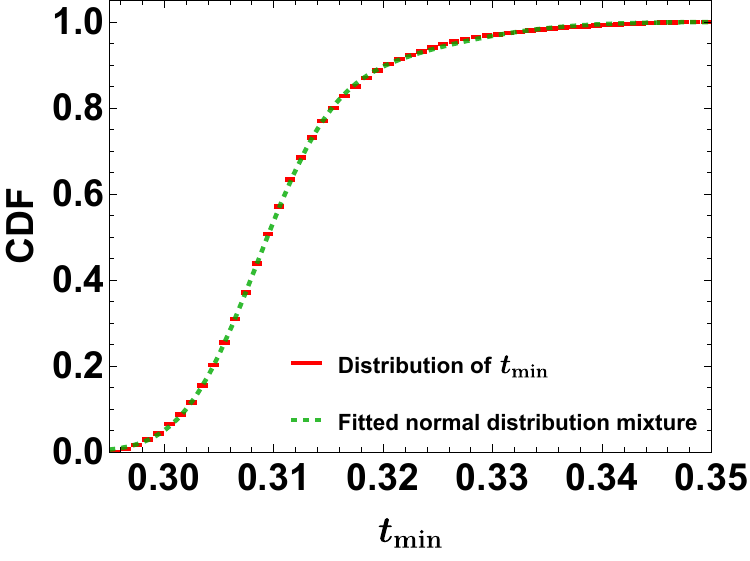} \label{fig:tmin-bulk-pdf-cdf-QQ-two-normal-mix:cdf}}
    \caption{\justifying {Panel (a): Normalized histogram of the bulk of the \(t_{\min}\) distribution in the interval \([0.295,0.350]\), together with the fitted two-normal mixture $\mathcal{N}_m$ defined in Eq.~\eqref{eq:two_normal}. Panel (b): Q-Q plot comparing the quantiles of the bulk of \(t_{\min}\) distribution in the interval \([0.295,0.350]\) denoted by $Q_{t_{\min}}$, with the quantiles of the fitted two-normal mixture $\mathcal{N}_m$ denoted by $Q_{t_{\min}; \mathcal{N}_m}$. The dashed diagonal is a reference line denoting perfect agreemen:t between the actual and fitted quantiles. Panel (c): CDF of the bulk \(t_{\min}\) data in the interval \([0.295,0.350]\), together with the CDF of the fitted two-normal mixture $\mathcal{N}_m$.}
    }
    \label{fig:tmin-bulk-pdf-cdf-QQ-two-normal-mix}
\end{figure*}

{For completeness, we also characterize the shape of the $t_{\min}$ distribution. We find that the bulk of the $t_{\min}$ distribution can be well approximated by a mixture of two normal distributions. In particular, for the bulk of the data in the range $[0.295,0.350]$, we fit the distribution with}
\begin{equation}\label{eq:two_normal}
{\mathcal{N}_m=w \, \mathcal{N}(\mu_1,\sigma_1)+(1-w) \, \mathcal{N}(\mu_2,\sigma_2),}
\end{equation}
{with fitted parameter values,}
\begin{align}
{w}&{=0.8840,\quad \mu_1=0.3088,\quad \sigma_1=0.005909,} \nonumber \\
{\mu_2}&{=0.3263,\quad \sigma_2=0.01565.}
\end{align}
{Here $\mathcal{N}(\mu_i,\sigma_i), \;  i=1,2$ denotes normal distribution with mean $\mu_i$ and standard deviation $\sigma_i$, respectively. The PDF overlay plot displaying together the normalized histogram of the bulk of the $t_{\min}$ distribution in the range $[0.295,0.350]$ along with the fitted distribution $\mathcal{N}_m$ is shown in Fig.~\ref{fig:tmin-bulk-pdf-cdf-QQ-two-normal-mix:pdf}. The corresponding Q-Q plot is shown in Fig.~\ref{fig:tmin-bulk-pdf-cdf-QQ-two-normal-mix:QQ}, and the CDFs of both the fitted $\mathcal{N}_m$ distribution and the distribution of the bulk of $t_{\min}$ in the range $[0.295,0.350]$ are shown in Fig.~\ref{fig:tmin-bulk-pdf-cdf-QQ-two-normal-mix:cdf}. These figures confirm that the bulk of the $t_{\min}$ data is well described by the fitted two-normal mixture.}

{The sampled data indicates that the CSQAR does not require significant fine tuning in order to exhibit transient refrigeration within the parameter regime considered here. Cooling of the cold qubit is observed throughout the sampled ensemble, while the corresponding first local-minimum time $t_{\min}$ remains confined to a comparatively narrow interval. Therefore, although the precise optimal cooling is quantitatively dependent on the specific parameter choice, the emergence of substantial transient cooling remains robust under broad variations of the system and bath energies.}

{\section{Distribution of the restricted BLP quantifier}
\label{app:blp-distribution}}

\begin{figure*}
    \centering
    \subfloat[]{\includegraphics[width=0.32\textwidth]{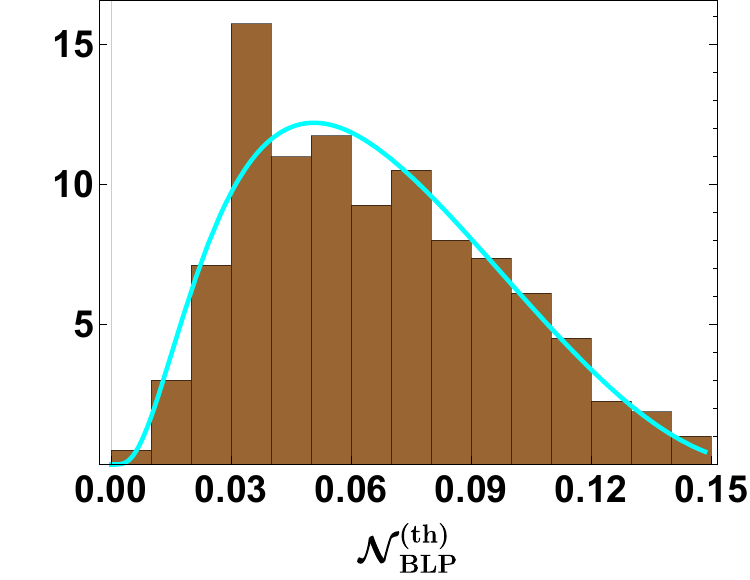} \label{fig:BLP-pdf-cdf-QQ-Johnson-SB:pdf}}
    \hfill
    \subfloat[]{\includegraphics[width=0.32\textwidth]{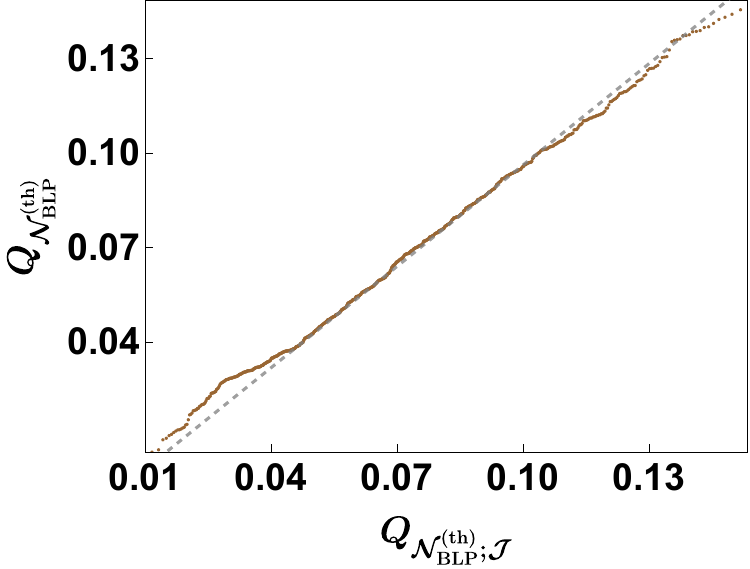} \label{fig:BLP-pdf-cdf-QQ-Johnson-SB:QQ}}
    \hfill
    \subfloat[]{\includegraphics[width=0.32\textwidth]{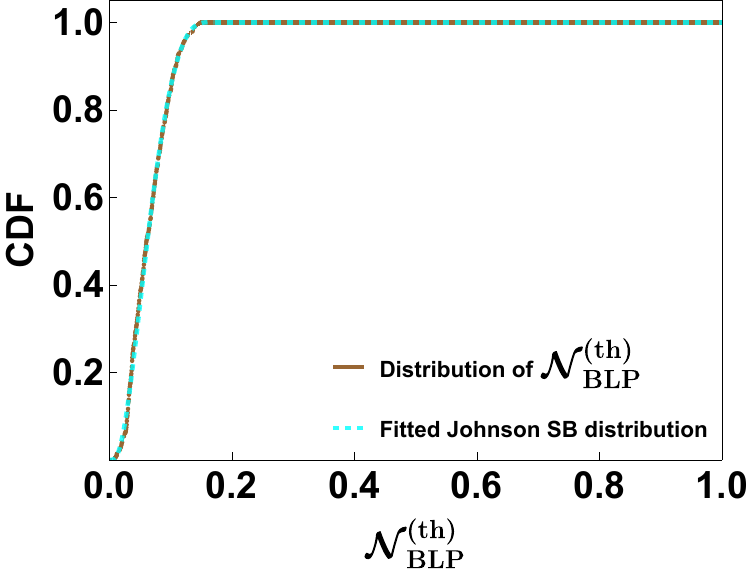} \label{fig:BLP-pdf-cdf-QQ-Johnson-SB:cdf}}
    \caption{\justifying {Panel (a): Normalized histogram of the restricted BLP quantifier $\mathcal{N}_{\mathrm{BLP}}^{(\mathrm{th})}$ for a sample of \(800\) randomly generated parameter sets, together with the fitted Johnson \(S_B\) distribution. The fitted parameters are \(\gamma=0.6610\), \(\delta=1.158\), \(\mu=0.0003671\), and \(\sigma=0.1713\). Panel (b): Q-Q plot comparing the quantiles of $\mathcal{N}_{\mathrm{BLP}}^{(\mathrm{th})}$ distribution denoted as $Q_{\mathcal{N}_{\mathrm{BLP}}^{(\mathrm{th})}}$ with the quantiles of the fitted Johnson \(S_B\) distribution denoted by $Q_{\mathcal{N}_{\mathrm{BLP}}^{(\mathrm{th})};\mathcal{J}}$. The dashed diagonal is a reference line denoting perfect agreement between the actual and fitted quantiles. Panel (c): CDF of $\mathcal{N}_{\mathrm{BLP}}^{(\mathrm{th})}$ distribution together with the CDF of the fitted Johnson \(S_B\) distribution.}
    }
    \label{fig:BLP-pdf-cdf-QQ-Johnson-SB}
\end{figure*}

{In this Appendix, we collect the details of the distributional analysis of the restricted BLP quantifier $\mathcal{N}_{\mathrm{BLP}}^{(\mathrm{th})}$ introduced in Sec.~\ref{sec:non-Markov}. The histogram of $\mathcal{N}_{\mathrm{BLP}}^{(\mathrm{th})}$ for a sample of size $800$ is shown in Fig.~\ref{fig:BLP_histo} in the main text. For this sample, the range of $\mathcal{N}_{\mathrm{BLP}}^{(\mathrm{th})}$ is}
\begin{equation}
{\mathcal{N}_{\mathrm{BLP}}^{(\mathrm{th})}\in[0.005002,\,0.1484].}
\end{equation}

{We find that the distribution of $\mathcal{N}_{\mathrm{BLP}}^{(\mathrm{th})}$ is reasonably well approximated by a Johnson $S_B$ distribution. For a continuous random variable $x$, the Johnson $S_B$ distribution is defined by}
\begin{equation}
{\mathcal{J}(x)=\mu+\frac{\sigma}{1+\exp\!\left[-\dfrac{X-\gamma}{\delta}\right]},}
\label{eq:Johnson-SB-dist}
\end{equation}
{where $X$ is normally distributed, $\gamma,\mu\in\mathbb{R}$, and $\delta,\sigma\in\mathbb{R}^{+}$ are parameters of the distribution.}

{Fitting the $\mathcal{N}_{\mathrm{BLP}}^{(\mathrm{th})}$ data to a Johnson $S_B$ distribution yields the best-fit parameter values,}
\begin{equation}
{\gamma=0.6610,\quad
\delta=1.158,\quad
\mu=0.0003671,\quad
\sigma=0.1713.}
\end{equation}

{The corresponding PDF overlay plot is shown in Fig.~\ref{fig:BLP-pdf-cdf-QQ-Johnson-SB:pdf}, where the normalized histogram of the actual $\mathcal{N}_{\mathrm{BLP}}^{(\mathrm{th})}$ distribution is displayed together with the fitted Johnson $S_B$ curve. The Q-Q plot~\cite{WilkGnanadesikan1968} is shown in Fig.~\ref{fig:BLP-pdf-cdf-QQ-Johnson-SB:QQ}. It is close to a straight line through the origin, indicating that the Johnson $S_B$ {distribution} 
provides a good overall description of the data, with only mild deviations near the tail. The CDF overlay plot is shown in Fig.~\ref{fig:BLP-pdf-cdf-QQ-Johnson-SB:cdf}.}

{To further assess the goodness-of-fit (GOF), we performed standard GOF tests~\cite{DAgostinoStephens1986,Stephens1974} for the fitted Johnson $S_B$ distribution. The resulting test statistics and $p$-values are listed in Table~\ref{tab:GOF-tests-BLP-JohnsonSB}. We find that the $p$-values are greater than $0.05$ for the Cram\'er--von Mises, Anderson--Darling, and Kolmogorov--Smirnov tests. Therefore, within the present sample size and {parameter ranges explored},
the Johnson $S_B$ distribution provides a reasonable description of the sampled values of $\mathcal{N}_{\mathrm{BLP}}^{(\mathrm{th})}$.}

\begin{table}[h!]
\centering
\begin{tabular}{|c|c|c|}
\hline
Test & Stastistic & p-value \\ 
\hline
Cram\'er-von Mises & 0.1555 &  0.3732 \\ 
\hline
Anderson-Darling & 1.0911 & 0.3127 \\ 
\hline
Kolmogorov-Smirnov & 0.0336 & 0.3216 \\
\hline
\end{tabular}
\caption{{GOF test statistics and $p$-values for fitting a Johnson $S_B$ distribution to the sampled values of $\mathcal{N}_{\mathrm{BLP}}^{(\mathrm{th})}$.}}
\label{tab:GOF-tests-BLP-JohnsonSB}
\end{table}

\section{Linblad operators and decay rates}\label{app:linblad}

The Lindblad operators, corresponding to the GKSL equation presented in Eq.~(\ref{eq:qref-M-qme}) with the dissipative term as specified in Eq.~(\ref{Lindblad}), are expressed as 

\begin{subequations}\label{eq:linblad}
    \begin{align}
        L_1^{\varepsilon_1} &= | 111 \rangle \langle 0 11 |  + |1 00 \rangle \langle 000 |  \\
        L_1^{\varepsilon_1 + g} &= \frac{1}{\sqrt{2}} \left( |110 \rangle \langle +|  + | - \rangle  | 001 \rangle \right) \\
         L_1^{\varepsilon_1 - g} &= \frac{1}{\sqrt{2}} \left( |+ \rangle \langle 001 |  - | 110 \rangle  | - \rangle \right) \\
         L_2^{\varepsilon_2} &= | 110 \rangle \langle 100 |  + |011 \rangle \langle 001 |  \\
          L_2^{\varepsilon_2 + g} &= \frac{1}{\sqrt{2}} \left( |111 \rangle \langle +|  - | - \rangle  | 000 \rangle \right) \\
          L_2^{\varepsilon_2 - g} &= \frac{1}{\sqrt{2}} \left( |+ \rangle \langle 000 |  + | 111 \rangle  | - \rangle \right) \\
          L_3^{\varepsilon_3} &= | 111 \rangle \langle 110 |  + |0 01 \rangle \langle 000 |  \\
           L_3^{\varepsilon_3 + g} &= \frac{1}{\sqrt{2}} \left( |011 \rangle \langle +|  + | - \rangle  | 100 \rangle \right) \\
           L_3^{\varepsilon_3 - g} &= \frac{1}{\sqrt{2}} \left( |+ \rangle \langle 100 |  - | 011 \rangle  | - \rangle \right) 
    \end{align}
\end{subequations}

where $|+ \rangle = \frac{1}{\sqrt{2}} ( |101 \rangle  + |010 \rangle ) $ and $|- \rangle = \frac{1}{\sqrt{2}} ( |101 \rangle  - |010 \rangle ) $

\begin{subequations}\label{eq:decay-rates}
    \begin{align}
        \gamma_i(\omega^{\prime}) &= J_i(\omega^{\prime})\left[ 1 + f(\omega^{\prime}, \beta_i)  \right] \qquad \omega^\prime > 0   \\
        &= J_i(|\omega^{\prime}|) f(|\omega^{\prime}|, \beta_i)  \qquad \quad \; \;  \omega^\prime < 0
    \end{align}
\end{subequations}

\bibliography{references} 

\end{document}